\documentclass[final,3p,times,onecolumn]{elsarticle}
\usepackage{ifpdf}
\usepackage[english]{babel}
\usepackage{color}
\usepackage{graphicx}
\usepackage{amssymb}
\usepackage{amsmath}
\usepackage[normalem]{ulem}
\usepackage{bm}
\usepackage{hyperref}
\usepackage{amsxtra}
\usepackage{latexsym}
\usepackage{dsfont}
\usepackage{caption}
\usepackage{subcaption}
\journal{\textbf{Chaos, Solitons and Fractals}}

\begin{document}
\begin{frontmatter}

\title{Controlling second-order rogue matter wave and line bright soliton dynamics in 2D Bose-Einstein Condensate with higher-order interactions and gain/loss atoms}
		
\bigskip
		
\author[udla]{Cyrille Edgard Nkenfack}
\author[udla]{Olivier Tiokeng Lekeufack*}
\author[ubra]{Subramaniyan Sabari}
\author[udla]{Rene Yamapi}
\author[botsw]{Timoleon Crepin Kofane}
\address[udla]{Pure physics Laboratory: Group of nonlinear physics and complex systems, Department of Physics, Faculty of Science, University of Douala, P.O. Box 24157, Douala, Cameroon}
\address[ubra]{Instituto de F\'\i sica Te\'{o}rica, Universidade Estadual Paulista (UNESP), 01140-070 S\~{a}o Paulo, SP, Brazil}
\address[botsw]{Department of Physics and Astronomy, Botswana International University of Science and Technology, Private Mail Bag 16, Palapye, Botswana }
\date{\today}
\begin{abstract}
We investigate the two-dimensional modified Gross-Pitaevskii equation, accounting for the effects of atom gain/loss and a time-independent isotropic confining potential, utilizing the Hirota's bilinear method. Through an appropriate bilinear form, we derive exact one-soliton and multi-soliton solutions. These solutions showcase two prominent phenomena:  the second-order rogue matter wave with spatio-temporal localization, and the line soliton with double spatial localization. We demonstrate the feasibility of controlling the soliton amplitude and the effects of gain/loss resulting in areas of collapse by suitably tuning the coefficient of higher-order interactions in the Bose-Einstein condensate. Additionally, by exploring the interaction dynamics of the multi-soliton solutions, we identify elastic-type interactions, claiming the intrinsic properties of solitons. The influence of higher-order interactions and gain/loss terms on the interaction dynamics is also thoroughly analyzed. These analyses demonstrate that, within the framework of Bose-Einstein condensates described by the two-dimensional modified Gross-Pitaevskii equation, higher-order interactions provide a means to control the properties of the generated rogue matter waves.

Intensive numerical simulations are performed and their convergence with theoretical predicted results then throw light about the emergent features of the chosen solutions. The exact analytical solutions derived in this study rigorously satisfy the original equation, which ensures their consistency with the numerical results and confirms their accuracy.
 Thus, our findings hold promise for potential future applications. 

\end{abstract}
		
\bigskip
		
\end{frontmatter}
	
*Corresponding Author E-mail: \emph{\textbf{lekeufackolivier@gmail.com}} 
	
\newpage
	
\section{Introduction}
In the field of ultracold atomic physics , it has been well established, for a sample of rubidium (Rb) gas, cooled to a critical temperature close to absolute zero, that almost all particles occupy the same state of energy, forming a giant wave called Bose-Einstein condensate (BEC) \cite{Anderson1995}. The first gas condensates in history obtained by Cornell and Wieman at Boulder and Ketterle at the MIT paved the way for experiments in various states of matter, hence revolutionizing research fields such as condensed matter physics \cite{Morsch2006}, atomic optics \cite{Lenz1993} and quantum computing \cite{McKeever2004}.
The dynamics of particles within a condensate consists of solving the nonlinear Gross-Pitaevskii (GP) equation  \cite{Gross1961, Pitaevskii1961}, in mean-field theory. The nonlinearity terms present in the GP equation result from the interactomic interaction whose coefficient is proportional to the density of the condensate and the scattering length, which can be controlled experimentally by the Feshbach resonance techniques \cite{Daifovo1999,Kohler2006}. In some regimes characterised by high density and governed by an Efimov resonance, the three-body interactions are necessarily taken into account, which makes it possible to achieve the GP equation with cubic and quintic nonlinearity \cite{Gammal2000,Lekeufack2015,Chen2014,Sabari2015}. Moreover, it has been reported that  higher-order (HO) interactions are considerable when the scattering length approches zero \cite{Zinner2009}, consequently, the macroscopic quantum tunneling rate can be modified for small condensate samples. On the other hand, experiments conducted on the surface of atomic chips and in atomic waveguides lead to a high compression of the traps, including a considerable increase in the density of the condensate \cite{Ping2009,He2010}. Beyond the various types of interactions taken into account for the study of condensates nowadays, there emerges a term (gain/loss) accounting for the gain or the loss of atoms in a condensate. Moreover, Strecker et al \cite{Strecter2002} have made it a remarkable study that has inspired several researches  taking into account a form of gain/loss atoms for areas of collapse and revival in a condensate \cite{Atre2006}. Thus, for a better description of the atom-atom interaction in the dynamics of a condensate, it is necessary to consider the modified GP equation with cubic-quintic nonlinearities, including HO interactions and the gain/loss terms  \cite{Nkenfack2025}.
	
The repulsive or attractive character of interactomic interactions present in the modified GP equation can give rise to nonlinear wave models including dark \cite{Burger1999} and bright solitons \cite{Strecter2002,Liang2005,Radha2007}.
Beyond the standard NLSE, the modified form has been widely explored in binary BECs, leading to studies on bright and dark quantum droplets \cite{Petrov2016,Edmonds2023}, the supersolid phase \cite{Parit2021}, and modulation instability \cite{Mithun2020,Sasi2023,Sabari2021}. Additionally, the existence of quantum droplets has been experimentally validated by Cabrera et al. \cite{Cabrera2018}.
Solitons with remarkable properties and a field of research for the description and understanding of many phenomena in both engineering and science. The solitons born from the balance between the effects of nonlinearity and dispersion \cite{Agrawal2013}, represent an essential pole for the understanding of fields such as nonlinear optics \cite{Yu2019, Yan2021}, fluid mechanics \cite{Gu1995}, hydrodynamics\cite{Kuznetsov1986}, plasma physics \cite{Congy2021} and BECs \cite{Wang2020,Chai2020,Potter2012,Lannig2020}. Occupying a place of choice as a major contribution in physics and mathematics, the solitonic solutions resulting from nonlinear equations such as the nonlinear Schr{\"o}dinger equation \cite{Liu2008}, the GP equation \cite{Nkenfack2025,Nkenfack2024} and the complex Ginzburg-Landau equation  \cite{Wong2015} constitute a fascinating pole of study in nonlinear physics. Since BEC is a remarkable medium for controlling dark and bright solitons and their dynamics, many experiments aimed at its control have been conducted in recent decades. Moreover, Khaykovich et al \cite{Khaykovich2002}, Cornish et al \cite{Cornish2006} have experimentally observed stable bright solitons in condensates consisting of $^7Li$ and $^{85}Rb$ atoms respectively via the Feshbach resonance technique. Many types of bright solitons have therefore been observed in the context of GP equation. Among others, there is line-soliton that exhibit double spatial localisation in the study of high-dimensional condensates \cite{Radha2010}.  Moreover, understanding of the phenomenon of interaction in order to control it, was of paramount interest in the interaction processes. To implement relevant theoretical results, one constructs multi-soliton solutions whose control over certain parameters allows the observation of multi-soliton interaction processes \cite{Wan2020}. This is a major step forward in the field of telecommunications.
	
In order to match experimental facts with theoretical results, solving the GP equation is an important axis through effective and practical methods such as, the F-expansion method, Darboux transformation, and  Hirota's bilinear method (HBM), to name just a few \cite{Wang2020,Liu2008,Li2021}. Out of these analytical methods, an investigation of modified GP equation was able to highlight solitary solutions whose accuracy in some numerical methods have proven \cite{Jena2020}. Work on the one-dimensional modified GP has also proven the existence of stable soliton, which has paved the way for the use of various external trapping potentials \cite{Yu2019,Su2016}. Moreover, the study of the two-dimensional GP equation remains today a pole of attraction given the difficulty in stabilizing the solitons resulting from such systems \cite{Guo2020}. Hence the need to theoretically study the GP equation in high dimension. 
A sample of such system could be made up of ultrawide dark solitons and the coexistence of droplet-solitons have been explored in dipolar BECs \cite{Kopycinski2023}. This topic is still of great interest for further investigations.
Thus, high-dimensional researches were conducted with the aim of obtaining stable glossy solitons in two dimensions \cite{Saito2003,Sabari2017}. The purpose and motivation of the present work is therefore to solve the modified GP equation in two dimensions, taking into account the gain/loss effects and a time-independent  trapping potential. This work will allows us, by fixing one of the constant spatial coordinates as carried out by Wang et al \cite{Wang2022}, to highlight the bright soliton  amplifier of second-order rogue matter wave (RMW) types on the one hand and the other hand to highlight the existence of line-soliton by fixing the constant temporal coordinate as carried out by Radha et al \cite{Radha2010}. We will make use of the HBM in order to construct the one-soliton  and the  multi-soliton solutions for which it is necessary to establish control through the HO parameter interaction and the gain/loss term from the modified GP equation in two dimension. Furthermore, we validated our analytical results through numerical simulations of the governing equations using the split-step Fourier method \cite{Sabari2021,Mani2024,Tamil2019a,Sabari2013,Sabari2020}, with spatial step sizes of $dx = dy = 0.001$ and a temporal step size of $dt = 0.01$. We are also confident that the novelty of our findings offers a valuable contribution to the deeper understanding of various phenomena arising in high-dimensional BECs.
	
The paper is organized as follows: in Sec.\ref{sec2}, we present the theoretical model of two-dimensional modified GP equation  that describes the condensates with HO nonlinearity and gain/loss term. In Sec.\ref{sec3}, an appropriate bilinear form of the modified Gross-Pitaevskii equation will be given.
Based on the given bilinear form, we will construct the one-soliton solution in Sec.\ref{sec4} and the multi-soliton solution in Sec.\ref{sec5}, where the corresponding interaction dynamics will also be presented, supported by the necessary numerical simulations. Finally in Sec.\ref{sec6} we will give a conclusion recalling all our main results obtained.

\section{ Modified GP equation with higher-order interaction and gain/loss terms } \label{sec2}
\setcounter{equation}{0}

In the mean-field theory, the dynamics of  macroscopic wave function for a Bose-Einstein Condensates (BEC) is governed by the Gross-Pitaevskii (GP) equation with two-body interaction. In the presence of high-density BEC, comprising the effects of three-body and HO interactions, leads to the modified GP equation given by \cite{Lekeufack2015,Nkenfack2025,Tamil2019,Sabari2022,Sabari2020,Qi2012}:
	
\begin{equation*}
i\hbar \frac{\partial }{\partial \tau}\zeta(\vec{r},\tau) =-\frac{\hbar^{2}}{2m}\nabla^{2}\zeta(\vec{r},\tau)+V_{ext}(\vec{r},\tau)\zeta(\vec{r},\tau)+g_{2}(\tau)|\zeta(\vec{r},\tau)|^{2} \zeta(\vec{r},\tau)   + g_{3}(\tau) |\zeta(\vec{r},\tau)|^{4}\zeta(\vec{r},\tau)+ 
\end{equation*}
\begin{equation}
+ {\eta(\tau)}\left(\nabla^{2}|\zeta(\vec{r},\tau)|^{2} \right)\zeta(\vec{r},\tau)   \label{a}
\end{equation}
	
where $\vec{r}$ and $\tau$ are respectively the spatial and temporal coordinates,  $\hbar$ is the reduced Planck's constant, $m$ is the mass of the boson,$V_{ext}(\vec{r},\tau)=\frac{m}{2}[\omega^{2}(X_0^{2}+Y_0^{2})+\omega_{Z_0}^{2}Z_0^{2}]$ is the external potential, with  $\omega_{Z_0}$ and $\omega$ being respectively the $Z_0$  and the common $(X_0,Y_0)$ frequencies of the external trap. The coefficient $g_{2}$ is  the strength of the two-body interatomic interactions defined by $g_{2}(\tau)= 4\pi\hbar^{2}a_s(\tau)/m$, with $a_s$ being the s-wave scattering length which can be attractive ($a_s<0$) or repulsive ($a_s>0$) and $g_{3}$, the strength of the three-body interatomic interaction of the order of $a_s^2$. The parameter $\eta$ is the higher-order (HO) scattering, considered absent on $Z_0$ axis, that depends on both the s-wave scattering length  and the effective range for collisions \cite{Zinner2009,bib47,Tamil2018}. This parameter reads $\eta=g_{2}g_4$, where $g_4=a_s^2/3-a_sr_e/2$, with $r_e$ being the effective range. The macroscopic wave function of the BEC  is given by :
	
	\begin{equation}
		\zeta(\vec{r},\tau)=\psi_0(Z_0,\tau)\psi(X_0,Y_0,\tau)
	\end{equation}
	with $\psi_0$ the ground state of the longitudinal part
	\begin{equation}
		\frac{-\hbar^2}{2m}\nabla_{Z_0}^{2 }\psi_0 + \frac{m}{2}\omega_{Z_0}^{2} Z_0^{2} \psi_0 = \frac{\hbar \omega_{Z_0}}{2} \psi_0,
	\end{equation}
 where the normalized $\psi_0$ given by
$\psi_0(Z_0,\tau)= (\pi a_{\perp}^2)^{-1/4} exp (-\frac{i\omega_{Z_0}\tau}{2}-\frac{Z_0^2}{2a_{\perp}^2})$,
and $a_{\perp}=\sqrt{\frac{\hbar}{m \omega_{Z_0}}}$.
	Then, multiplying both sides of the GP Eq.(\ref{a}) by $\psi_0^{\ast}$
	and integrating over $Z_0$, we include the gain/loss rate of atoms and obtain a quasi two-dimensional  modified GP equation with time-independent harmonic potential  given by:
	
	\begin{equation}
		i \hbar\frac{\partial }{\partial \tau} \psi=\left[-\frac{\hbar^{2}}{2m}\left(\frac{\partial^{2} }{\partial X_0^{2}}+\frac{\partial^{2} }{\partial Y_0^{2}}\right)+\frac{1}{2}\omega^{2}(X_0^{2}+Y_0^{2})+iM(\tau) +g_2(\tau)\tilde{\eta}|\psi|^{2} 
		+g_{3}(\tau)\tilde{\eta_1}|\psi|^{4}	+\eta(\tau)\tilde{\eta} \left(\frac{\partial^{2} }{\partial X_0^{2}}|\psi|^{2}+\frac{\partial^{2} }{\partial Y_0^{2}}|\psi|^{2} \right)\right]\psi. 	  \label{ao}
	\end{equation}

	where $\tilde{\eta}= \int |\psi_0(Z_0,\tau)|^4 dZ_0 = \frac{1}{a_{\perp}\sqrt{2\pi}} $ and $\tilde{\eta_1}=\int |\psi_0(Z_0,\tau)|^6 dZ_0 = \frac{1}{\pi a_{\perp}^2 \sqrt{3}}$. 
	
	Here, $M$ is the term of gain/loss and can be positive(gain) or negative (loss). $M>0$ characterises a mechanism for charging the thermal cloud in the BEC by optical pomping of the atoms of the external source.  $M<0$ describes the dynamics of a condensate continuously exhausted due to the loss of atoms \cite{Atre2006}.
	
Work on the two-dimensional GP equation with two-body interactions and a time-dependent potential has been previously conducted in Refs. \cite{Wang2022} and \cite{Wu2020}, leading to exact solutions derived via HBM. Unlike the models considered in these studies, which focus solely on the effects of two-body interactions, the present framework takes into account the effects of three-body interactions, HO interactions, and atom gain/loss mechanisms. In this work, we adopt the approach used in Ref. \cite{Wang2022} to seek  for exact solutions of Eq.(\ref{ao}) using the bilinear method. Furthermore, we emphasize on the role played by HO interactions and the gain/loss term in shaping the system's dynamics. 
	
	\section{Appropriate bilinear form of the modified GP equation} \label{sec3}
	The application of the HBM requires first of all a dimensionless form of Eq.(\ref{ao}) that we achieve by using the following variable transformation : 	 
	
	\begin{equation}
		\phi(x,y,t)=\frac{\psi(X_0,Y_0,T_0)}{a_\perp \sqrt{4\pi a_0 \tilde{\eta}}}; \hspace{0.3cm}  X_0=a_{\perp}x ; \hspace{0.3cm}  Y_0=a_{\perp}y; \hspace{0.3cm}  T_0=t/\omega_{Z_0}; \hspace{0.3cm} a_{\perp}=\sqrt{\hbar/m\omega_{Z_0}} .
	\end{equation}
Here, we can choose a constant length $a_0$, which is used to measure the scattering length of s-wave with time-dependent. In terms of new variables, the 2D modified GP equation reads :

	\begin{equation*}
	i \frac{\partial }{\partial t} \phi(x,y,t)=\left[-\frac{1}{2}\left(\frac{\partial^{2} }{\partial x^{2}}+\frac{\partial^{2} }{\partial y^{2}}\right)+\frac{1}{2}K (x^{2}+y^{2}) +iM(t)+G_0(t)|\phi(x,y,t)|^{2}
	+\overline{\chi_0}(t)|\phi(x,y,t)|^{4} \right]\phi(x,y,t)
\end{equation*}
\begin{equation}
	+N_0(t)\left[ \left(\frac{\partial^{2} }{\partial x^{2}} +\frac{\partial^{2} }{\partial y^{2}}\right)|\phi(x,y,t)|^{2}  \right]\phi(x,y,t).  \label{b1}
\end{equation}
where
	\begin{equation*}
	K=\pm\omega^2/\omega_{Z_0}^2;\hspace{0.2cm}	G_0(t)=a_s(t)/a_0;\hspace{0.2cm}  \overline{\chi_0}(t)=\frac{g_3 \tilde{\eta_1}m^2 \omega_{Z_0}}{\hbar^3(4 \pi a_0\tilde{\eta})^2}; \hspace{0.2cm}		N_0(t)=\frac{\tilde{\eta} m^2\omega_{Z_0}}{\hbar^3(4 \pi a_0)}.
	\end{equation*}
The potential parameter $K$ measures the strength of the magnetic trap, where for $K<0$, we are in the presence of a confining potential, and for $K>0$, a repulsive one. The parameters $G_0$, $\overline{\chi_{0}}$ and $N_0$  are the new coefficients of interaction with two-, three-body and HO, respectively.
	
In order to switch the potential and the gain/loss terms in Eq.(\ref{b1}), we employ a transformation similar to the one earlier used in refs. \cite{Wang2022} and \cite{Wu2020}:  
	
	\begin{equation} \label{71}
		\phi(x,y,t)=\phi_0(\tilde{x},\tilde{y},t) exp\left[-\frac{i}{2}\frac{d U}{dt}(x^2+y^2)+U(t)  +\int M(t) dt\right],  
	\end{equation}
where
\begin{equation}\label{K}
	K=\frac{d^2 U}{dt^2}-\left(\frac{dU}{dt}\right)^2,
\end{equation}
with
\begin{equation}\label{K9}
	 \tilde{x}=e^{U(t)}x \hspace{0.3cm} and  \hspace{0.3cm} \tilde{y}=e^{U(t)}y.                       
\end{equation} 

 By introducing transformation (\ref{71}) and Eqs.(\ref{K}) and (\ref{K9}) into Eq.(\ref{b1}), we obtain:

	\begin{equation}
	i \frac{\partial }{\partial t} \phi_0=\left[-\frac{1}{2} e^{2U}\left(\frac{\partial^{2} }{\partial \tilde{x}^{2}}+\frac{\partial^{2} }{\partial \tilde{y}^{2}}\right)+g_{01}(t) e^{2U}|\phi_0|^{2} 
	+\chi_{01}(t) e^{4U}|\phi_0|^{4}	+\eta_{01}(\tilde{t}) e^{4U}\left(\frac{\partial^{2} }{\partial \tilde{x}^{2}} +\frac{\partial^{2} }{\partial \tilde{y}^{2}}\right)|\phi_0|^{2}\right]\phi_0. \label{b}
   \end{equation}
   where
   \begin{equation}
   		g_{01}(t)= G_{0}(t)e^{2\int M(t) dt},   \hspace{0.5cm}  \chi_{01}(t)=\overline{\chi_0}(t) e^{4\int M(t) dt}, \hspace{0.5cm} \eta_{01}(t)=N_{0}(t)e^{2\int M(t) dt}
   \end{equation}
	
Let's now rewrite Eq.(\ref{b}) according to Hirota formalism by first defining its D-operators as follows \cite{bib18}:
	\begin{equation}
		D_x^{l}D_y^{m}D_t^{n}(g \cdot f)
	=\left(\frac{\partial}{\partial x}-\frac{\partial}{\partial x_1}\right)^{l}\left(\frac{\partial}{\partial y}-\frac{\partial}{\partial y_1}\right)^{m}\left(\frac{\partial}{\partial t}-\frac{\partial}{\partial t_1}\right)^{n}g(x,y,t)f(x_1,y_1,t_1)|_{x=x_1,y=y_1,t=t_1}.
	\end{equation}
	
	Moreover, an exact solution of Eq.(\ref{b}) can then be searched in the form \cite{bib10}:
	\begin{equation}
		\phi_0(\tilde{x},\tilde{y},t)=\frac{g(\tilde{x},\tilde{y},t)}{f(\tilde{x},\tilde{y},t)}    \label{7}
	\end{equation}
	with $g(\tilde{x},\tilde{y},t)$ a complex differentiable function and $f(\tilde{x},\tilde{y},t)$ a real one. The terms of Eq.(\ref{b}) can be expressed according to the Hirota oprators as follows :
	\begin{equation}
		\frac{\partial }{\partial t} \phi_0=\frac{g_{t}f-f_{\tilde{t}}g}{f^2}=\frac{D_{t}  (g \cdot f)}{f^2}              \label{wa}
	\end{equation}
\begin{equation}
	\left(\frac{\partial^{2} }{\partial \tilde{x}^{2}}+\frac{\partial^{2} }{\partial \tilde{y}^{2}}\right) \phi_0=\frac{(D_{\tilde{x}}^{2}+D_{\tilde{y}}^{2}) (g \cdot f)}{f^2}-\frac{(D_{\tilde{x}}^{2}+D_{\tilde{y}}^{2})(f \cdot f) }{f^2}\frac{g}{f}   \label{wo}
\end{equation}
\begin{equation}
	\left(\frac{\partial^{2} }{\partial \tilde{x}^{2}} +\frac{\partial^{2} }{\partial \tilde{y}^{2}}\right)|\phi_0|^{2}=\frac{(D_{\tilde{x}}^{2}+D_{\tilde{y}}^{2})(|g|^{2}\cdot f^{2})}{f^{4}}  - 2 \frac{(D_{\tilde{x}}^{2}+D_{\tilde{y}}^{2})(f\cdot f)}{f^4}|g|^{2}.  \label{wu}
\end{equation}
	
	 Eq.(\ref{b}) yields
	 
	\begin{equation*}
		-\left[iD_{t}+\frac{1}{2}e^{2U}(D_{\tilde{x}}^{2}+D_{\tilde{y}}^{2})\right]g \cdot f+\frac{1}{2}e^{2U}( D_{\tilde{x}}^{2}+D_{\tilde{y}}^{2})(f \cdot f) \frac{g}{f} + g_{01}(t) e^{2U}|g|^2\frac{g}{f} + \chi_{01}(t)e^{4U}|g|^{4}\frac{g}{f^3}
	\end{equation*}
	\begin{equation}
         	+\eta_{01}(t) e^{4U}\left[     (D_{\tilde{x}}^{2}+D_{\tilde{y}}^{2})(|g|^{2}\cdot f^{2})\frac{1}{f^{2}}  - 2 (D_{\tilde{x}}^{2}+D_{\tilde{y}}^{2})(f\cdot f)\frac{|g|^{2}}{f^{2}}    \right]\frac{g}{f}=0	\label{p}
	\end{equation}
	
	and in appropriate bilinear form : 

		\begin{equation}\label{c}
				-\left[iD_{t}+\frac{1}{2}e^{2U}(D_{\tilde{x}}^{2}+D_{\tilde{y}}^{2})\right] (g \cdot f) =0,	
	    \end{equation}
		\begin{equation}\label{d}
			(D_{\tilde{x}}^{2}+D_{\tilde{y}}^{2})(f\cdot f)=\frac{2g_{01} f^{2} |g|^2 +2\chi_{01}e^{2U}|g|^{4}+2\eta_{01}e^{2U}(D_{\tilde{x}}^{2}+D_{\tilde{y}}^{2})(|g|^{2}\cdot f^{2})}{4\eta_{01}e^{2U}|g|^2-f^2}  
	    \end{equation}
    
For $\chi_{01}=0$ and $\eta_{01}=0$, we recover the bilinear form earlier obtained in \cite{Wang2022}.
	Thus, to solve Eqs.(\ref{c}) and (\ref{d}), we introduce an expansive power series of $g$ and $f$ :
	\begin{equation}
		g(\tilde{x}, \tilde{y}, t)=\epsilon g_1(\tilde{x}, \tilde{y}, t) +\epsilon^{3}g_3(\tilde{x}, \tilde{y},t)+\epsilon^{5}g_5(\tilde{x}, \tilde{y}, t) + ...+ \epsilon^{2n-1}g_{2n-1}(\tilde{x}, \tilde{y}, t),\hspace{0.5cm}  n=1,2,3,...
	\end{equation}
	\begin{equation}
		f(\tilde{x}, \tilde{y}, t)=1+\epsilon^{2} f_2(\tilde{x}, \tilde{y}, t) +\epsilon^{4}f_4(\tilde{x}, \tilde{y}, t)+\epsilon^{6}f_6(\tilde{x}, \tilde{y}, t) + ...+\epsilon^{2n}f_{2n}(\tilde{x}, \tilde{y},t), \hspace{0.5cm}  n=1,2,3,...
	\end{equation}
	
Next, we consider a time-independent harmonic potential analogous to that employed by Khaykovich et al \cite{Khaykovich2002} for creating bright solitons in the context of BEC. Specifically, we adopt a confining potential with the trapping parameter $K$ defined as $K=-\omega^2/\omega_{Z_0}^2=-2\kappa^2$ 
 (with $\kappa \simeq 0.05 $) for the choice of experiments where  $\omega_{Z_0}=2 \pi \times 7 Hz$ and  $\omega=2\pi \times 700 Hz$. Herein, $\omega>>\omega_{Z0}$, and $K=10^4$. We shall then use the same expression of $K$ as part of our 2D study.
	
	Thus, the choice of expressions of the parameters $M(t), G_0(t), \overline{\chi_0}(t)$ and $N_0(t)$ adopted in our work is as follows :
	\begin{equation}
		M(t)=-M_0sin(2t), \hspace{0.3cm} G_0(t)=g_{02}exp\left[-2M_0cos(2t)\right], \hspace{0.3cm} \overline{\chi_0}(t)=\chi_0exp\left[-4M_0cos(2t)\right]=\pm |\Gamma|g_{02}exp\left[-4M_0cos(2t)\right]   \label{d9}
	\end{equation}
and 
\begin{equation}
	 N_0(t)=\eta_0\left[e^{-6M_0cos(2t)}/3-e^{-4M_0cos(2t)}r_e/2\right].  \label{d10}
\end{equation}

From the work of Atre et al \cite{Atre2006} on the one hand and Strecker et al \cite{Strecter2002} more generally, the chosen form of $M(t)$ in which the linear part over time has been neglected, makes it possible to generate collapse and revival of the condensate. We hereby mainly focus on the trigonometric part that is clearly responsible for the collapse and revival effects observed in Atre et al work;  $G_0(t)$, $\overline{\chi_0}(t)$ and $N_0(t)$ are respectively the time-dependent two-, three-body and HO interaction coefficients;  $g_{02}, \Gamma, \chi_0$ and $\eta_0$ are taken in constant form. Note that through the expressions of $G_0(t)$, $\overline{\chi_0}(t)$ and $N_0(t)$ given  by Eqs.(\ref{d9}) and (\ref{d10}), we can readily get the expressions of $g_{01}(t)$, $\chi_{01}(t)$ and $\eta_{01}(t)$.   

Regarding the parameter $U(t)$, we consider in our study that it is linear over time, and owing to that, Eq.(\ref{K}) becomes :
\begin{equation}
	\left(\frac{dU}{dt}\right)^2=2\kappa^2=U_0^2.
\end{equation}

It therefore follows that:
\begin{equation}
	U(t)=U_0t + U_1
\end{equation}
where $U_0=\kappa \sqrt{2}$ and $U_1$ is a constant considered zero in the rest.

\section{One-soliton solution } \label{sec4}

In this  section,  we determine the one-soliton solution by introducing the forms of $g$ and $f$ chosen in Eqs.(\ref{c}) and (\ref{d}). Additionally, we analyze the role of gain/loss and HO interactions, on the dynamics of one-soliton solutions. Once the bright-soliton solutions are constructed, we will consider two distinct case studies. The first case involves fixing the spatial coordinate $y=1$, as demonstrated in the work of Haotian Wang et al \cite{Wang2022}, while the second one involves fixing the temporal coordinate t, as explored in the work of Radha et al \cite{Radha2010}.

\begin{figure}[!h]
	\centering
	\includegraphics[width=12cm,height=4cm]{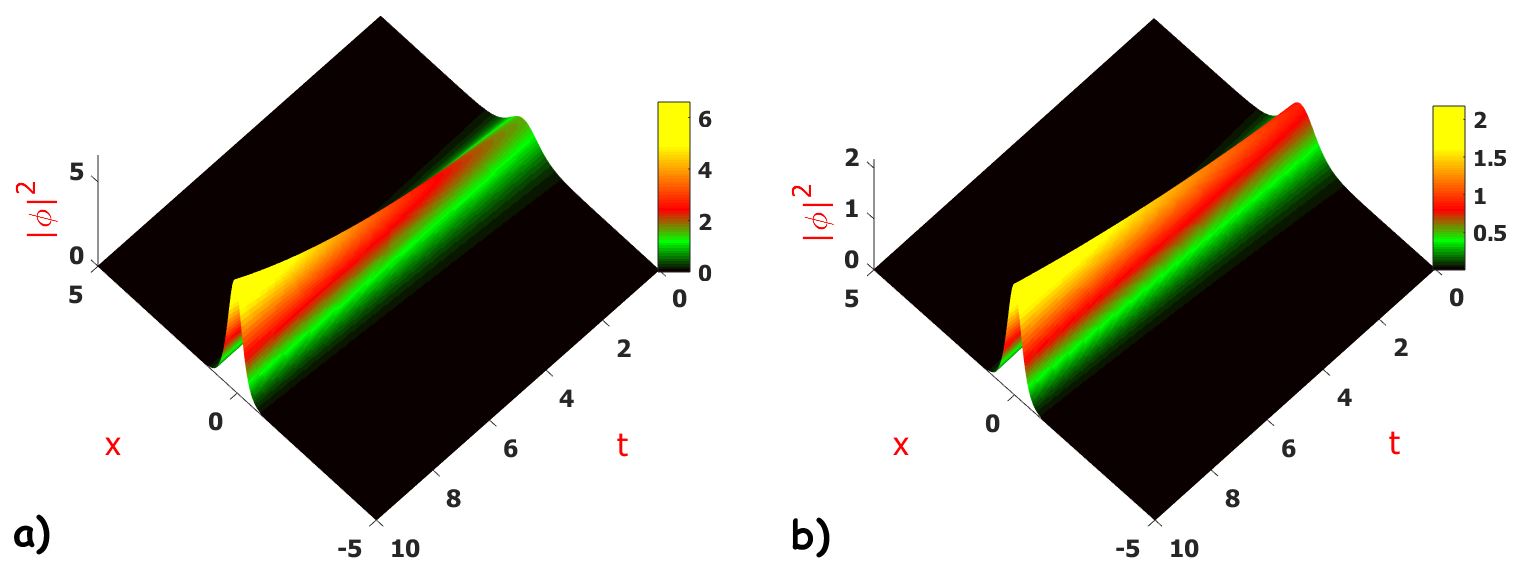}
	\caption{  Evolution of one-soliton solution (\ref{g1}) characterized by second-order rogue matter wave for the parameters $\beta_{+}=1.5$, $\beta_{-}=0$, $\beta_{0+}=-0.4$, $\beta_{0-}=0$, $M_0=0$, $k_{0r}=1$, $C=C_-$, $y=1$  with \textbf{a)} $g_{02}=-1.5$, $\chi_{0}=0$ and \textbf{b)} Effects of three-body interaction for $g_{02}=-1.5$, $\chi_{0}=0.46g_{02}$.}
	\label{fig:Nk1}
\end{figure}

\begin{figure}[!h]
	\centering
	\includegraphics[width=12cm,height=4cm]{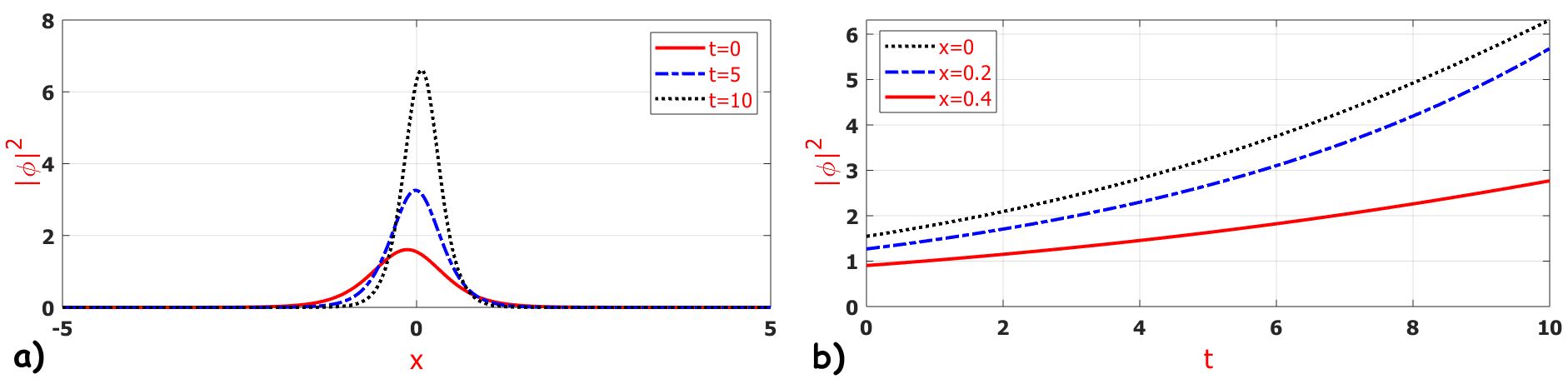}
	\caption{   Spatiotemporal evolution of the second-order rogue matter wave soliton for Fig.(\ref{fig:Nk1}\textbf{a)}  with \textbf{a)} Spatial evolution for t= 0, t=5, t=10 and \textbf{b)} Temporal evolution for x= 0, x=0.2, x=0.4.}
	\label{fig:Nk01}
\end{figure}

\begin{figure}[!h]
	\centering
	\includegraphics[width=12cm,height=3cm]{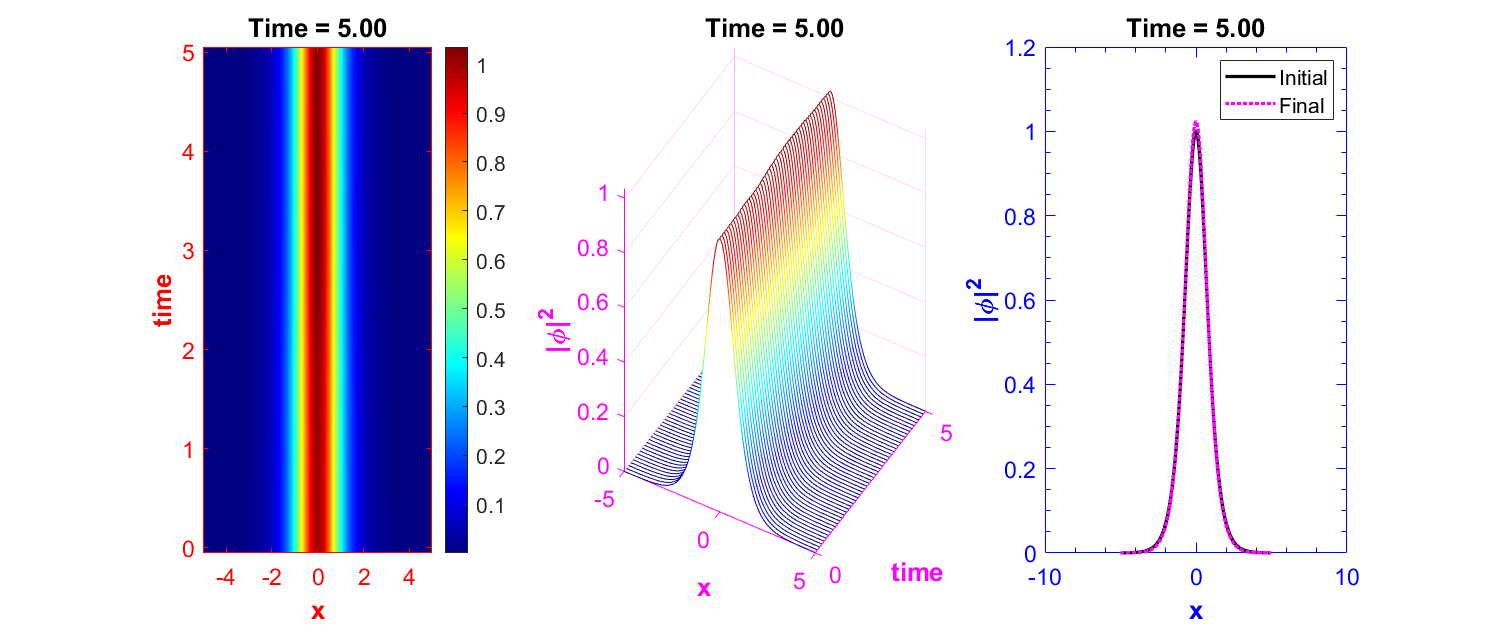}
	\caption{Numerical evolution of the one-soliton solution in fig.(\ref{fig:Nk1}) in the presence of two-body interaction term alone}
	\label{NumFig1}
\end{figure}
\begin{figure}[!h]
	\centering
	\includegraphics[width=12cm,height=3cm]{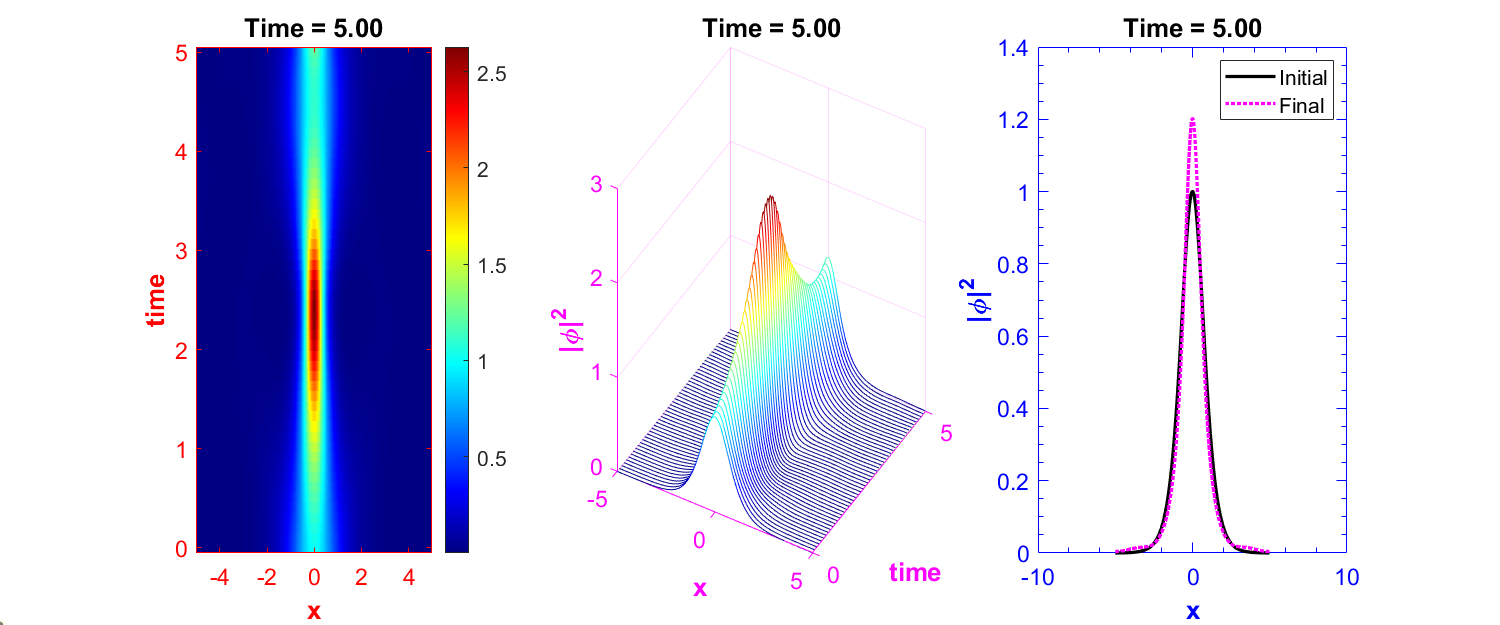}
	\caption{Numerical evolution of the one-soliton solution in fig.(\ref{fig:Nk1}) in the presence of two- and three-body interactions}
	\label{NumFig1b}
\end{figure}

The construction process of one-soliton solution for Eq.(\ref{ao}) uses the following form of $g(\tilde{x},\tilde{y},t)$ and $f(\tilde{x},\tilde{y},t)$ earlier predefined by Hirota \cite{Nkenfack2025,Nkenfack2024}: 
	\begin{equation}
		g(\tilde{x},\tilde{y},t)=\epsilon C_1 e^{\theta(\tilde{x},\tilde{y},t)},        \label{e}
	\end{equation}
	
	
	\begin{equation}
		f(\tilde{x},\tilde{y},t)=1+\epsilon^2 C(\tilde{t})e^{\theta+\theta^*},         \label{f}
	\end{equation}
	where te time-dependent parameter $C(\tilde{t})$ is to be determined as a function of the parameters of the system $g_{01}(t)$, $\chi_{01}(t)$ and $\eta_{01}(t)$.
	
	The phase $\theta(\tilde{x},\tilde{y},t)$ is given by :
	
	\begin{equation}
		\theta(\tilde{x},\tilde{y},t)=\Theta(t)+\beta \tilde{x}+ \beta_0 \tilde{y} +k_0
	\end{equation}
 where, $\Theta(t)$ is a differentiable function to be determined; $\beta=\beta_{+}+i\beta_{-}$, $\beta_0=\beta_{0+}+i\beta_{0-}$ and $k_0=k_{0r}+ik_{0i}$ are the arbritary complex parameters.\\
	
First, by replacing Eq.(\ref{e}) and Eq.(\ref{f}) in Eq.(\ref{c}), we obtain the expression of the differentiable function $\Theta$ given by : 
	
	\begin{equation}
		\Theta(t)=\frac{i}{2}(\beta^{2}+\beta_0^{2}) \int e^{2U(t)} dt.
	\end{equation}
		where $\Theta$ is obtained by considering the  first order of $\epsilon$ in Eq.(\ref{c}).\\
	
Finally, by	substituting Eq.(\ref{e})  and Eq.(\ref{f}) in Eq.(\ref{d}), we obtain by considering the  power four of $\epsilon$ the time-dependent parameter $C$ given by : 
	
\begin{equation}\label{f1}
C_{\pm}(t)=\frac{|C_1|^2} {2B} \left[ \left(2\eta_{01}(t) e^{2U} B-g_{01}(t)\right)\pm \sqrt{\left[2\eta_{01}(t)e^{2U}B-g_{01}(t) \right]^2-2\chi_{01}(t)e^{2U}B}\right]  
\end{equation}
With 
$B=(\beta+\beta^*)^2+(\beta_0+\beta_0^*)^2$ and	$	\Delta=\left[2\eta_{01}(\tilde{t})e^{2U}B-g_{01} \right]^2-2\chi_{01}(t)e^{2U}B$.
	
Therefore, 
\begin{equation}
f_2(\tilde{x},\tilde{y},t)=\frac{|C_1|^2}{2B}  \left[ \left(2\eta_{01}(t)e^{2U}B-g_{01}(t)\right)\pm \sqrt{\Delta} \right] exp(\theta+\theta^*)
\end{equation}

Note that for $\chi_{01}=0$ and $\eta_{01}=0$, we have: 	
\begin{equation}\label{f2}
C(t)=-|C_1|^2\frac{g_{01}(t)}{B}.    
\end{equation} 

This is acurately similar to the results of ref.\cite{Wang2022} in their work on two dimensional GP equation with only  two-body interaction.

\begin{figure}[!h]
	\centering
	\includegraphics[width=12cm,height=3cm]{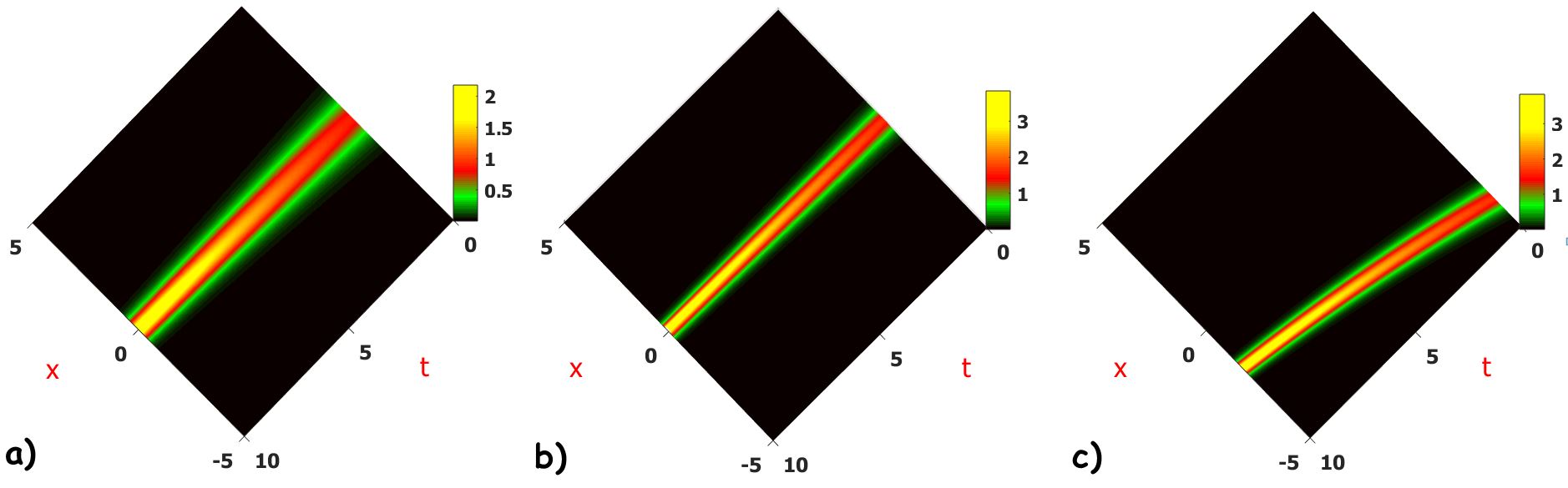}
	\caption{  Space-time evolution of the second-order rogue matter waves   for the parameters  $\beta_{-}=0$, $\beta_{0+}=-0.4$, $\beta_{0-}=0$,  $M_0=0$, $g_{02}=-1.5$, $\chi_{0}=0.46g_{02}$, $C=C_-$, $y=1$ \textbf{a)} $\beta_{+}=1.5$, $k_{0r}=1$,  \textbf{b)} Effect of  the parameter $\beta_{+}$ for $\beta_{+}=2.5$, $k_{0r}=1$ and \textbf{c)} Effect of the parameter $k_{0r}$ for $\beta_{+}=2.5$, $k_{0r}=10$.}
	\label{fig:Nk2}
\end{figure}
\begin{figure}[!h]
	\centering
	\includegraphics[width=12cm,height=3cm]{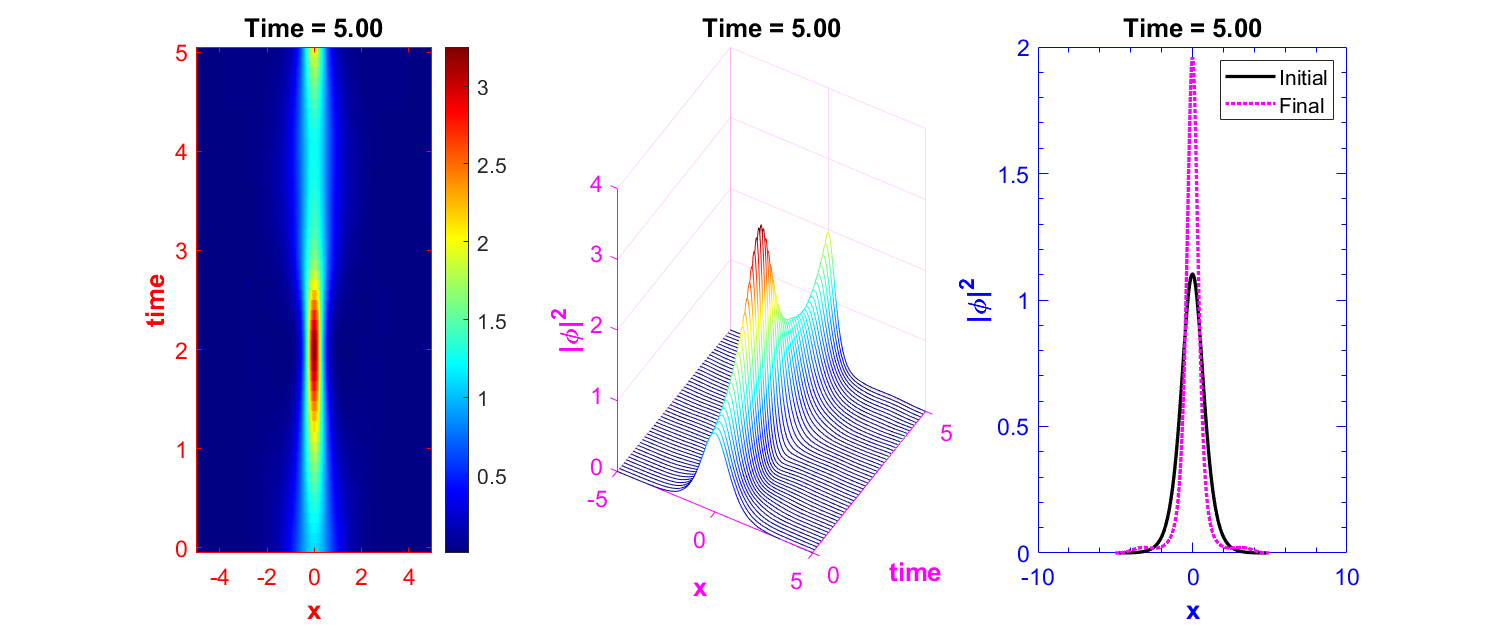}
	\caption{Numerical evolution of one-soliton solution in eq.(\ref{fig:Nk2}) with absence of gain/loss effects.}
	\label{NumFig3b}
\end{figure}
	
\begin{figure}[!h]
	\centering
	\includegraphics[width=12cm,height=3cm]{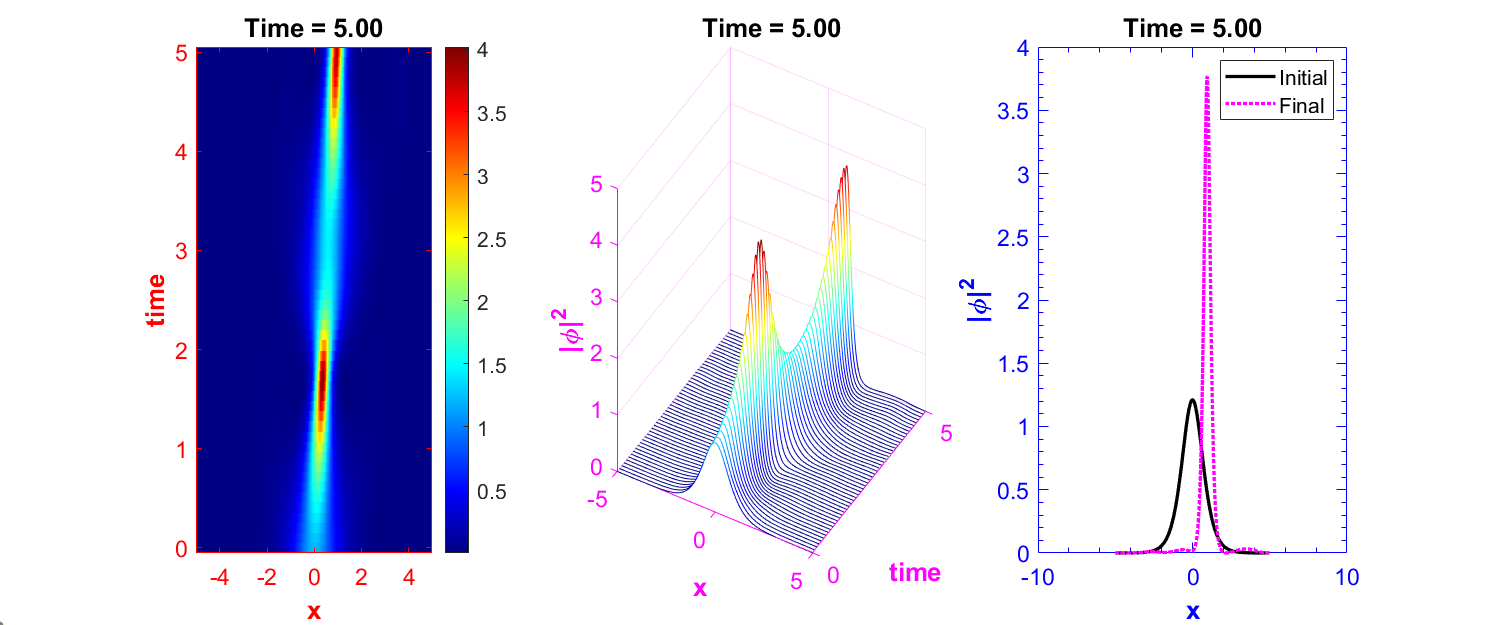}
	\caption{Numerical evolution of one-soliton solution in eq.(\ref{fig:Nk2}) with absence of gain/loss effects.}
	\label{NumFig3c}
\end{figure}

	Henceforth, the one-bright soliton solution explicity reads as:
	
		\begin{equation}\label{g}
			\phi_0(x,y,t)=|C_1|^2 \epsilon exp(\theta) \left[1+	\epsilon^2\frac{\left[2\eta_{01}(t)e^{2U}B-g_{01}(t)\right]\pm \sqrt{\Delta}}{2B}exp(\theta + \theta^{*})\right]^{-1}      
	\end{equation}
	where 
	\begin{equation}
	\theta(x,y,t)=	\frac{i}{2}(\beta^{2}+\beta_0^{2}) \int e^{2U(t)} dt+\beta e^{U} x+ \beta_0 e^{U}y +k_0.
	\end{equation}

	Rewrite one-soliton solution Eq.(\ref{g}) of Eq.(\ref{b})  as
	
	\begin{equation}
		\phi_0(x,y,t)=\frac{C_1} {2\sqrt{C}}exp\left[\Theta_{im}(t)+\beta_{-}e^U x +\beta_{0-}e^U y+k_{0i}\right] sech\left[A(x,y,t)\right]
	\end{equation}
with
$A(x,y,t)=\Theta_{r}(t)+\beta_{+}e^{U(t)} x +\beta_{0+}e^{U(t)} y+k_{0r}+\frac{1}{2}ln C$ and $\epsilon=1$.

    By transformation Eq.(\ref{71}), the one-soliton solution of two-dimensional GP equation Eq.(\ref{b1})  can be written as
    
    \begin{equation}\label{g0}
    	\phi(x,y,t)=\frac{C_1 }{2\sqrt{C}}exp\left[U(t)+\int M(t) dt+iE(x,y,t)\right]sech\left[A(x,y,t)\right]
    \end{equation}
where
\begin{equation*}
	E(x,y,t)=\Theta_{im}(t)+\beta_{-}e^U x +\beta_{0-}e^U y+k_{0i}-\frac{1}{2}\frac{dU(t)}{dt}(x^2+y^2),
\end{equation*}
\begin{equation*}
 \Theta_r(t)=-(\beta_+\beta_- +\beta_{0+}\beta_{0-})\int e^{2U(t)} dt, 
\end{equation*}
\begin{equation*}
	 \Theta_{im}(t)=\frac{1}{2}(\beta_+^2-\beta_-^2 +\beta_{0+}^2-\beta_{0-}^2)\int e^{2U(t)} dt
\end{equation*}
and
\begin{equation}
	C(t)=\frac{g_{01}(t)}{2B}\left[\frac{2 \eta_{01}(t)B  e^{2U}}{g_{01}(t)}-1\pm \sqrt{\left(\frac{2 \eta_{01}(t) B e^{2U}}{g_{01}(t)}-1\right)^2-\frac{2\chi_{01}(t)e^{2U}B}{g_{01}^2 (t)}}\right]     \label{ho}
\end{equation}
 
In the rest of the paper, we will study the one-soliton solution  Eq.(\ref{g0})  in the cases of spatio-temporal localization and double localization respectively. We will take $\epsilon=1$ and limit our study to the case where the condensate is of attractive type, i.e.  $g_{02}<0, \chi_0<0$ and $\eta_0<0$.
	
	\subsection{\textbf{Study of the one-soliton solution for $\phi(x,y,t)=\phi(x,y=1,t)$ : second-order rogue matter waves}}
	
	\subsubsection{Case study: $\eta_0=0$ (Absence of HO interactions)}
	
In a situation where the effects of HO interactions are negligible, i.e. $\eta_{01}(t)=0$, the one-soliton solution Eq.(\ref{g0}) reads :
		 \begin{equation}\label{g1}
		\phi(x,y=1,t)=\frac{C_1 }{2\sqrt{C(t)}}exp\left[U_0t+\frac{M_0}{2}cos(2t)+iE(x,1,t)\right]sech\left[A(x,1,t)\right]
	\end{equation}
where the amplitude of soliton is  $\left|\frac{C_1 }{2\sqrt{C(t)}}exp\left[U_0t+\frac{M_0}{2}cos(2t)\right]\right|$
with
$
	C(t)=|C_1|^{2}g_{01}(t)\left[\left(-1\pm \sqrt{1-\frac{2 \chi_{01}(t)e^{2U}B}{g_{01}^2(t)}}\right)/2B\right].
$

The choice of $g_{01}(t)$ and $\chi_{01}(t)$ must then obey the condition:
		\begin{equation}
			g_{01}^2(t) > 2\chi_{01}(t)e^{2U}B.
	\end{equation}

We now perform an analysis of the one-soliton solution Eq.(\ref{g1}) for different values of the parameters  $\beta$, $\beta_0$, $U(t)$, $K$, $k_{0r}$, $g_{01}(t)$ and $\chi_{01}(t)$. For this, we consider only the case where two- and three-body interactions are of attractive type, i.e. $g_{02}<0$ and $\chi_0<0$. 

Figs.(\ref{fig:Nk1}\textbf{a}) and (\ref{fig:Nk1}\textbf{b})  describe the evolution of the one-soliton solution Eq.(\ref{g1})  for various values of $g_{02}$ and $\chi_0$. This solution characterises a second-order rogue matter wave with properties almost similar to those found in Kengne's work (obtained in the case of one-dimensional GP equation) \cite{Kengne2020}.  It appears  in Fig.(\ref{fig:Nk1}\textbf{a}) that in the presence of two-body interaction alone, the second-order rogue matter wave sees its amplitude increased over time and whose quasi-centralisation of the trajectory of the latter is due to the value of the real part of the initial phase $k_{0r}=1$.

Looking at the addition of the three-body interaction term in the condensate ($g_{02}=-1.5, \chi_{0}=0.46g_{02}$), we have two observations in Fig.(\ref{fig:Nk1}\textbf{b}) : the second-order rogue matter wave sees its maximum amplitude decrease and its spatial location increase. This is due to the high density in the condensate. 

Fig.(\ref{fig:Nk01}) describes the spatio-temporal location of the second-order rogue matter wave obtained in Fig.(\ref{fig:Nk1}\textbf{a}) whose spatial location is observed for  the time values $t=0$, $t=5$ and $t=10$. When time increases, the soliton sees its amplitude increase. In addition, the temporal localization was observed for $x=0$, $x=0.2$ and $x=0.4$. We realise, of course, in Fig.(\ref{fig:Nk01}\textbf{b}) that the temporal location  induces an exponential increase of amplitude.
These observations are confirmed by results in figs.(\ref{NumFig1} and \ref{NumFig1b}) obtained by numerical integrations. The black colored curve stands for initial state while the final state is colored pink. The time is fixed at $t=5.00$ and the evolution of the soliton shows an amplitude increase by actions of two- and three-body interaction terms, as mentionned in analytical results. It appears that both approached provide convergent results.
Looking now at the effects of the parameters $\beta$ and $k_{0r}$ of the phase $\theta$, we observe in Fig.(\ref{fig:Nk2}\textbf{b)} that when increasing the value of $\beta_{+}$ (from $\beta_+=1.5$ to $\beta_+=2.5$), the second-order rogue matter wave undergoes a spatial compression and thus an increase in amplitude. Moreover, by changing real part of the initial phase $k_{0r}$ from $k_{0r}=1$ to $k_{0r}=10$, the second-order rogue matter wave sees its trajectory modified in this case and this without changing the amplitude as indicated in Fig.(\ref{fig:Nk2}\textbf{c)}.

In figs.(\ref{NumFig3b} and \ref{NumFig3c}) obtained by numerical integrations during fixed time at $t=5.00$ the evolution of the soliton shows how the base is compressed from initial state (black curve) to final state (pink curve). This is accompaigned by an amplitude increase. The analytical results are then confirmed by numerical ones.

	\begin{figure}[!h]
		\centering
		\includegraphics[width=12cm,height=3cm]{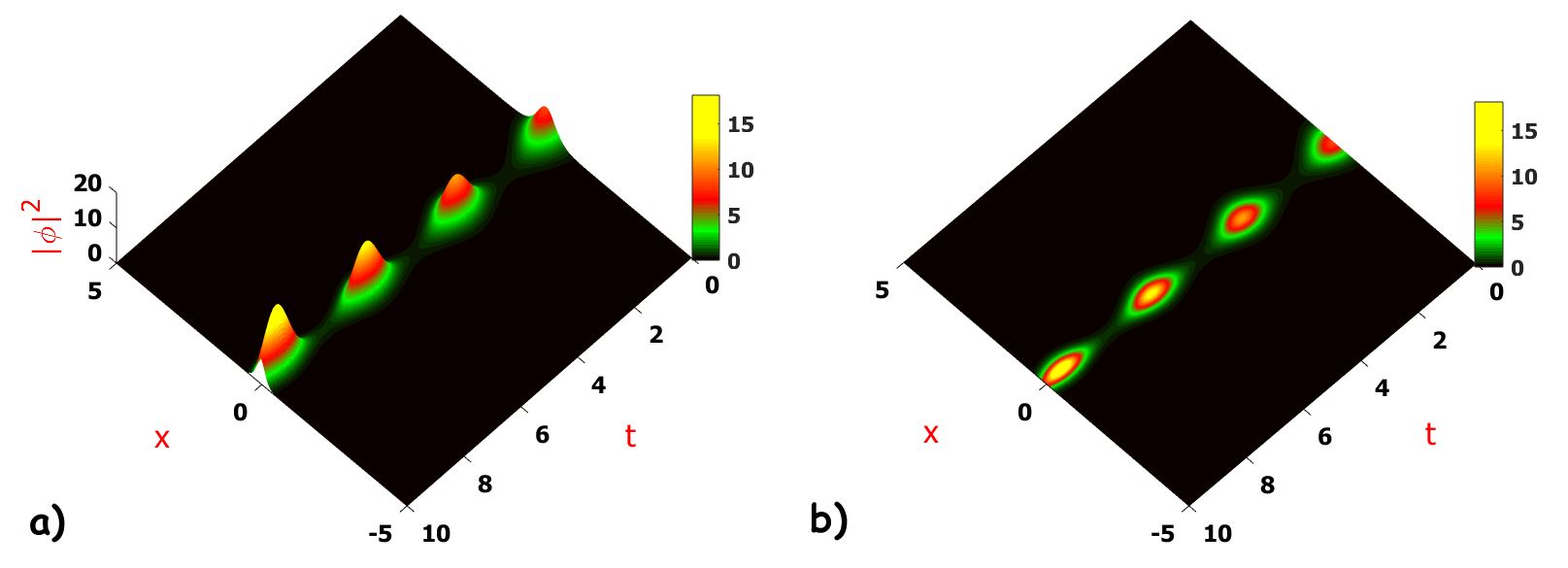}
		\caption{  Evolution of the second-order rogue matter wave with effect of gain/loss term for  the parameters $\beta_{+}=1.5$, $\beta_{-}=0$, $\beta_{0+}=-0.4$, $\beta_{0-}=0$, $k_{0r}=1$, $g_{02}=-1.5$, $\chi_{0}=0.46g_{02}$, $C=C_-$, $y=1$ with \textbf{a)} $M_{0}=0.8$ and \textbf{b)} Top view.}
		\label{fig:Nk3}
	\end{figure}

\begin{figure}[!h]
	\centering
	\includegraphics[width=12cm,height=3cm]{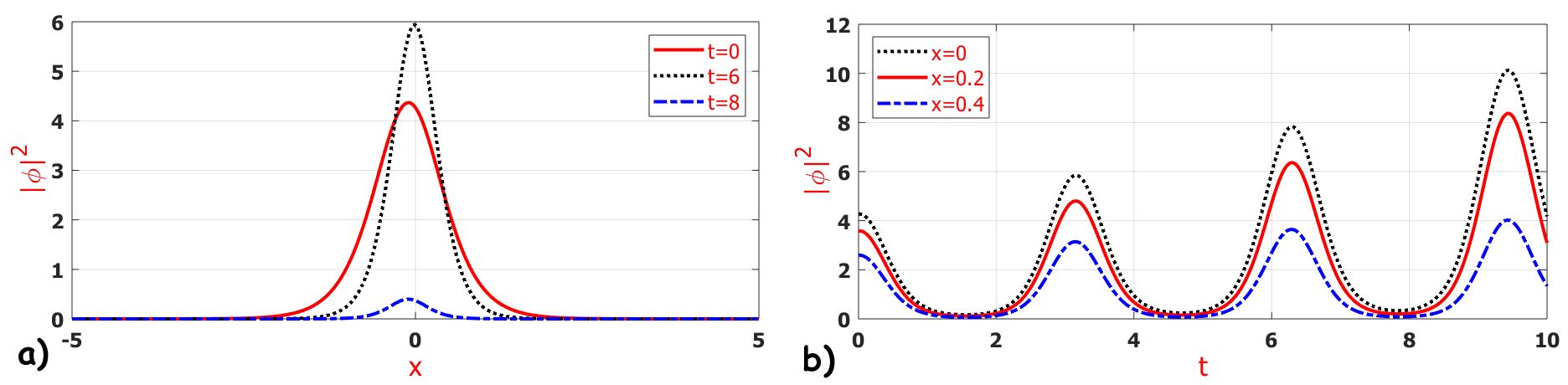}
	\caption{  Spatiotemporal evolution of the second-order rogue matter wave for Fig.(\ref{fig:Nk3}\textbf{a)}  with effect of gain/loss term for \textbf{a)} Spatial evolution for t= 0, t=6, t=8 and \textbf{b)} Temporal evolution for x= 0, x=0.2, x=0.4.}
	\label{fig:Nk30}
\end{figure}
\begin{figure}[!h]
	\centering
	\includegraphics[width=12cm,height=4cm]{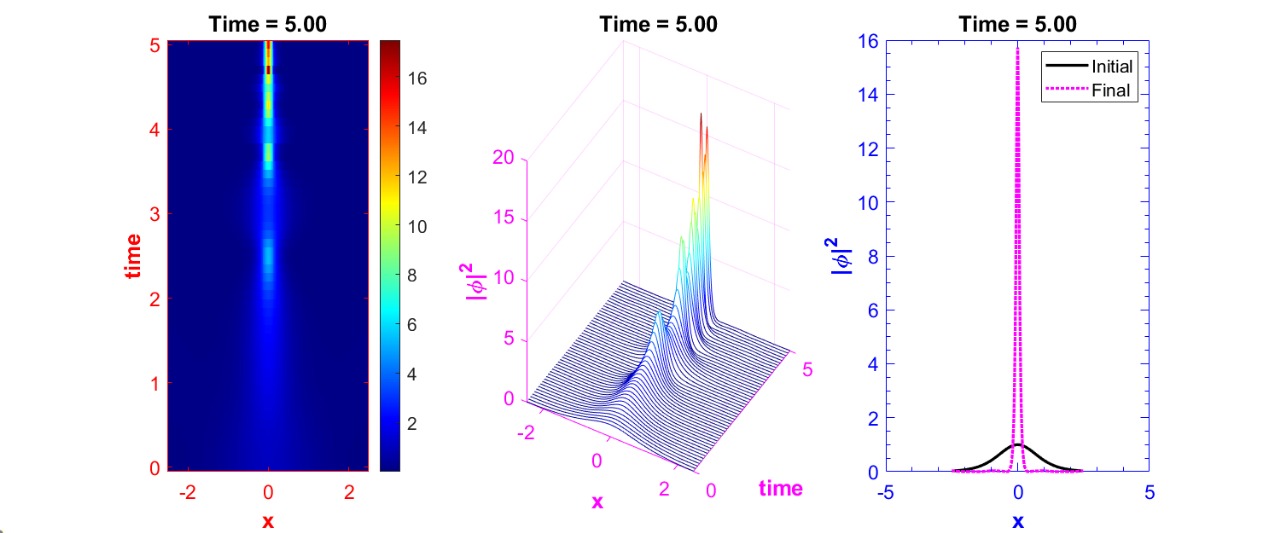}
	\caption{Numerical evolution of one-soliton solution in eq.(\ref{fig:Nk3}) with the effects of gain/loss term.}
	\label{NumFig4}
\end{figure}

Fig.(\ref{fig:Nk3}) illustrates the effects of gain/loss $M(t)$ on the dynamics of the second-order rogue matter wave obtained in Fig.(\ref{fig:Nk1}\textbf{b)}. We observe that the form of the gain/loss $M(t)=-M_0sin(2t)$ causes areas of collapse with change in amplitude over time. Fig.(\ref{fig:Nk30}\textbf{a)} and (\ref{fig:Nk30}\textbf{b)} respectively describe the spatial and temporal evolutions of the second-order RMW for various values of the temporal coordinate $t$ and spatial $x$. It can be seen in Fig.(\ref{fig:Nk30}\textbf{a)} that between $t=0$ and $t=6$, the matter wave sees its amplitude increase while between $t=6$ and $t=8$, it rather sees its amplitude decrease. This caraterises that we are in the collapse zone.
The observations in fig.(\ref{NumFig4}) obtained by numerical integrations then confirm the analytical predictions. The evolution of the soliton shows a deep amplitude increase when gain/loss term is activated.

It emerges from these interpretations of the previous figures that the quintic parameter $\chi_{0}$ causes a decrease in the amplitude of the second-order rogue matter wave unlike the gain/loss parameter which, beyond creating areas of collapses and revival, leads to an amplification the matter wave. It should also be said that to control the compression and the trajectory of the wave of matter, it will be necessary to act on the parameter $\beta_+$ of the phase and the initial phase $k_0$ respectively.

\subsubsection{Case study: $\eta_0 \neq 0$ (Presence of HO interactions)}

	We now study the effects of HO interactions on the dynamics of the second-order rogue matter wave.  In this case, the one-soliton solution reads :
	
	\begin{equation}\label{g4}
		\phi(x,y=1,t)=\frac{C_1 }{2\sqrt{C(t)}}exp\left[U_0 t+\frac{M_0}{2}cos(2t)+iE(x,1,t)\right]sech\left[A(x,1,t)\right],
	\end{equation}
where the amplitude of the soliton is given by  $\left|\frac{C_1 }{2
	\sqrt{C(t)}}exp\left[U_0 t+\frac{M_0}{2}cos(2t)\right]\right|$ with
$C(t)$ expressed by relation (\ref{ho}). In this case, the representation of the square modulus of  the one-soliton solution Eq.(\ref{g4}) must verify the condition:
	\begin{equation}
		\chi_{01}(t) < \frac{\left[2\eta_{01}(t)e^{2H}B-g_{01}(t) \right]^2}{2e^{2H}B}
	\end{equation}

\begin{figure}[!h]
	\centering
	\includegraphics[width=12cm,height=3cm]{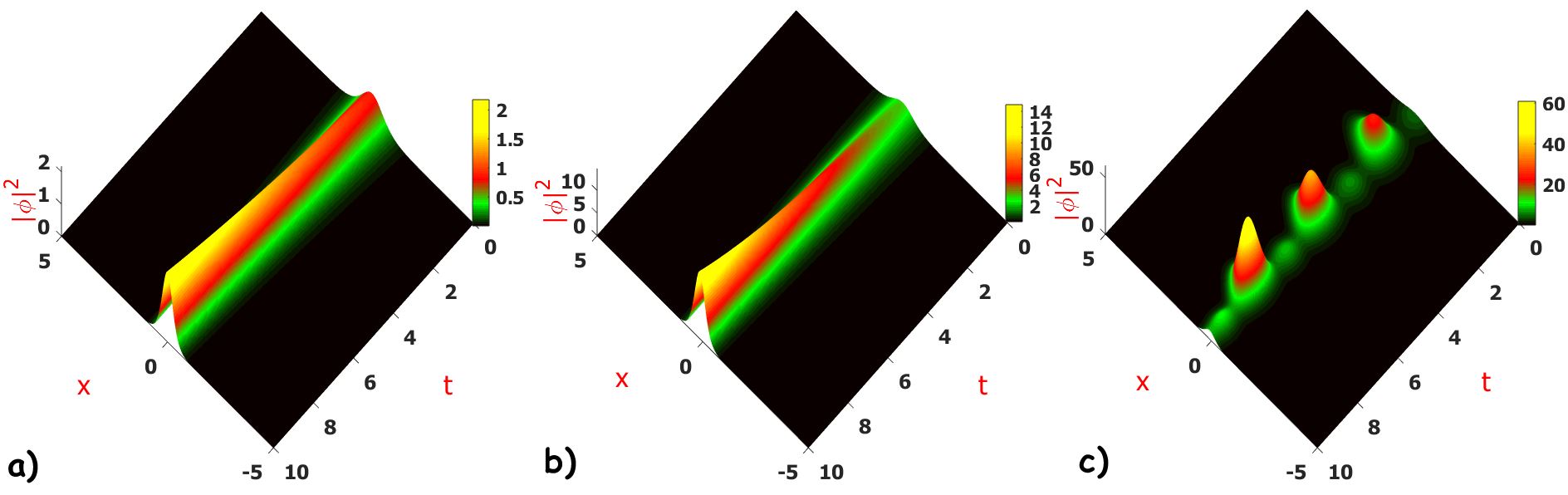}
	\caption{  Space-time evolution of (\ref{g4}) characterized by second-order rogue matter wave for the parameters  $\beta_{+}=1.5$, $\beta_{-}=0$, $\beta_{0+}=-0.4$, $\beta_{0-}=0$, $k_{0r}=1$, $g_{02}=-1.5$, $\chi_{0}=0.46g_{02}$, $C=C_-$, $y=1$ with \textbf{a)} Absence of HO interaction and gain/loss $\eta_{0}=0$, $M_0=0$  \textbf{b)} Effects of HO interaction for  $\eta_{0}=-0.9$, $M_0=0$ and \textbf{c)} Effects of gain/loss term for  $\eta_{0}=-0.9$, $M_0=0.8$.}
	\label{fig:Nk6}
\end{figure}

The square module of solution Eq.(\ref{g4}) makes it possible to obtain Fig.(\ref{fig:Nk6}\textbf{b)} and (\ref{fig:Nk6}\textbf{c)} for  $\beta_{+}=1.5$, $\beta_{-}=0$, $\beta_{0+}=-0.4$, $\beta_{0-}=0$, $k_{0r}=1$, $g_{02}=-1.5$, $\chi_{0}=0.46g_{02}$, $C=C_-$, $\eta_{0}=-0.9$ and $M_0=0.8$. We observe in Fig.(\ref{fig:Nk6}\textbf{b)}  that when the effects of the HO interaction are taken into account, the second-order rogue matter wave undergoes an increase in amplitude as well as reduction in this spatio-temporal location. This is partly due to the consideration of the case of attractive HO interaction, i.e. $\eta_{0}<0$. Taking into account the effects of the gain/los term ($M_0=0.8$) in the presence of the HO interaction, we also see in this case the creation of collapse and revival zones in the dynamics of propagation of the matter wave as shown in Fig.(\ref{fig:Nk6}\textbf{c)} accompagnied by an increase in its amplitude.

\begin{figure}[!h]
	\centering
	\includegraphics[width=12cm,height=3cm]{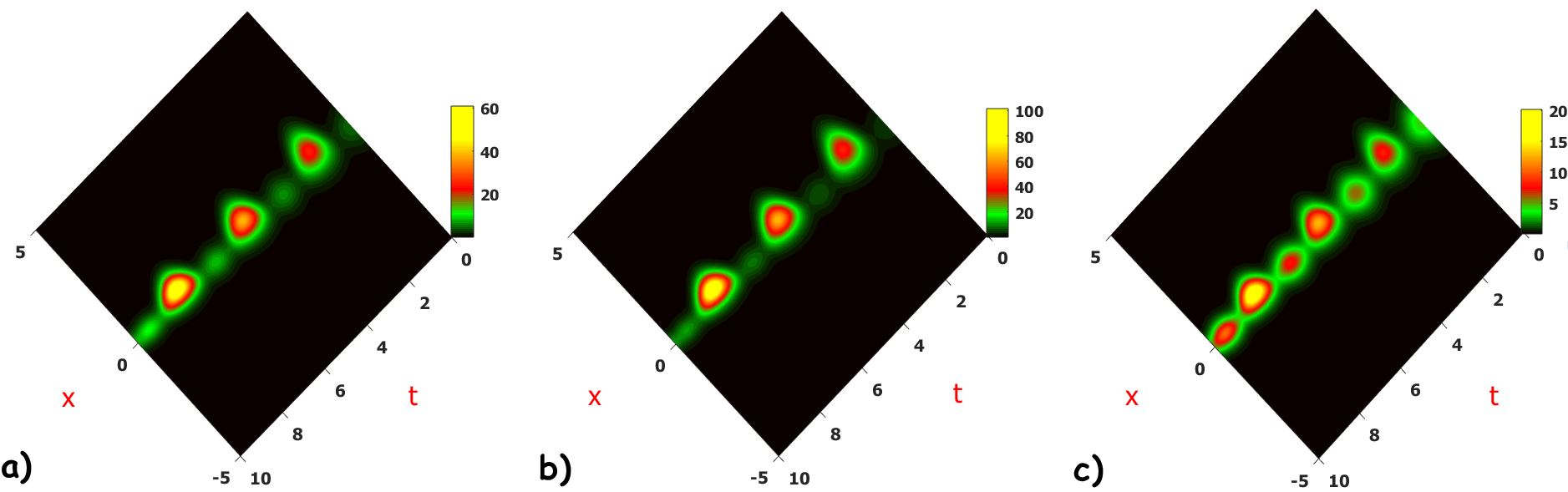}
	\caption{  Effect of HO interaction on evolution of second-order rogue matter wave  (\ref{g4})  in presence of gain/loss term for the parameters $\beta_{+}=1.5$, $\beta_{-}=0$, $\beta_{0+}=-0.4$, $\beta_{0-}=0$, $k_{0r}=1$, $g_{02}=-1.5$, $\chi_{0}=0.46g_{02}$, $M_0=0.8$, $C=C_-$, $y=1$   with \textbf{a)} $\eta_{0}=-0.9$, \textbf{b)} $\eta_{0}=-1.5$ and \textbf{c)} $\eta_{0}=-0.3$.}
	\label{fig:Nk7}
\end{figure}

By decreasing the value of the HO parameter of interaction in absolute value from $\eta_{0}=-0.9$ to $\eta_{0}=-2.9$, the following two facts are observed in Fig.(\ref{fig:Nk7}\textbf{b)} : the second-order rogue matter wave not only undergoes an increase in its amplitude of propagation but also tends to expand the areas of collapse due to the effects of gain/loss. Moreover, by increasing the value of the HO parameter from $\eta_0=-0.9$ to $\eta_0=-0.3$, the matter wave sees its amplitude certainly decrease but an annihilation of the collapse zone is also observed in Fig.(\ref{fig:Nk7}\textbf{c)}.	

Thus, it emerges from these interpretations that to control the amplitude and the effects of the gain/loss that generates the collpase zones, it would be wise to act properly on the HO parameter either by increasing its value to attenuate the collapse zones or by decreasing its value to cause an increase in the amplitude of the second-order rogue matter wave during its spatiotemporal propagation.
	
		\subsection{\textbf{Study of the one-soliton solution for $\phi(x,y,t)=\phi(x,y,t=0.2)$ : Line-soliton solution }}
	
	\subsubsection{Case study: $\eta_0=0$ (Absence of HO interactions)}
	
We now look at the double spatial localization of the one-soliton solution Eq.(\ref{g1}). For that, we set a time value at $t=0.2$.

\begin{figure}[!h]
	\centering
	\includegraphics[width=12cm,height=4cm]{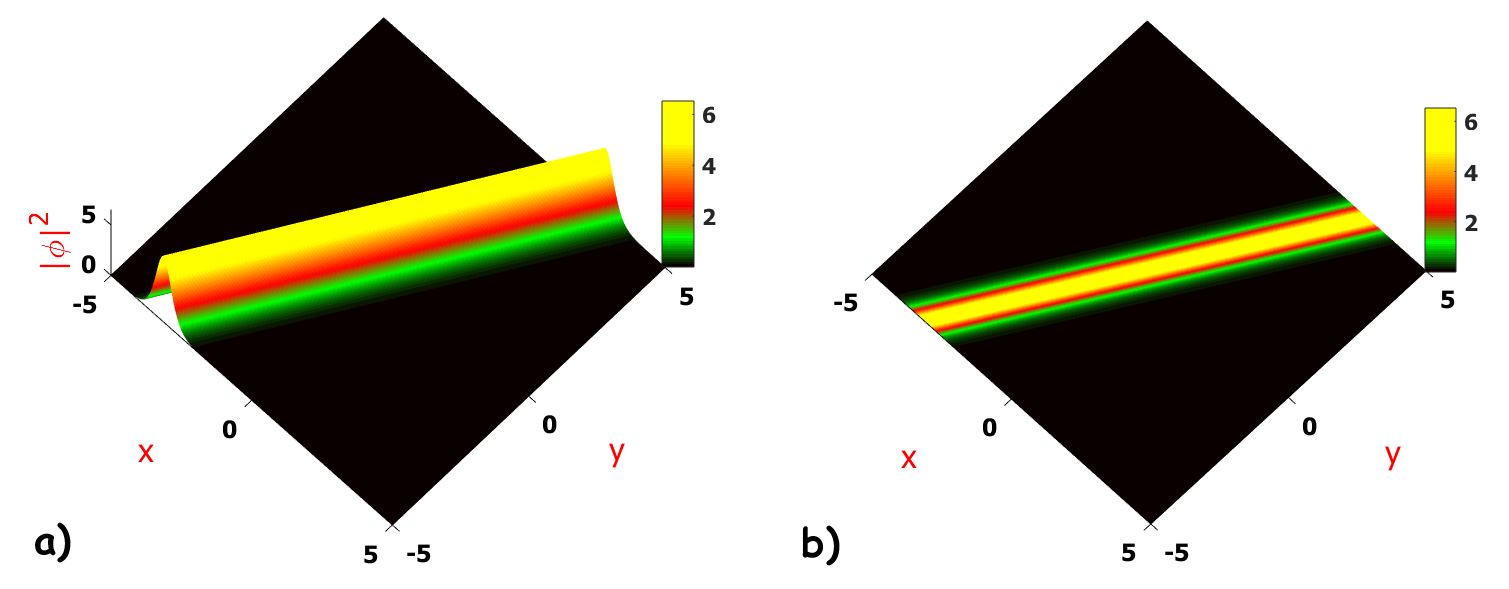}
	\includegraphics[width=12cm,height=4cm]{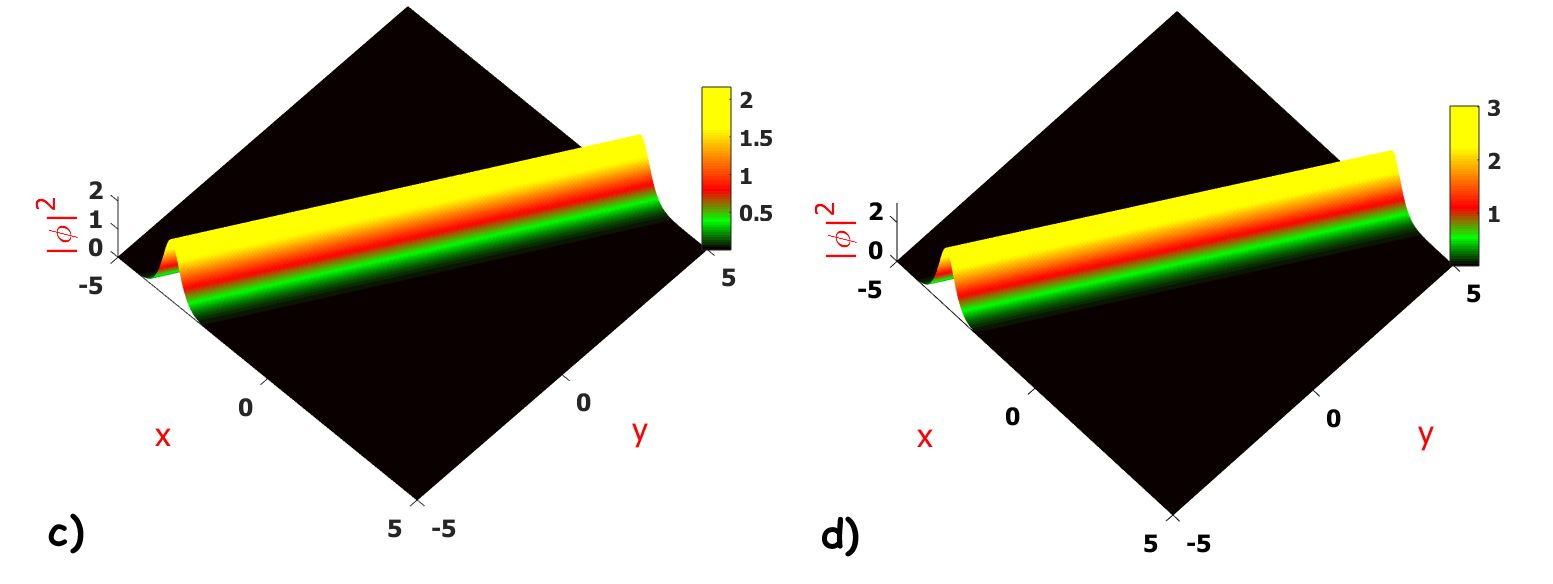}
	\caption{ Evolution of one-soliton solution Eq.(\ref{g1}) characterized by Line-soliton for the parameters $\beta_{+}=2.5$, $\beta_{-}=0$, $\beta_{0+}=-1.5$, $\beta_{0-}=0$, $k_{0r}=1$, $C=C_-$  with \textbf{a)} $g_{02}=-1.5$, $\chi_{0}=0$, $M_0=0$,  \textbf{b)} Top view of Fig.(\ref{fig:Nk4}\textbf{a)},  \textbf{c)} Effect of three-body interaction for $g_{02}=-1.5$, $\chi_{0}=-0.7$, $M_0=0$ and \textbf{d)} Effect of gain/loss term for $g_{02}=-1.5$, $\chi_{0}=0.46g_{02}$, $M_0=0.8$.}
	\label{fig:Nk4}
\end{figure}

\begin{figure}[!h]
	\centering
	\includegraphics[width=12cm,height=4cm]{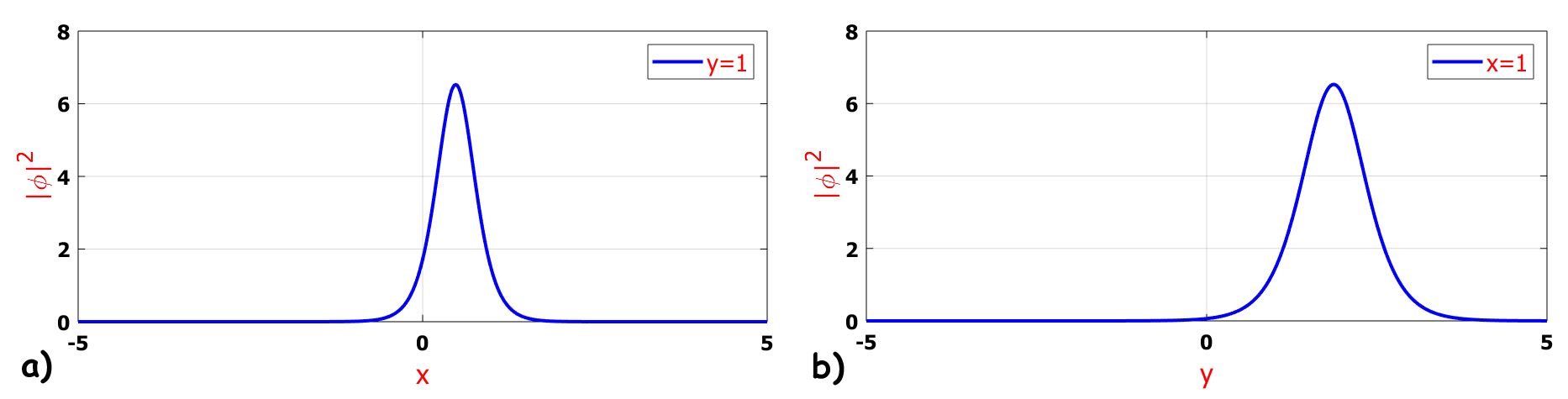}
	\caption{ Double spatial localization of Line-soliton of  Fig.(\ref{fig:Nk4}\textbf{a)})  with \textbf{a)} Spatial evolution for $y=1$ and \textbf{b)} Spatial evolution $x=1$.}
	\label{fig:Nk04}
\end{figure}

Fig.(\ref{fig:Nk4}) describes the dynamic evolution of the condensate with cubic and quintic nonlinearities and gain/loss characterizing a Line-soliton. This profile represents a Line-soliton by its double spatial localization properties as observed in the ref. \cite{Radha2010}. We note that by considering the quintic nonlinearity in the condensate, Fig.(\ref{fig:Nk4}\textbf{c)} informs, as in the case of the second-order rogue matter wave, that this nonlinearity leads to a decrease of amplitude. Moreover, in the case of the Line-soliton, the profile does not undergo an increase in its double spatial location.

By considering the effects of gain/loss, we see that the Line-soliton does not suffer the collapse effect unlike the second-order rogue matter wave. But on the other hand, the Line-soliton undergoes an increase in amplitude as shown in Fig.(\ref{fig:Nk4}\textbf{d)}.

Fig.(\ref{fig:Nk04}) highlights the double spatial location of the Line-soliton. We see in Figs.(\ref{fig:Nk04}\textbf{a)} and  (\ref{fig:Nk04}\textbf{b)} that the propagation  of the Line-soliton keeps the same amplitude for $y$ and $x$ fixed but is not identical in all aspects. Indeed, by fixing the spatial coordinates, the Line-soliton has a more compressed spatial localisation interval for $y=1$, i.e. in $\left[-0.8,1.6\right]$ and less compressed for $x=1$, i.e. in $\left[0,4\right]$.

		\subsubsection{Case study: $\eta_0 \neq 0$ (Presence of HO interactions)}

In this subsection, we highlight the role of the HO  interaction parameter on the propagation dynamics of the Line-soliton in double spatial localization. In this case, the one-soliton solution Eq.(\ref{g0}) reads :
	
		 \begin{equation}\label{g6}
			\phi(x,y,t=0.2)=\frac{C_1 }{2\sqrt{C(0.2)}}exp\left[0.2U_0+\frac{M_0}{2}cos(0.4) +iE(x,y,0.2)\right]sech\left[A(x,y,0.2)\right]
		\end{equation}
where the amplitude of soliton is given by  $\left|\frac{C_1 }{2\sqrt{C(0.2)}}exp\left[0.2U_0+\frac{M_0}{2}cos(0.4)\right]\right|$ with
$C(0.2)$ expressed by :	
	\begin{equation}
		C(0.2)=\frac{g_{01}(0.2)}{2B}\left[\frac{2 \eta_{01}(0.2)B  e^{2U}}{g_{01}(0.2)}-1\pm \sqrt{\left(\frac{2 \eta_{01}(0.2) B e^{2U}}{g_{01}(0.2)}-1\right)^2-\frac{2\chi_{01}(0.2)e^{2U}B}{g_{01}^2 (0.2)}}\right]. 
	\end{equation}
	
\begin{figure}[!h]
		\centering
		\includegraphics[width=12cm,height=4cm]{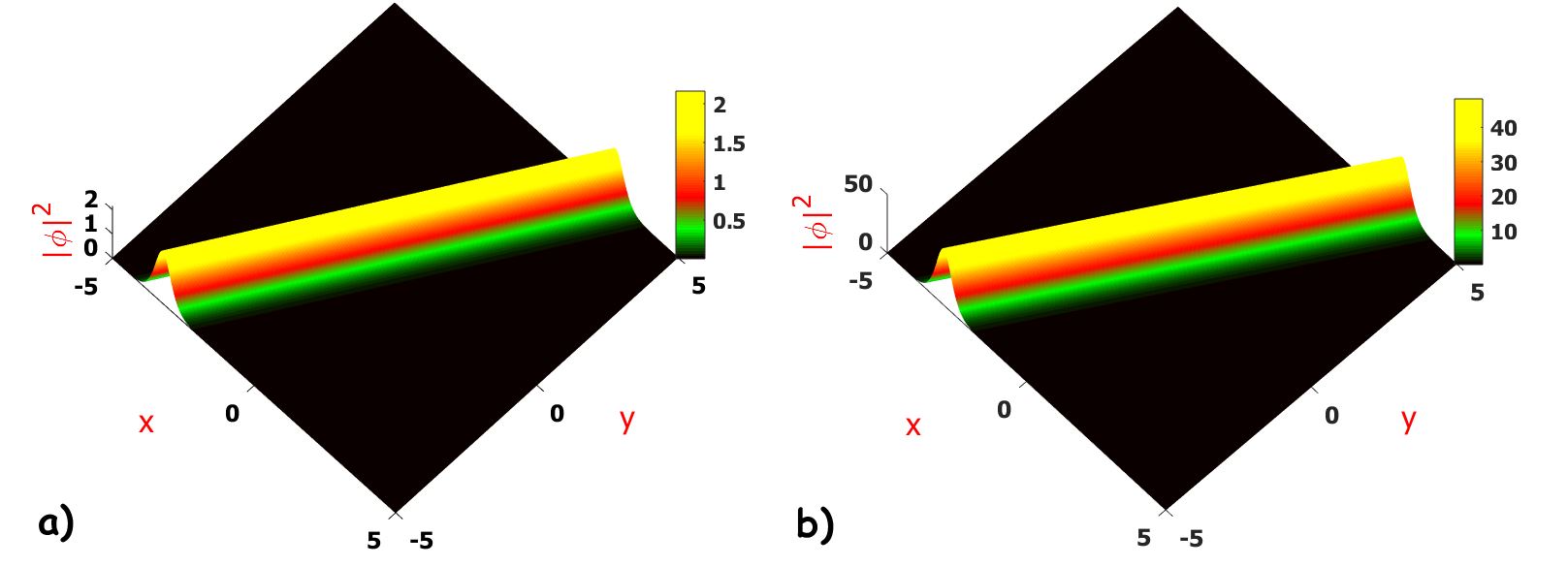}
		\caption{ Effect of HO interaction on evolution of Line-soliton for the parameters $\beta_{+}=2.5$, $\beta_{-}=0$, $\beta_{0+}=-1.5$, $\beta_{0-}=0$, $k_{0r}=1$, $g_{02}=-1.5$, $\chi_{0}=0.46g_{02}$, $M_0=0$, $C=C_-$ with \textbf{a)} $\eta_{0}=0$  and \textbf{b)} $\eta_{0}=-2.9$.}
		\label{fig:Nk8}
	\end{figure}

	The representation of the square module of the solution Eq.(\ref{g6}) reveals that by considering the effects of the HO  interaction parameter from $\eta_0=0$ to $\eta_0=-2.9$, the Line-soliton undergoes an increase in amplitude from $2$ to $50$ as indicated in Fig.(\ref{fig:Nk8}). This result confirms the insight of HO interaction in favour of the emergence of large amplitude solitons \cite{Nkenfack2025}.
	
	\begin{figure}[!h]
	\centering
	\includegraphics[width=12cm,height=4cm]{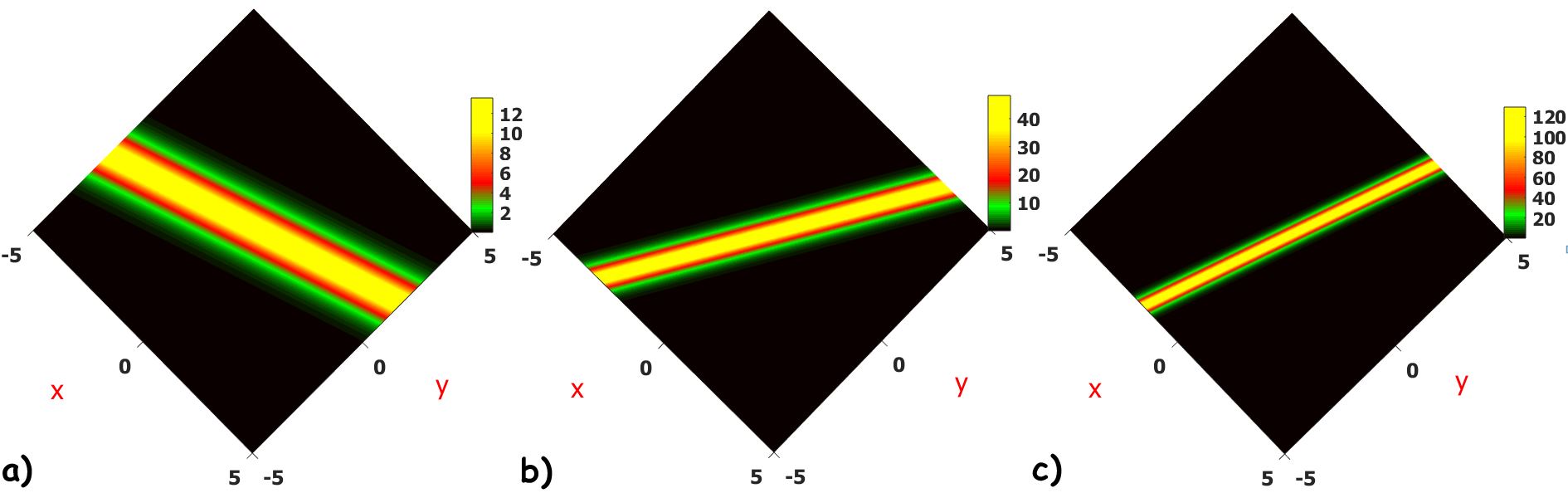}
	\caption{  Effect of $\beta_{+}$ on evolution of Line-soliton for the parameters $\beta_{-}=0$, $\beta_{0+}=-1.5$, $\beta_{0-}=0$, $k_{0r}=1$, $g_{02}=-1.5$, $\chi_{0}=0.46g_{02}$, $\eta_{0}=-2.9$, $M_0=0$, $C=C_-$ with \textbf{a)} $\beta_{+}=0.5$, \textbf{b)} $\beta_{+}=2.5$  and \textbf{c)} $\beta_{+}=4.5$.}
	\label{fig:Nk80}
\end{figure}

\begin{figure}[!h]
		\centering
		\includegraphics[width=12cm,height=4cm]{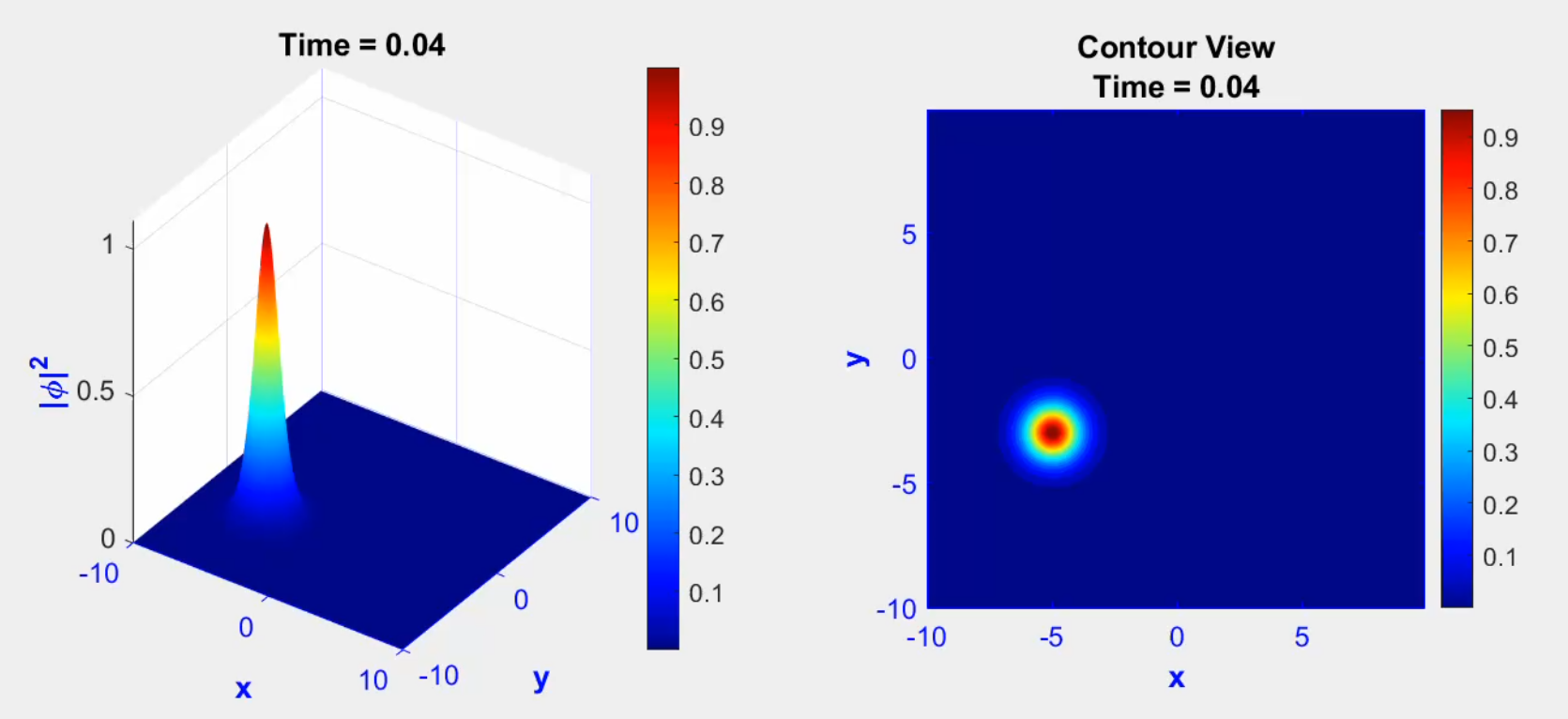}
		\includegraphics[width=12cm,height=4cm]{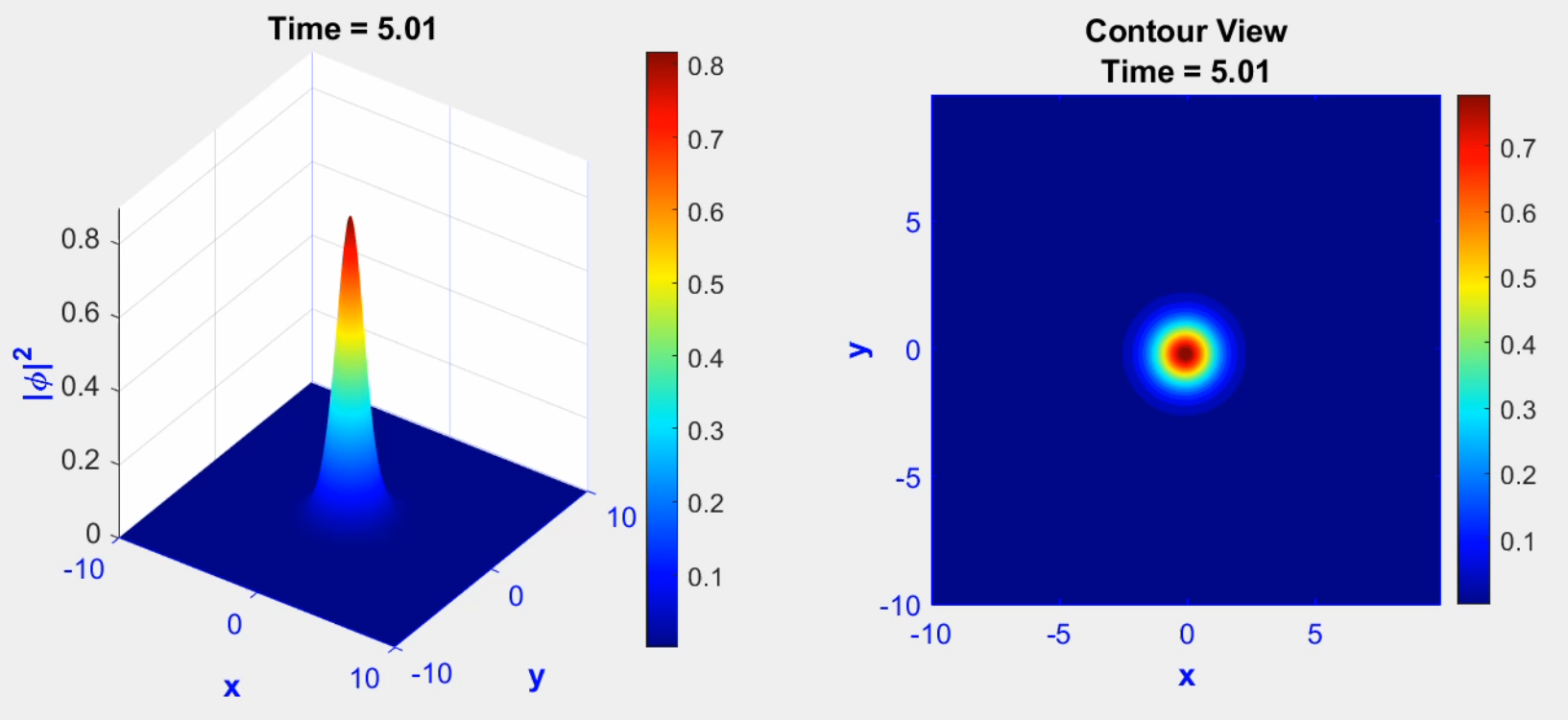}
		\includegraphics[width=12cm,height=4cm]{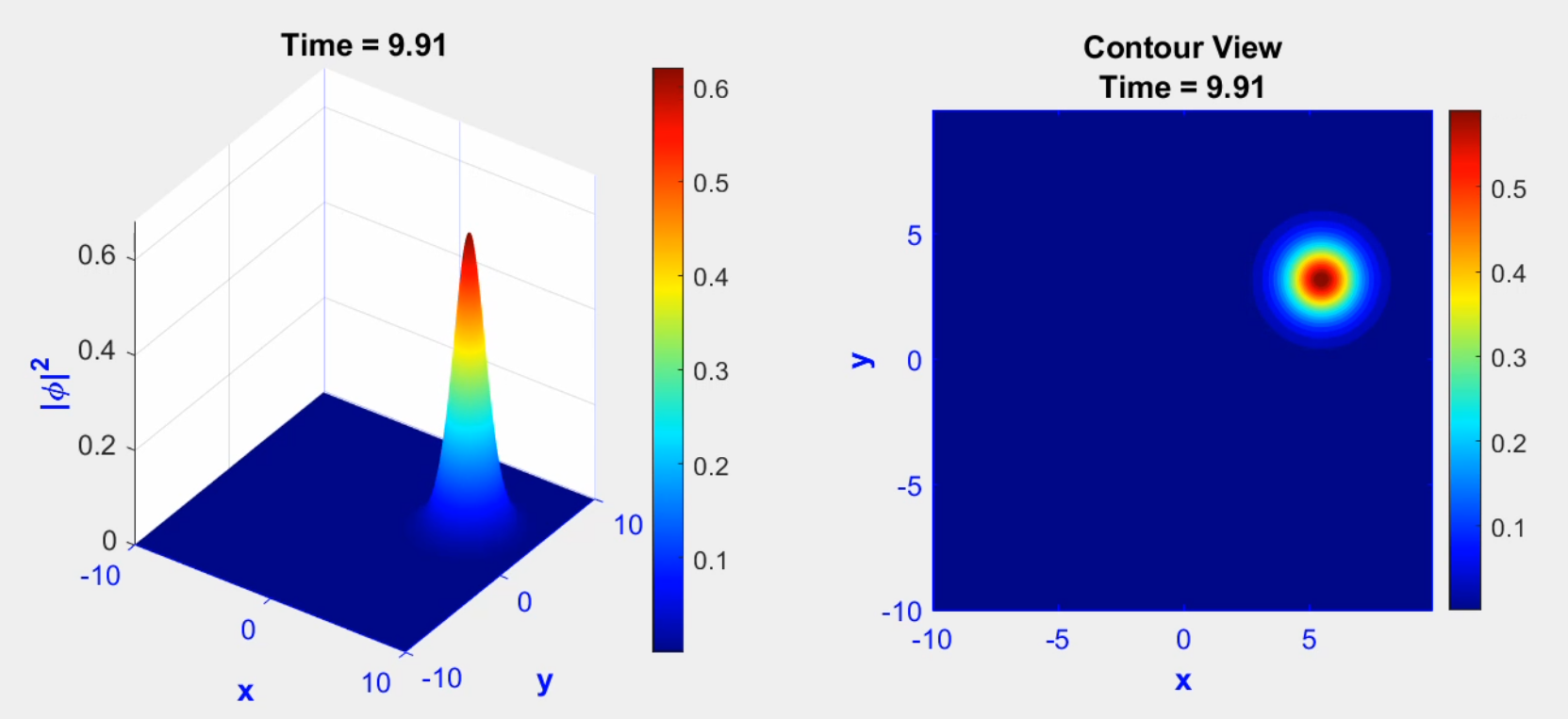}
		\caption{ Numerical dynamics of one-soliton solution through numerical integration of Eq.(\ref{b1}). First column is the square module in 2D space, and second column is the corresponding contour view, all for time varying from time=0.04 (top) to time=9.91 (bottom). (The full dynamics of this figure is available as movie among the supplemental material)}
		\label{1sol2d}
	\end{figure}

Direct numerical integration of Eq.(\ref{b1}) presented in Fig.(\ref{1sol2d}) shows how the 2D one-soliton solution is obtained and evolves in time varying from $t=0.04$ to $t=9.91$. It appears that for some values of the set of parameters, the one-soltion moves according to anylitical predictions and hence the confirmation on the acuracy of our theoretical results.
	
Fig.(\ref{fig:Nk80}) highlights the role of the parameter $\beta_+$ of the phase on the evolution of the Line-soliton. It can be seen by observing the Fig.(\ref{fig:Nk80}) that the parameter $\beta_+$ first plays a role of compression of the spatial localization interval. Moreover, by increasing the value of $\beta_+$, the Line-soliton sees its amplitude increase and its location interval decrease by appearing to leave a spatial location in $y$ to a spatial location in $x$. This result highlights this property of the parameter $\beta_+$ already observed in the refs. \cite{Radha2010}.

In the following section, we will study the interaction of the different solitons obtained previously in different cases. In addition, we will highlight the role of the HO interaction and the term gain/loss in the interaction process of second-order rogue matter wave on the one hand and Line-solitons on the other hand.
	
	\section{Two-soliton solutions}  \label{sec5}
	
	
One of the  special features of HBM is the generation of multi-soliton solutions that lead to the description of their interacting processes. In this section,  we construct the two-soliton solutions and describe their  interactions by distinguishing the two cases of spatio-temporal localization on the one hand and double spatial localization on the other hand. The two-soliton solution is obtained for (we take $\epsilon= 1$ so as not to lose generality)

\begin{figure}[!h]
	\centering
	\includegraphics[width=12cm,height=3.3cm]{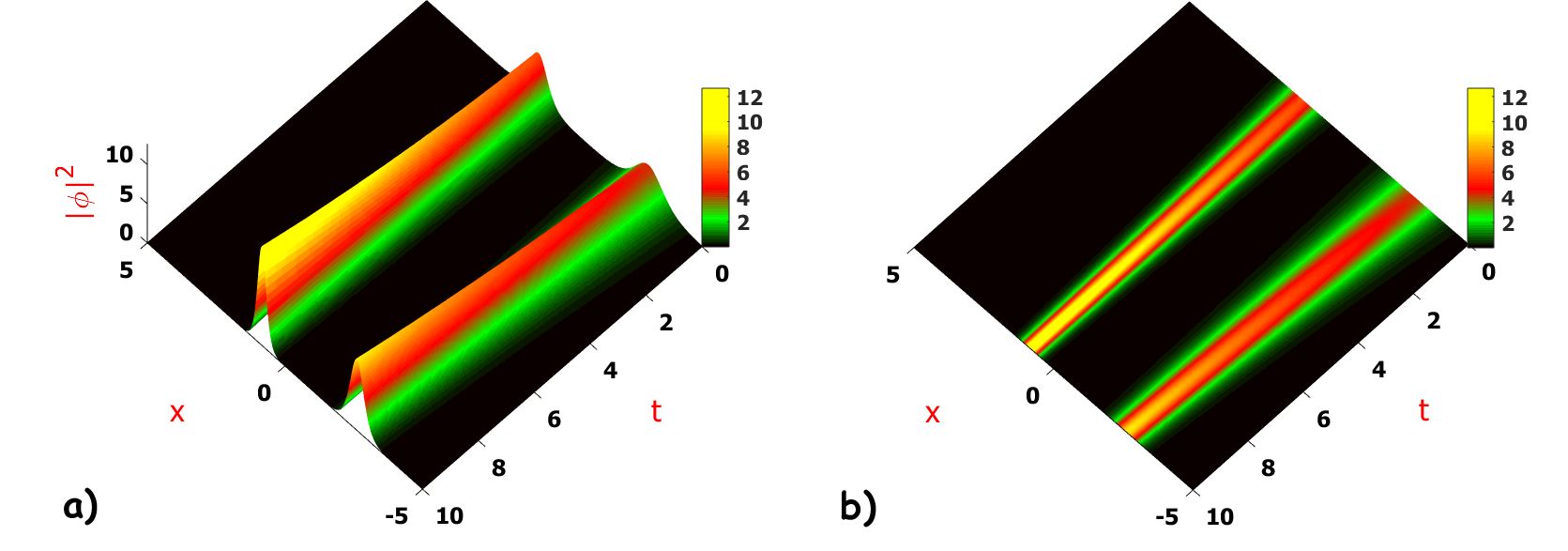}
	\includegraphics[width=12cm,height=3.3cm]{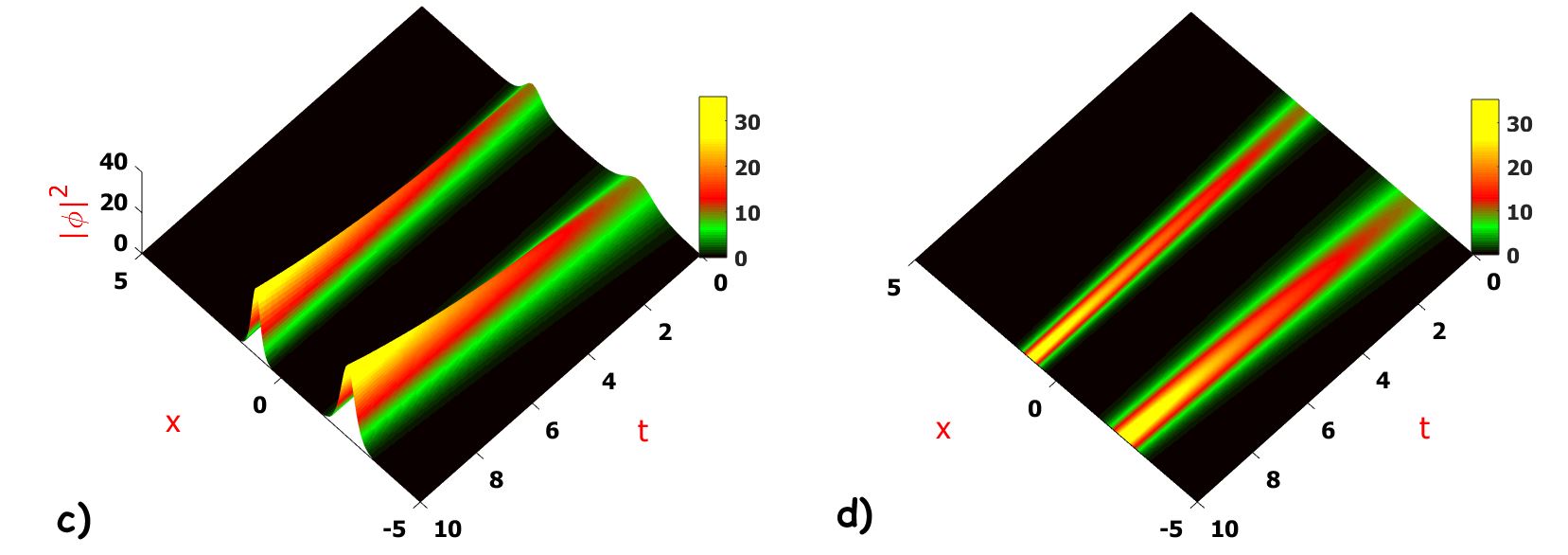}
	\caption{ Evolution of two second-order parallel RMWs for $\beta_{1}=1.5$, $\beta_{2}=2.4$, $\beta_{3}=3$, $\beta_4=-2$,  $g_{02}=-0.5$, $\chi_{0}=0.5g_{02}$, $k_{01}=3$, $k_{02}=1$ and $C(t)=C_+(t)$ with \textbf{a)} $\eta_0=0$, \textbf{b)} Top view, \textbf{c)} $\eta_0=-0.15$ and \textbf{d)} Top view.}
	\label{fig:Nk90}
\end{figure}
\begin{figure}[!h]
	\centering
	\includegraphics[width=0.8\linewidth]{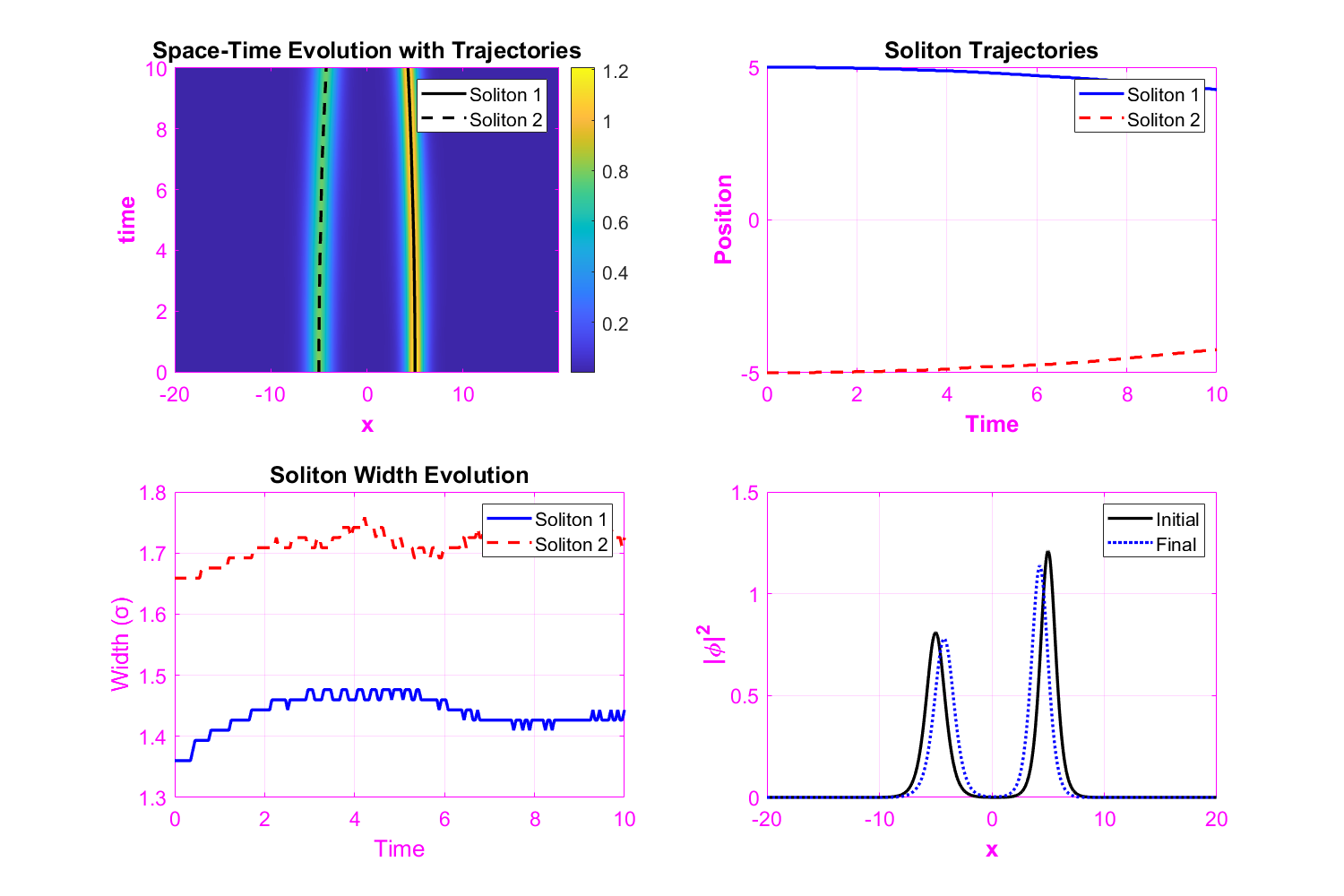}
	\caption{Numerical evolution of space-time, width and soliton trajectories for two second-order RMWs in fig.(\ref{fig:Nk90}).}
	\label{NumFig12}
\end{figure}
	
	\begin{equation}
	g(\tilde{x},\tilde{y},t)=g_1(\tilde{x},\tilde{y},t)+g_3(\tilde{x},\tilde{y},t)
	\end{equation}
	and
	\begin{equation}
	f(\tilde{x},\tilde{y},t)=1+f_2(\tilde{x},\tilde{y},t)+f_4(\tilde{x},\tilde{y},t)
	\end{equation}
	with
	\begin{equation*}
	g_1(\tilde{x},\tilde{y},t)=\sum_{k=1}^{2}C_ke^{\theta_k},    \hspace{0.5cm}         \theta_j(\tilde{x},\tilde{y},t)=\Theta_j(t)+\beta_j \tilde{x}+ \beta_{j+2} \tilde{y}+k_{0j}(j=1,2)
	\end{equation*}
	\begin{equation*}
	g_3(\tilde{x},\tilde{y},t)=\sum_{k=1}^{2}A_{2k}(t)e^{\theta_1+\theta_2+\theta_k^{*}}
	\end{equation*}
	
	where $A_{21}$ and $A_{22}$ are  to be determined. In this case we get:
	\begin{equation}
		\Theta_j(t)=\frac{i}{2}(\beta_j^{2}+\beta_{j+2}^{2}) \int e^{2U(t)} dt , \hspace{0.5cm}  (j=1,2)
	\end{equation}
	\begin{equation}
	\theta_j(\tilde{x},\tilde{y},t)=\frac{i}{2}(\beta_j^{2}+\beta_{j+2}^{2}) \int e^{2U(t)} dt +\beta_j \tilde{x}+ \beta_{j+2} \tilde{y} +k_{0j}.
\end{equation} 
	
	and the two-soliton solution obtained by solving the resulting linear partial differential equations recursively, can be written in its explicit form as (with $C_1=C_2=1$):
	\begin{equation}
		\phi_0(\tilde{x},\tilde{y},t)=\frac{e^{\theta_1}+e^{\theta_2}+g_3(\tilde{x},\tilde{y},t)}{1+f_2(\tilde{x},\tilde{y},t)+f_4(\tilde{x},\tilde{y},t)}  \label{g41}
	\end{equation}
	where

		\begin{equation}
		f_2(\tilde{x},\tilde{y},t)= E_{11}(t)e^{\theta_{1}+\theta_{1}^{*}}+	 E_{12}(t)e^{\theta_{2}+\theta_{1}^{*}}+ E_{13}(t)e^{\theta_{1}+\theta_{2}^{*}}
		+E_{14}(t)e^{\theta_{2}+\theta_{2}^{*}}
	\end{equation}
and 
\begin{equation}
	f_4(\tilde{x},\tilde{y},t)=N_{21}(t)e^{\theta_1+\theta_2+\theta_1^{*}+\theta_2^{*}}.
\end{equation}

Our main goal here is to determine the different parameters of the functions $g_3, f_2$ and $f_4$. By replacing the expression of the functions in Eq.(\ref{d}) of the bilinear form we obtain : 

 \begin{equation}
 	\left[f_{\tilde{x}\tilde{x}}f-f_{\tilde{x}}^2+f_{\tilde{y}\tilde{y}}f-f_{\tilde{y}}^2 \right]\left[ 4\eta_{01} |g|^2-f^2\right]=g_{01}|g|^2 f^2+ \chi_{01}|g|^4 + \eta_{01}\left[f^2 \left(|g|_{\tilde{x}\tilde{x}}^2 +|g|_{\tilde{y}\tilde{y}}^2\right)-2\left(|g|_{\tilde{x}}^2f_{\tilde{x}}^2 + |g|_{\tilde{y}}^2f_{\tilde{y}}^2 \right) + |g|^2 \left(f_{\tilde{x}\tilde{x}}^2+ f_{\tilde{y}\tilde{y}}^2\right)\right]    \label{ky}
 \end{equation}

Considering the different expressions of the function  $g_3, f_2$ and $f_4$ in Eq.(\ref{ky}) and proceeding by identification, we obtain the following parameters :

		\begin{equation}
			E_{1p}(t)=\frac{\left[2\eta_{01}(t)e^{2U}a_{1p}-g_{01}(t)\right]\pm \sqrt{\Delta_p}}{2a_{1p}},  \hspace{0.5cm} p=1,2,3,4
	\end{equation}
where $a_{11}=(\beta_1+\beta_1^*)^2+(\beta_3+\beta_3^*)^2$, $a_{12}=(\beta_1+\beta_2^*)^2+(\beta_3+\beta_4^*)^2$,  $a_{13}=(\beta_2+\beta_1^*)^2+(\beta_4+\beta_3^*)^2$,  $a_{14}=(\beta_2+\beta_2^*)^2+(\beta_4+\beta_4^*)^2$  and $\Delta_p=\left[2\eta_{01}(t)e^{2\alpha}a_{1p}-g_{01}(t)\right]^2-2\chi_{01}(t)e^{2\alpha}a_{1p}$,
together with

		\begin{equation}
			A_{12}(t)=K_0\left[ E_{11}^2E_{13}a_{15}+2\eta_{01}(t) e^{2U}E_{11}E_{13}(a_{11}-a_{15})-\frac{\chi_{01}(t)}{2}E_{13}e^{2U}\right], 
	\end{equation}
    \begin{equation}
			A_{22}(t)=K_1\left[ E_{14}^2E_{12}a_{15}+2\eta_{01}(t) e^{2U}E_{14}E_{12}(a_{14}-a_{15})-\frac{\chi_{01}t)}{2}E_{12}e^{2U}\right], 
	\end{equation}

	where
$a_{15}=(\beta_1-\beta_2)^2+(\beta_3-\beta_4)^2$, $a_{16}=(\beta_2+\beta_1^*)^2+(\beta_4+\beta_3^*)^2$, $K_0=\left[\eta_{01}(t) E_{11}e^{2U}(2a_{11}-a_{16})-g_{01}(t)E_{11}-\chi_{01}(t)e^{2U}\right]^{-1}$, $K_1=\left[\eta_{01}(t) E_{14}e^{2U}(2a_{14}-a_{17})-g_{01}(t)E_{14}-\chi_{01}(t)e^{2U}\right]^{-1}$, $a_{17}=a_{16}^*$\\
and
		\begin{equation*}
			N_{21}(t)=K_2 |A_{12}(t)|^2\left[ g_{01}(t)E_{11}(t)+\chi_{01}(t)e^{2U} -\eta_{01}(t)e^{2U}E_{11}(t)(2a_{11}-a_{18})\right]
	\end{equation*}	
with
	\begin{equation*}
			K_2=\left(\eta_{01}(t)E_{11}(t)e^{2U}(2a_{14}-2a_{11}-a_{18})-2E_{11}^2(t)a_{14}+\frac{\chi_{01}(t)}{2}e^{2U}\right)^{-1}. \hspace{0.2cm} a_{18}= (\beta_1+\beta_1^*+\beta_2+\beta_2^*)^2 + (\beta_3+\beta_3^*+\beta_4+\beta_4^*)^2
	\end{equation*}
	
	  By transformation Eq.(\ref{71}), the two-soliton solution of two-dimensional GP equation Eq.(\ref{b})  can be written as :
	  
\begin{equation}
\phi(x,y,t)=\frac{e^{\theta_1}+e^{\theta_2}+g_3(x,y,t)}{1+f_2(x,y,t)+f_4(x,y,t)}  exp\left[-\frac{i}{2}\frac{d U}{dt}(x^2+y^2)+U(t) +\int M(t) dt\right]  \label{uy}
\end{equation}\\
  In this form we can now look at profiles and possible interactions, while observing the effects of some key parameters.

\begin{figure}[!h]
	\centering
	\includegraphics[width=0.7\linewidth]{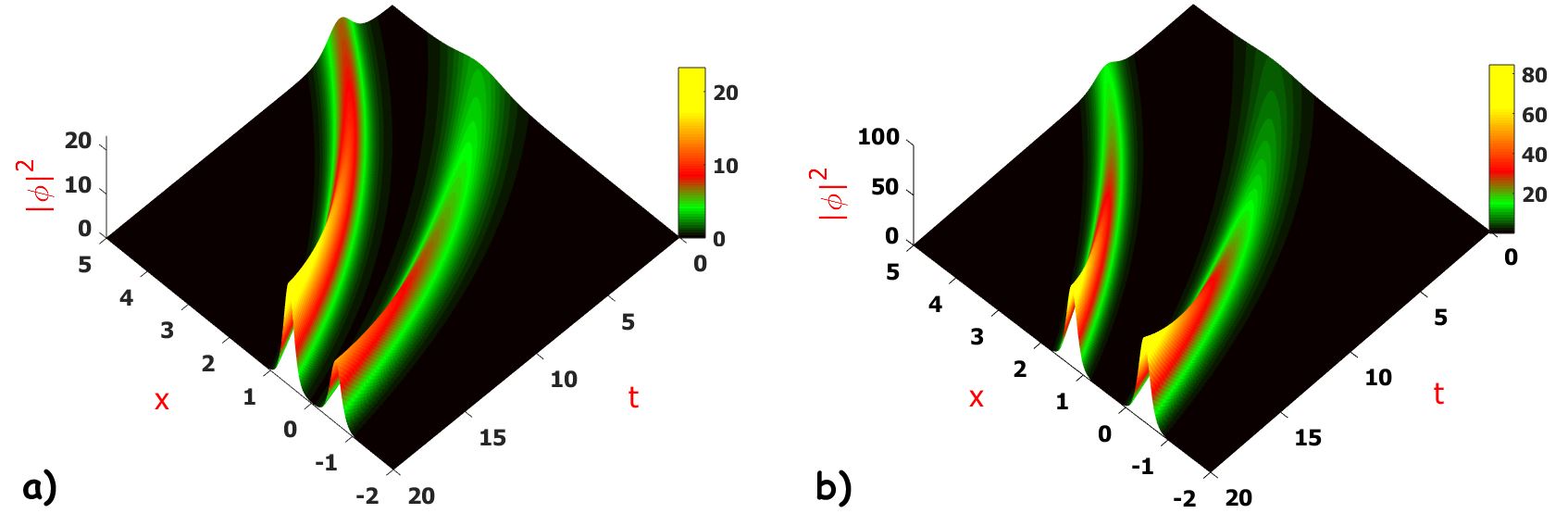}
	\includegraphics[width=0.7\linewidth]{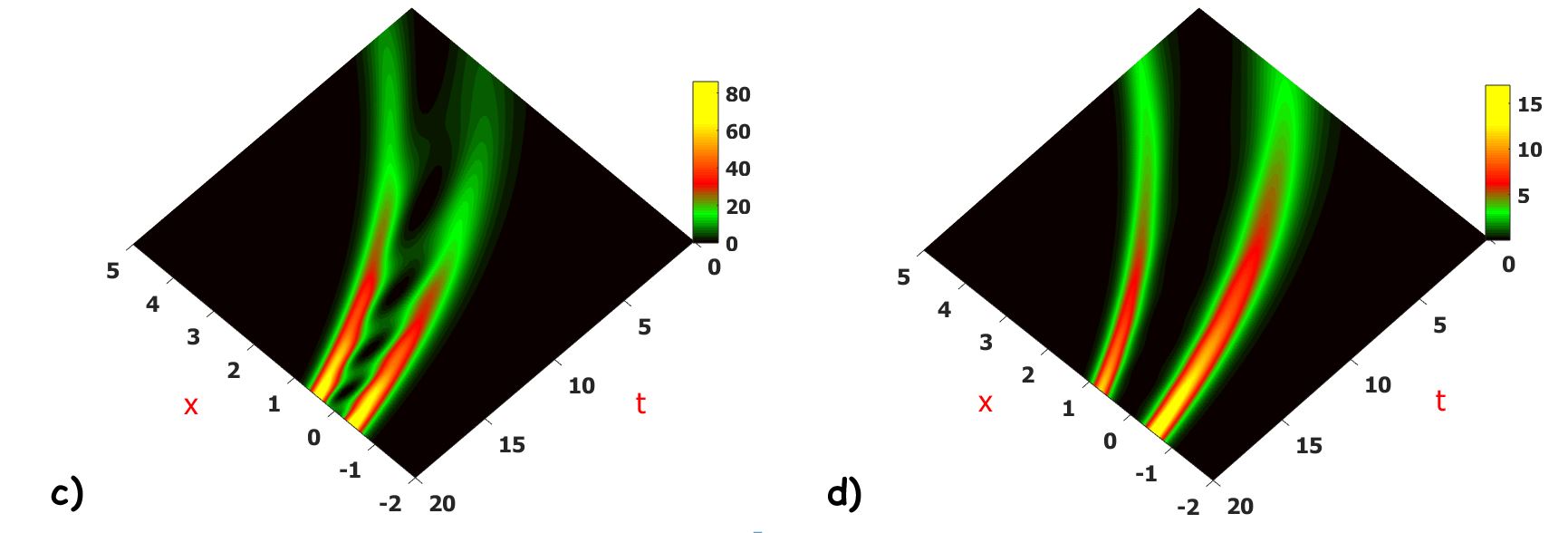}
	\caption{ Evolution of two-soliton solution  (\ref{u1}) characterized by two second-order rogue matter wave for the parameters $\beta_{2}=1.4$, $\beta_4=2.2$, $g_{02}=-1$, $\chi_{0}=0.25g_{02}$ and $C(t)=C_+(t)$  with \textbf{a)} Absence of HO interaction effects and effect of the values of the initial phases $k_{01}$ and $k_{02}$ for  $k_{01}=-10$, $k_{02}=-5$, $\beta_{1}=1.5$, $\beta_{3}=2$, $\eta_{0}=0$, \textbf{b)} Effects of the HO interaction on two second-order rogue matter wave of curved trajectory for $k_{01}=-10$, $k_{02}=-5$, $\beta_{1}=1.5$, $\beta_{3}=2$, $\eta_{0}=-0.15$,  \textbf{c)}  Effects of the imaginary parts of $\beta_{1}$ and $\beta_{3}$ for $k_{01}=-10$, $k_{02}=-5$, $\beta_{1}=1.5+3.5i$, $\beta_{3}=2-2.5i$, $\eta_{0}=-0.15$  and \textbf{d)}  Effects of the HO interaction on the interaction of two second-order rogue matter wave obtained in Fig.(\ref{fig:Nk91}\textbf{c)} for $k_{01}=-10$, $k_{02}=-5$, $\beta_{1}=1.5+3.5i$, $\beta_{3}=2-2.5i$, $\eta_{0}=-0.01$.}
	\label{fig:Nk91}
\end{figure}

\begin{figure}[!h]
	\centering
	\includegraphics[width=12cm,height=4.3cm]{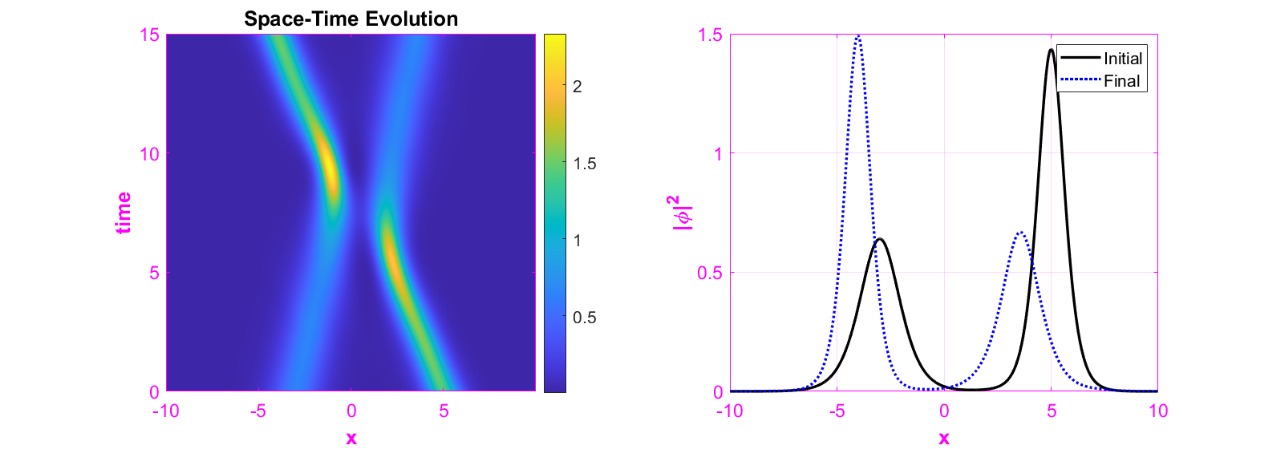}
	\caption{Numerical space-time evolution of two second-order RMWs in fig.(\ref{fig:Nk91}). (The full dynamics of this figure is available as movie among the supplemental material)}
	\label{NumFig13}
\end{figure}

  	\subsection{\textbf{Study of the two-soliton solution for $\phi(x,y,t)=\phi(x,y=1,t)$ : interaction of two second-order rogue matter waves}}
  
We focus in this subsection on studying the two-soliton solution when the spatial coordinate $y=1$. We then study the interaction process of the solitons generated. The two-soliton solution Eq.(\ref{uy}) is therefore put in the form : 	

 \begin{equation}
	\phi(x,y=1,t)=\frac{e^{\theta_1}+e^{\theta_2}+g_3(x,1,t)}{1+f_2(x,1,t)+f_4(x,1,t)}  exp\left[-\frac{i}{2}U_0(x^2+1)+U_0t +\frac{M_0}{2}cos(2t)\right]  \label{u1}
\end{equation}

\begin{figure}[!h]
	\centering
	\includegraphics[width=12cm,height=3.3cm]{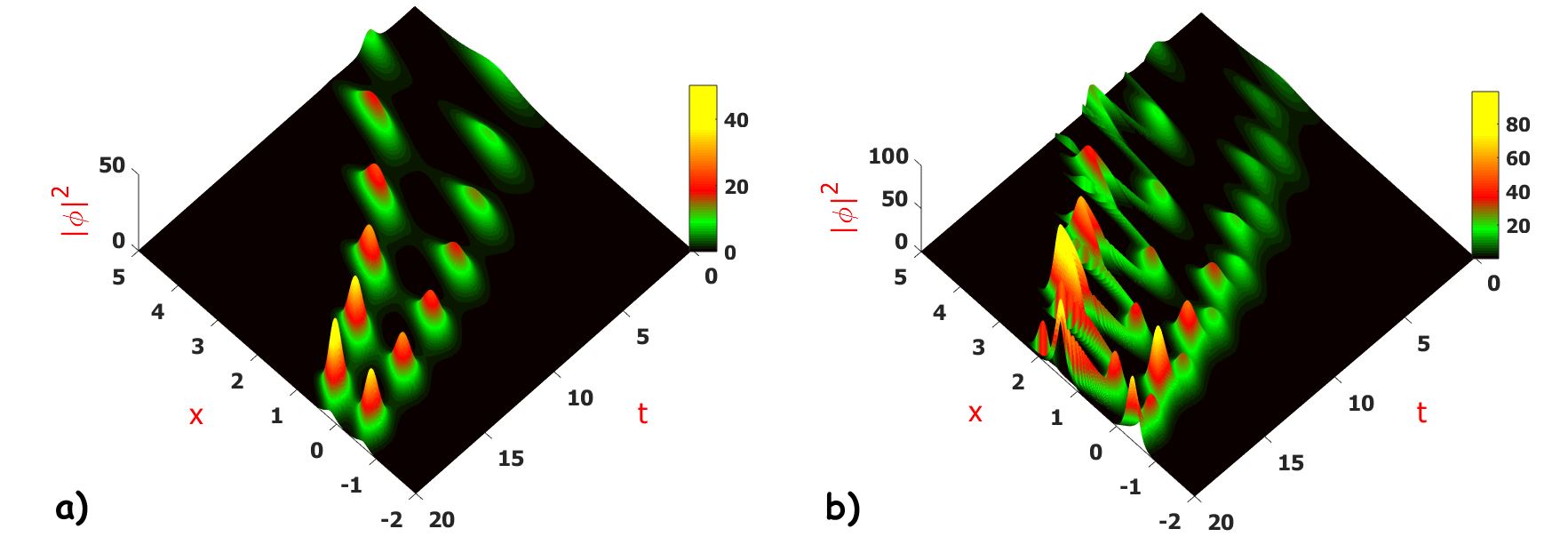}
	\includegraphics[width=12cm,height=3.3cm]{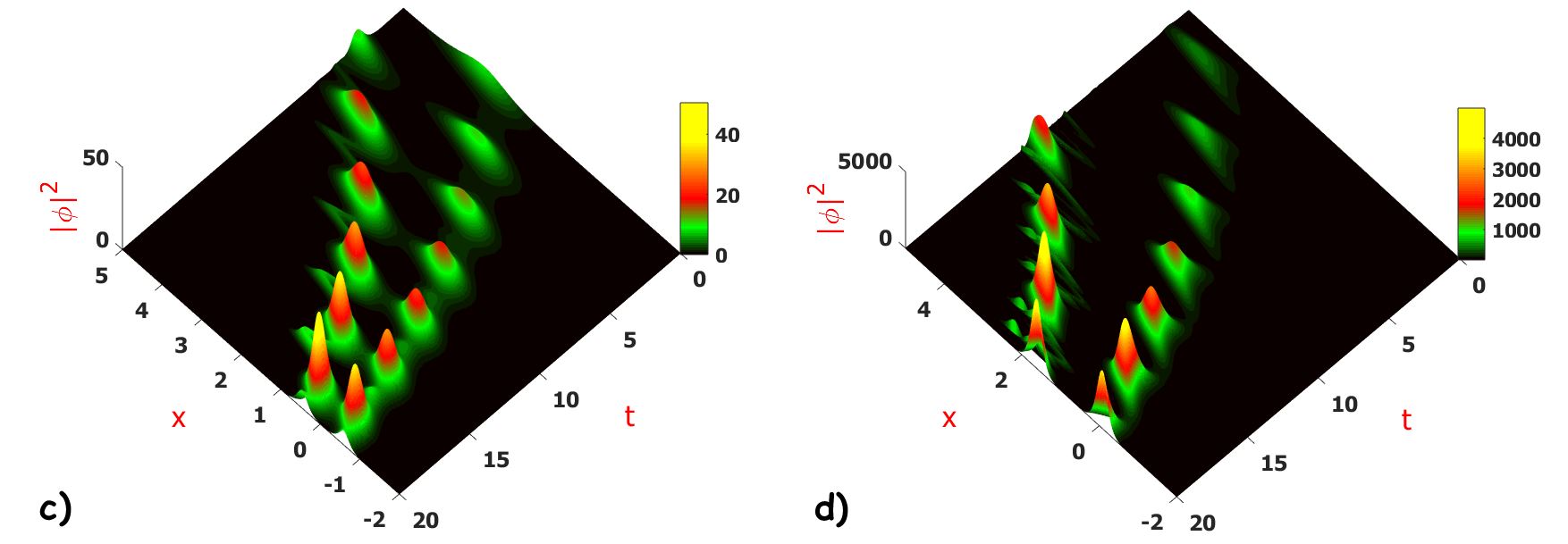}
	\caption{ Evolution of two-soliton solution  (\ref{u1}) for $\beta_{1}=1.4$, $\beta_{2}=1.4$, $\beta_{3}=2.5$, $\beta_4=2.2$,  $M_0=0$, $g_{02}=-1.5$, $\chi_{0}=0.33g_{02}$, $k_{01}=-10$, $k_{02}=-5$ and $C(t)=C_+(t)$ with \textbf{a)} $M_0=0.7$, $\eta_0=0$, \textbf{b)} $M_0=0.7$, $\eta_0=-0.1$, \textbf{c)} $M_0=0.7$, $\eta_0=-0.01$ and \textbf{d)} $M_0=0.7$, $\eta_0=-5$.}
	\label{fig:Nk20}
\end{figure}

\begin{figure}[!h]
	\centering
	\includegraphics[width=12cm,height=7cm]{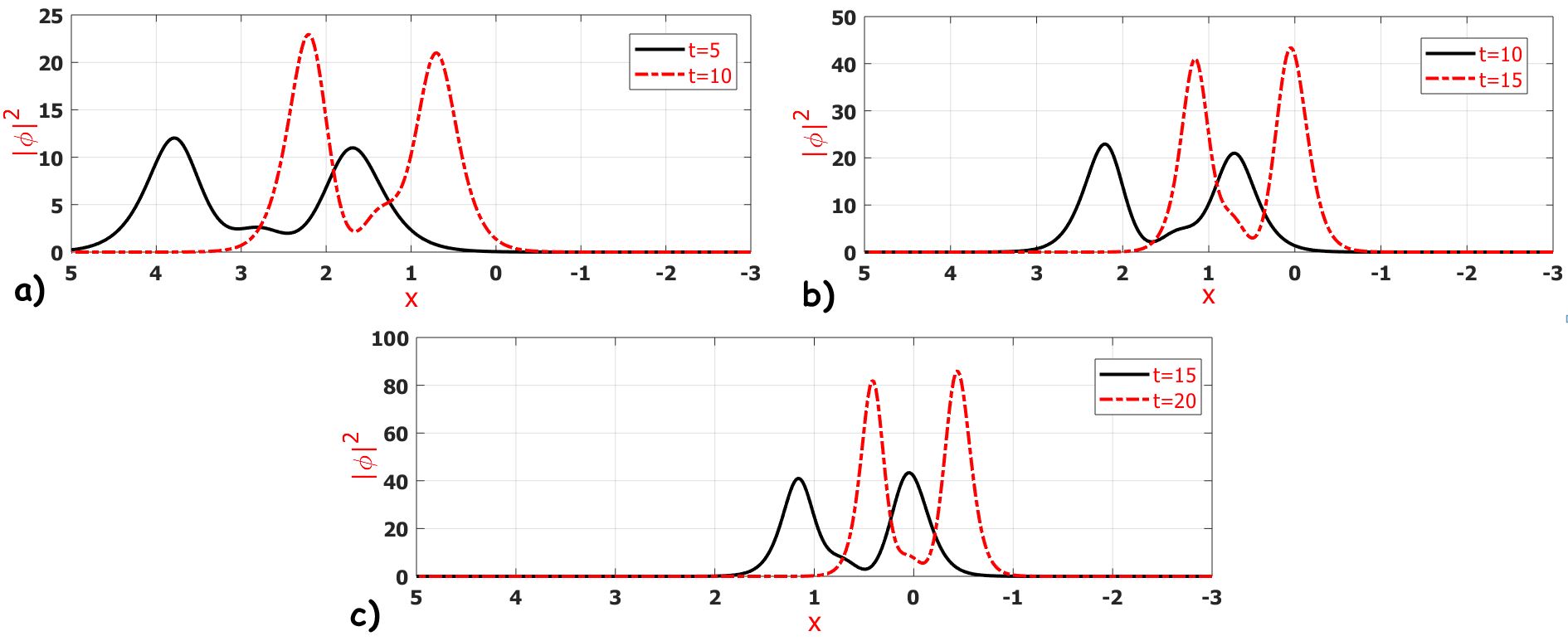}
	\caption{  \textbf{a)}, \textbf{b)} and \textbf{c)} interaction process of two second-order rogue matter wave obtained in Fig.(\ref{fig:Nk91}\textbf{c)}  starting  from $t=5$ to $t=20$. }
	\label{fig:Nk93}
\end{figure}

The square module of solution Eq.(\ref{u1}), is found in Figs.(\ref{fig:Nk90})-(\ref{fig:Nk11}) 
for both analytical and numerical results. In general, there is a convergence between the two results as to confirm our findings. Especially in figs.(\ref{NumFig12} and \ref{NumFig13}) are presented the space-time evolution of analytical and numerical trajectories of the two solitons showing their initial and final states amplitudes. The time is fixed at $t=15.00$ and the rich dynamics is clearly  evidenced in confirmation of analytical presented results. It always appears that both approached provide convergent results.

\subsubsection{Spatio-temporal evolution of two second-order parallel RMWs}

For different values of the parameters $\beta_{1}=1.5$, $\beta_{2}=2.4$, $\beta_{3}=3$, $\beta_4=-2$,  $M_0=0$, $g_{02}=-0.5$, $\chi_{0}=0.5g_{02}$, $\eta_0=-0.15$, $k_{01}=3$, $k_{02}=1$ and $C(t)=C_+(t)$, we can obtain Fig.(\ref{fig:Nk90}\textbf{a)} and  Fig.(\ref{fig:Nk90}\textbf{c)} characterizing the spatio-temporal evolution of two second-order rogue matter wave. This figure highlights the evolution of the two second-order RMWs by keeping almost the same separation over time. We also note that the values of $\beta_1=1.5$ and $\beta_3=3$ control the space occupation interval, which allows to know what parameter controls one of the two solitons for possible experimental analyses.

In addition, Fig.(\ref{fig:Nk90}\textbf{a)} represents the spatiotemporal evolution of two second-order rogue matter wave in the absence of HO interaction effects. Considering the effects of HO, an increase the amplitude of the two second-order rogue matter wave is observed in Fig.(\ref{fig:Nk90}\textbf{c)}. It is thus possible through the HO interaction to control the amplitude of two second-order rogue matter wave as in the case of one second-order rogue matter wave.

\subsubsection{Effect of the HO interactions and gain/loss on the interaction of two second-order rogue matter waves of curved trajectory}

Fig.(\ref{fig:Nk91}) describes the evolution of two second-order rogue matter waves whose trajectory is modified over time due to the effects of the initial values of phases $\theta_1$ and $\theta_2$, and this for the values of the parameters  $\beta_{1}=1.5$, $\beta_{2}=1.4$, $\beta_{3}=2$, $\beta_4=2.2$,  $M_0=0$, $g_{02}=-1$, $\chi_{0}=0.25g_{02}$, $\eta_0=-0.15$, $k_{01}=-10$, $k_{02}=-5$ and $C(t)=C_+(t)$. 	

By considering the imaginary parts of the parameters $\beta_1$ and $\beta_3$ of the phase $\theta_1$, we observe in Fig.(\ref{fig:Nk91}\textbf{c)} that for $\beta_1=1.5+3.5i$ and $\beta_3=2-2.5i$, the two second-order RMWs realize an elastic type interaction over time without change in amplitude. By then focusing on the effects of the HO interaction on the interaction dynamics, it is observed in Fig.(\ref{fig:Nk91}\textbf{d)} that by increasing the value of the parameter $\eta_0$ from $-0.15$ to $-0.01$, the following two observations can be revealed : the amplitude of the two second-order rogue matter wave decreases on the one hand and on the other hand the phenomenon of elastic interaction is dissipated, and the second-order rogue matter wave set continues its propagation without interaction.

\begin{figure}[!h]
	\centering
	\includegraphics[width=12cm,height=3.3cm]{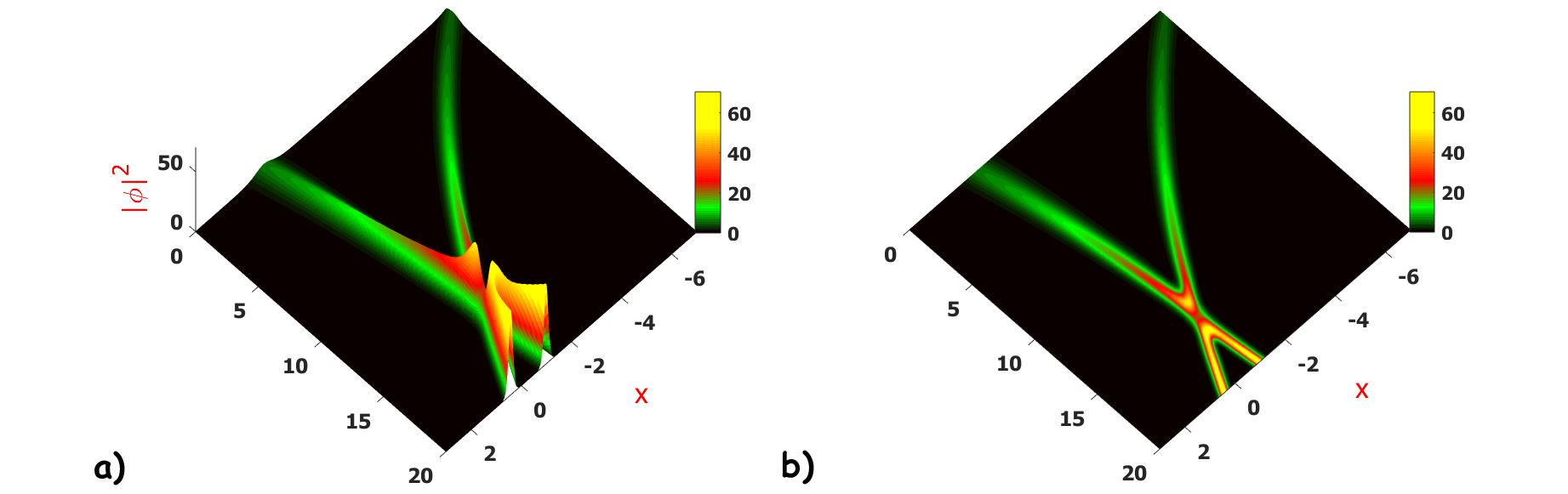}
	\caption{  Evolution of two-soliton solution  Eq.(\ref{u1}) characterized by interaction of two second-order rogue matter wave  with \textbf{a)} $\beta_{1}=2.2+5.5i$, $\beta_{2}=2.2+6i$, $\beta_{3}=2.2-8i$, $\beta_4=-2.2-5.2i$,  $M_0=0$, $g_{02}=-2$, $\chi_{0}=0.15g_{02}$, $\eta_0=-0.1$, $k_{01}=-2.1$, $k_{02}=20.5$ and $C(t)=C_+(t)$ and \textbf{b)} Top view.}
	\label{fig:Nk10}
\end{figure}

\begin{figure}[!h]
	\centering
	\includegraphics[width=12cm,height=3.3cm]{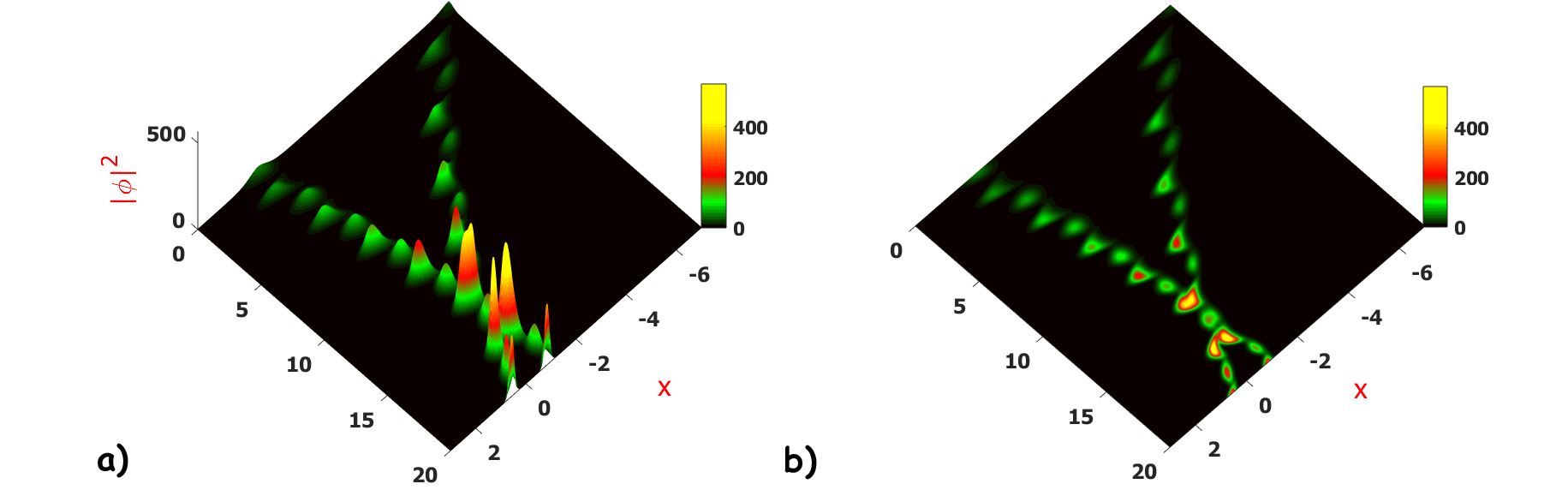}
	\caption{ \textbf{a)} Evolution of two-soliton solution  Eq.(\ref{u1}) characterized by interaction of two second-order rogue matter wave with effect of gain/loss for the parameters $\beta_{1}=2.2+5.5i$, $\beta_{2}=2.2+6i$, $\beta_{3}=2.2-8i$, $\beta_4=-2.2-5.2i$,  $M_0=1.2$, $g_{02}=-2$, $\chi_{0}=0.15g_{02}$, $\eta_0=-0.1$, $k_{01}=-2.1$, $k_{02}=20.5$ and \textbf{b)} Top view.}
	\label{fig:Nk11}
\end{figure}

Fig.(\ref{fig:Nk20}) highlights respectively the effects of the gain/loss term and that of the HO parameter in the presence of the latter. In the absence of the HO interaction, it is observed in Fig.(\ref{fig:Nk20}\textbf{a)} that the term gain/loss also leads in this case to the formation of areas of collapse and revival. Therefore, taking into account the effects of the parameter $\eta_0$, the following facts are observed in Fig.(\ref{fig:Nk20}\textbf{b)} : the amplification of the two second-order rogue matter wave as well as a repulsive interaction. By increasing the value of $\eta_0$ from $-0.1$ to $5$, a decrease the amplitude of the two second-order rogue matter wave is observed followed by a dissipation of the interaction phenomenon as shown in Fig.(\ref{fig:Nk20}\textbf{c)}. By decreasing the value of $\eta_0$ from $-0.08$ to $-2$, the two second-order rogue matter wave in interaction sees in this case its amplitude increase and a total dissipation of the repulsive interaction phenomenon leaving room for a clear observation of the areas of collapse and revival over time as shown in Fig.(\ref{fig:Nk20}\textbf{d)}.

Fig.(\ref{fig:Nk93}) thus illustrates the interaction process obtained in Fig.(\ref{fig:Nk91}\textbf{c)}. It is well observed that for $t$ varying from $0$ to $15$, the two solitons interact and then separate each continuing its propagation, which thus evidence the elastic type interaction process.

It is clear from all these interpretations carried out that the control of the amplitude of the two second-order rogue matter waves can be done through a suitable  value of the HO parameter. Moreover, it can also shown from these analyses that there is an interplay between the imaginary parts of the parameters $\beta_1$ and $\beta_3$, and the HO parameter interaction. Indeed, as much as  the imaginary parts of $\beta_1$ and $\beta_3$ favour the interaction of the two second-order rogue matter waves, the HO parameter interaction can favour a dissipation of the interaction phenomenon. This could be beneficial in various processes of interaction of information in telecommunications.

\subsubsection{Effect of gain/loss on the interaction of two second-order rogue matter waves of different trajectories}

	\begin{figure}[!h]
	\centering
	\includegraphics[width=12cm,height=3.3cm]{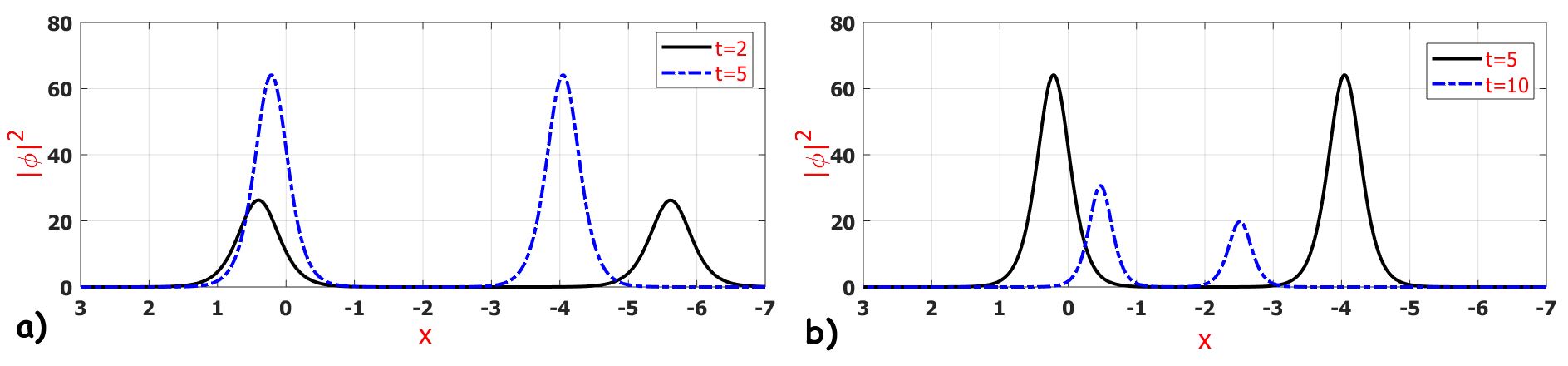}
	\includegraphics[width=12cm,height=3.3cm]{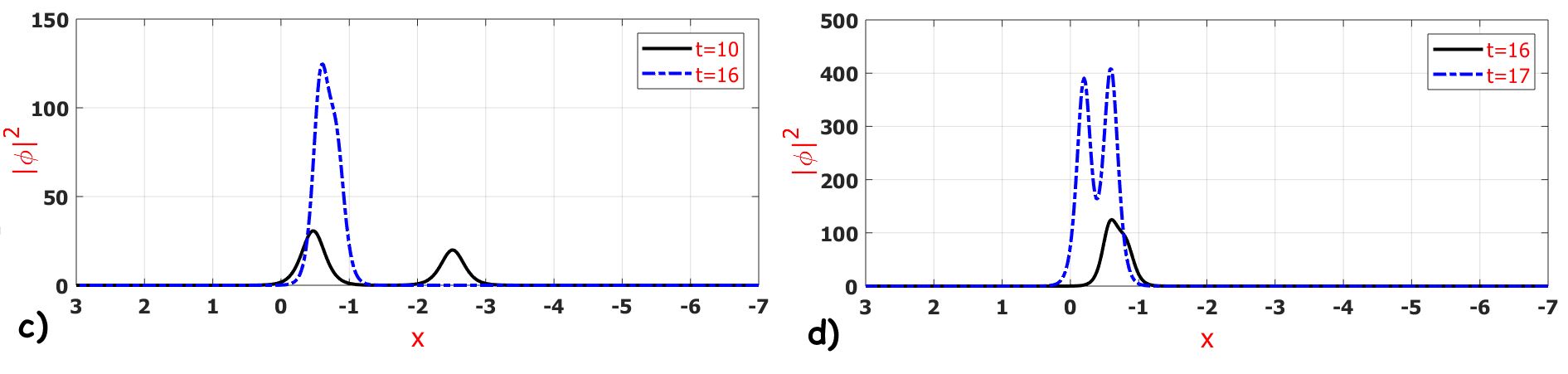}
	\includegraphics[width=12cm,height=3.3cm]{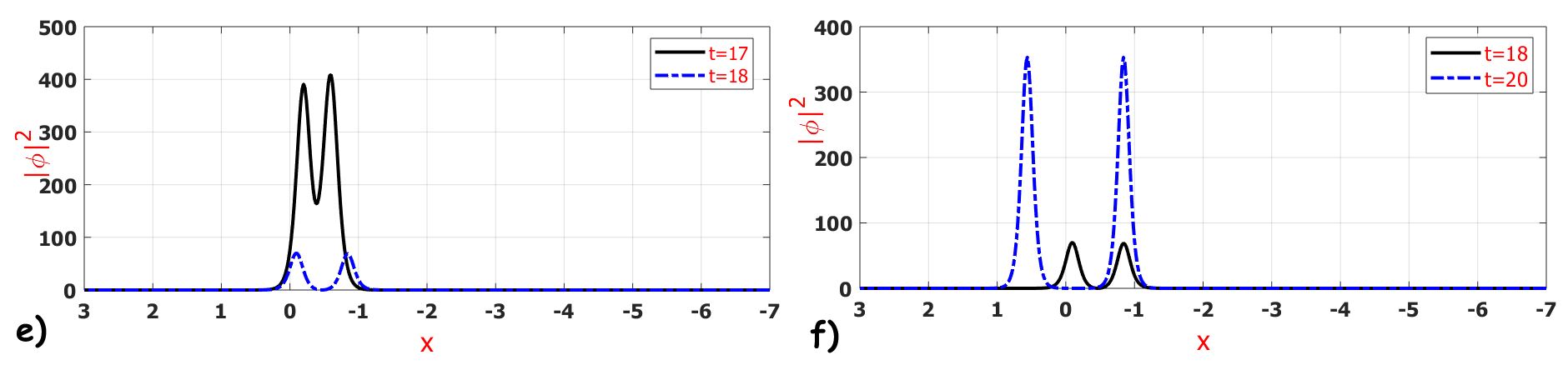}
	\caption{  \textbf{a)}, \textbf{b)}, \textbf{c)}, \textbf{d)}, \textbf{e)} and \textbf{f)} interaction process of two second-order rogue matter wave obtained in Fig.(\ref{fig:Nk11}\textbf{a)}  starting  from $t=0$ to $t=20$.}
	\label{fig:Nk13}
\end{figure}
Fig.(\ref{fig:Nk10}) illustrates the spatiotemporal evolution of two second-order rogue matter wave, one of which is of almost centered trajectory and the other of curved trajectory in time. The values of the chosen parameters are given by  $\beta_{1}=2.2+5i$, $\beta_{2}=2.2+6i$, $\beta_{3}=2.2-8i$, $\beta_4=-2.2-5.2i$, $g_{02}=-2$, $\chi_{0}=0.15g_{02}$, $\eta_0=-0.1$, $k_{01}=-2.1$, $k_{02}=-20.5$ and $C(t)=C_+(t)$. 

Taking into account the effects of gain/loss ($M_0=1.2$) on the interaction dynamics obtained in Fig.(\ref{fig:Nk10}\textbf{a)}) there is a creation of collapse zones and this with variation of the maximum amplitude ($500$). By representing the interaction process we can obtain the Figs.(\ref{fig:Nk11}\textbf{a)} - (\ref{fig:Nk11}\textbf{f)}. We observe in Fig.(\ref{fig:Nk11}\textbf{a)} that at $t=0.8$ we are in a collapse zone characterised by an almost zero amplitude of the two second-order rogue matter wave. Between $t=0$ and $t=8$, the second-order rogue matter wave of curved trajectory approcimates that of almost centered trajectory so as to interact between $t=1.6$ to $t=16$ as shown in Fig.(\ref{fig:Nk11}\textbf{e)}. After interaction, the two solitons separate and continue their propagation as observed in Fig.(\ref{fig:Nk11}\textbf{f)} from $t=18$ to $t=20$.

\begin{figure}[!h]
	\centering
	\includegraphics[width=12cm,height=3.3cm]{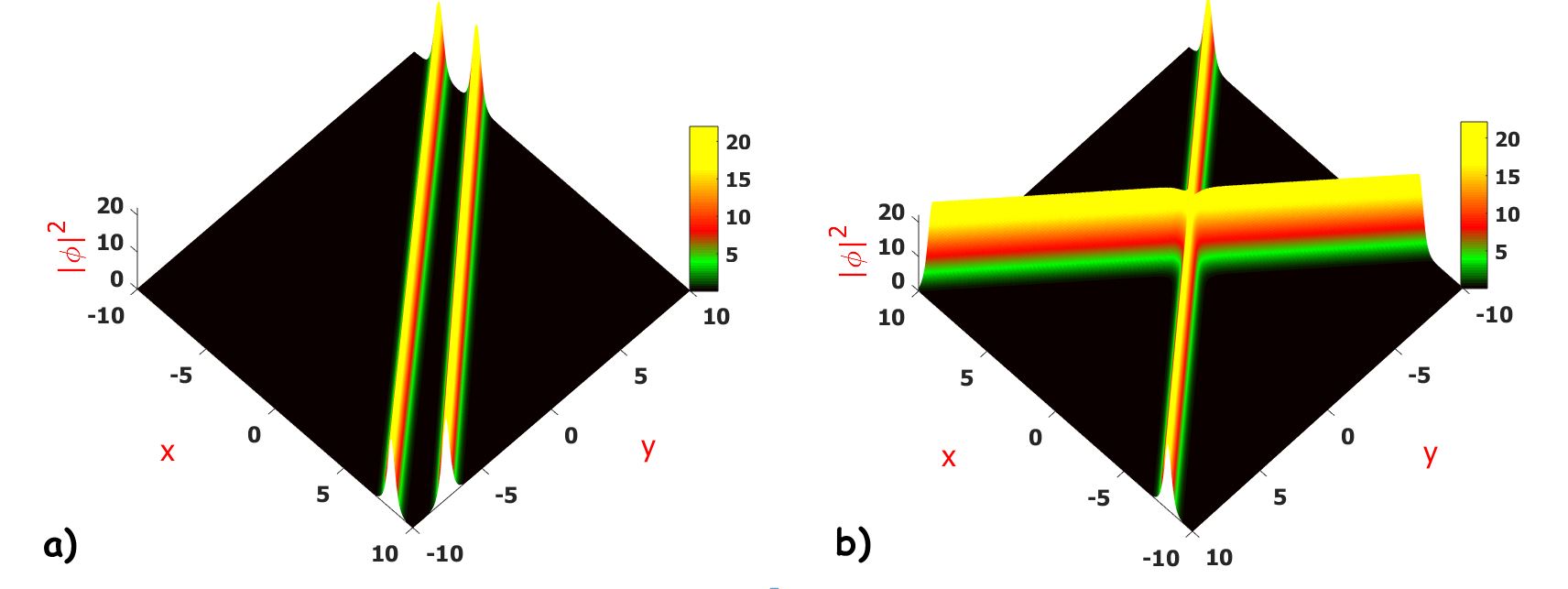}
	\includegraphics[width=12cm,height=3.3cm]{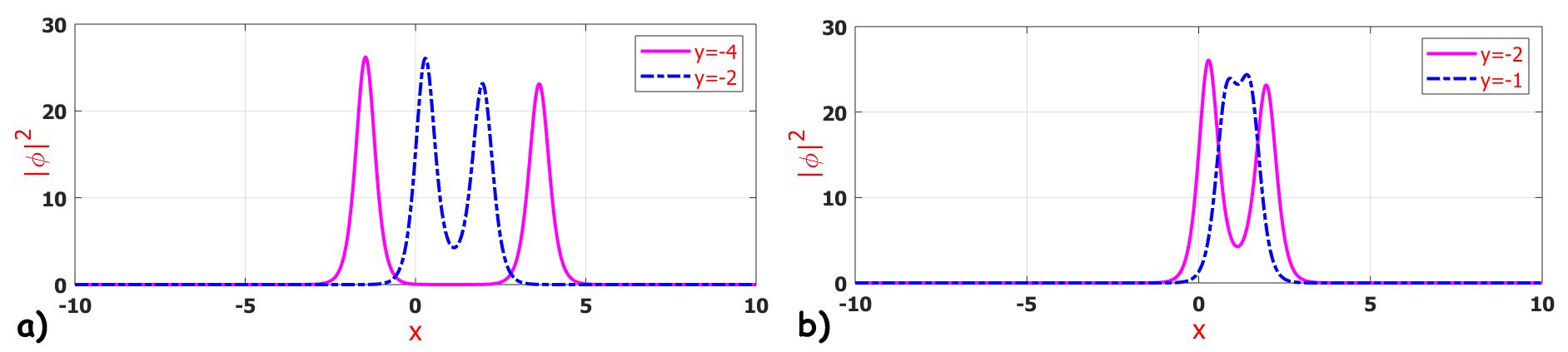}
	\includegraphics[width=12cm,height=3.3cm]{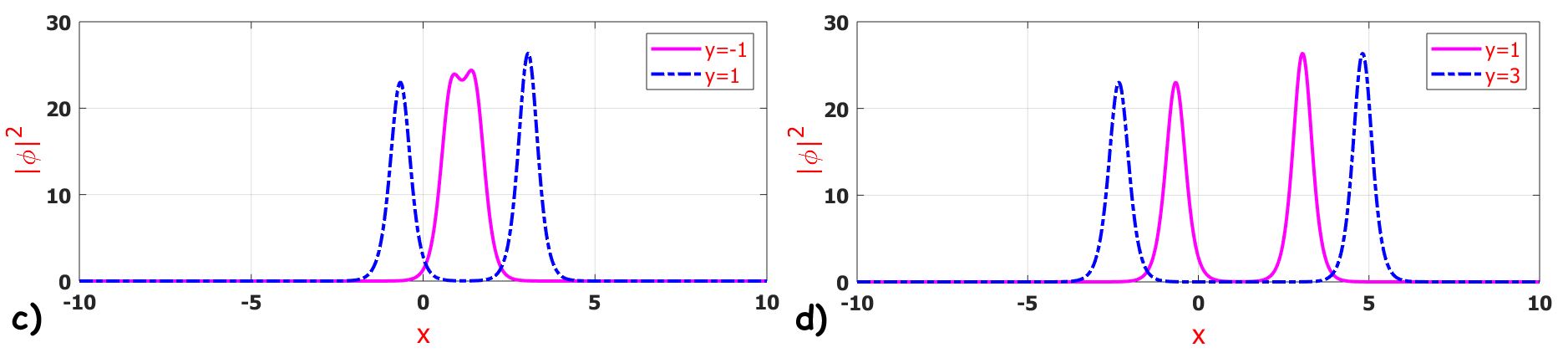}
	\caption{  \textbf{a)} Evolution of two-soliton solution  (\ref{u2}) characterized by two Line-soliton for the parameters $\beta_{1}=2.5$, $\beta_{2}=2.4$, $\beta_{3}=2.2$, $\beta_4=2$,  $M_0=1$, $g_{02}=-1$, $\chi_{0}=0.25g_{02}$, $\eta_0=-0.15$, $k_{01}=-3$, $k_{02}=2$, \textbf{b)}  Evolution of two-soliton solution  Eq.(\ref{u2}) characterized by the interaction of two Line-soliton for the same used in Fig.(\ref{fig:Nk12}\textbf{a)}  except $\beta_3=-2.2$, \textbf{c)}, \textbf{d)}, \textbf{e)} and \textbf{f)}  interaction process of two Line-soliton obtained in Fig.(\ref{fig:Nk12}\textbf{b)}  starting  from $y=-4$ to $y=3$.}
	\label{fig:Nk12}
\end{figure}

\begin{figure}[!h]
	\centering
	\includegraphics[width=12cm,height=3cm]{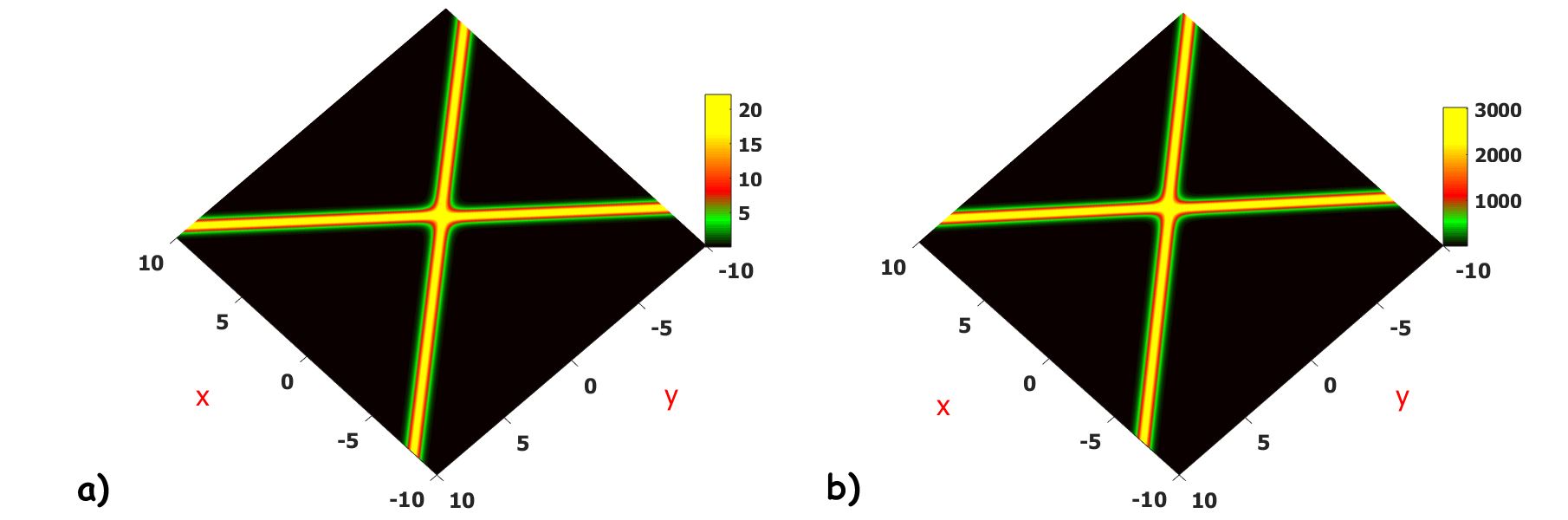}
	\caption{  Effects of HO interaction on evolution of interaction of two Line-soliton obtained in Fig.(\ref{fig:Nk12}\textbf{b)} for the parameters  $\beta_{1}=2.5$, $\beta_{2}=2.4$, $\beta_{3}=2.2$, $\beta_4=2$,  $M_0=1$, $g_{02}=-1$, $\chi_{0}=0.25g_{02}$,  $k_{01}=-3$, $k_{02}=2$ \textbf{a)} $\eta_0=-0.15$  and \textbf{b)} $\eta_{01}=-20.15$.}
	\label{fig:Nk18}
\end{figure}

\subsection{\textbf{Study of the two-soliton solutions for $\phi(x,y,t)=\phi(x,y,t=1)$ : interaction of two Line-solitons}}

We focus in this subsection on studying the two-soliton solution when we fix the temporal coordinate $t=1$. We then study the interaction process of the solitons obtained. The two-soliton solution Eq.(\ref{uy}) is therefore put in the form : 	

\begin{equation}
	\phi(x,y,t=1)=\frac{e^{\theta_1}+e^{\theta_2}+g_3(x,y,1)}{1+f_2(x,y,1)+f_4(x,y,1)}  exp\left[-\frac{i}{2}U_0(x^2+y^2)+U_0 +\frac{M_0}{2}cos(2)\right]  \label{u2}
\end{equation}\\

\begin{figure}[!h]
	\centering
	\includegraphics[width=12cm,height=3.3cm]{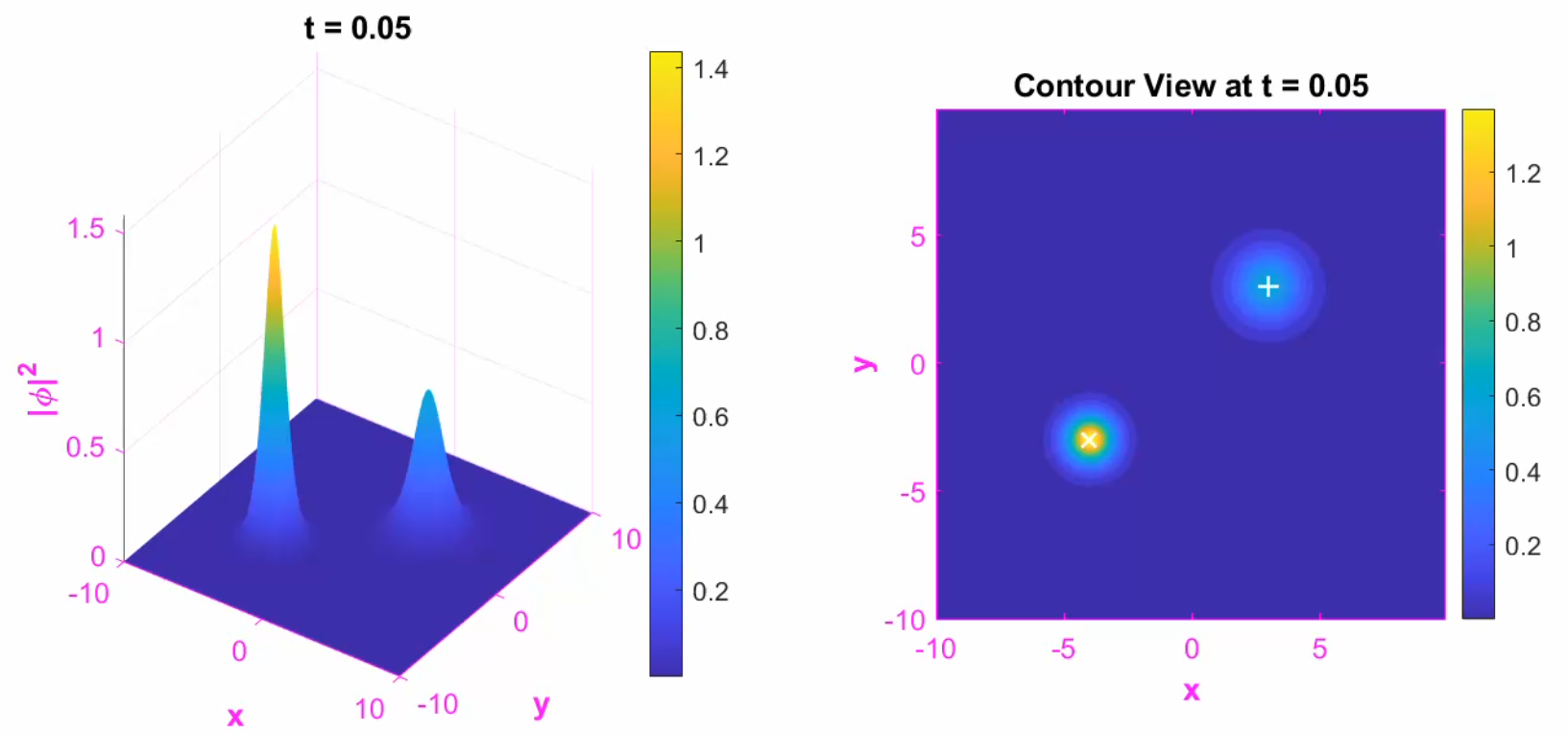}
	\includegraphics[width=12cm,height=3.3cm]{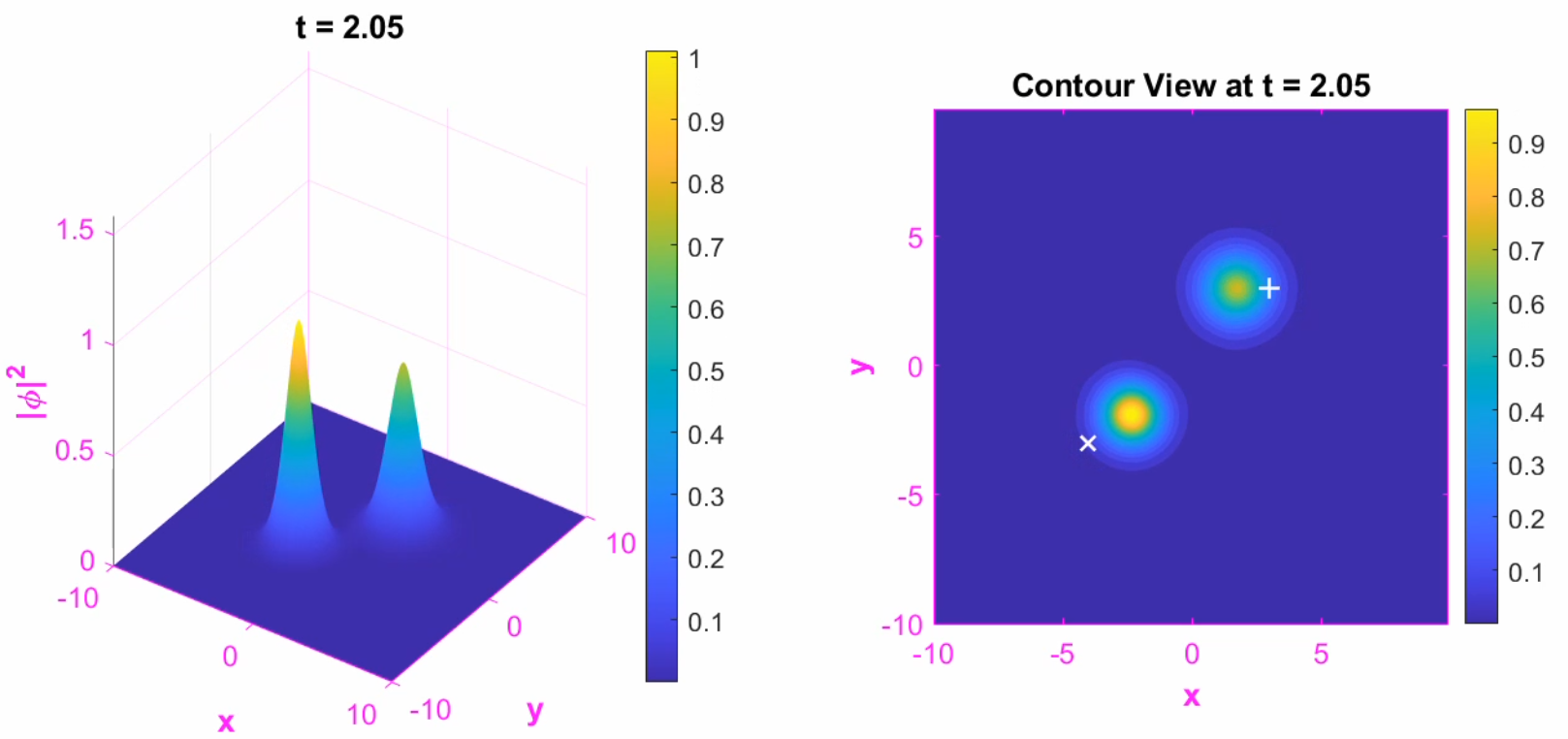}
	\includegraphics[width=12cm,height=3.3cm]{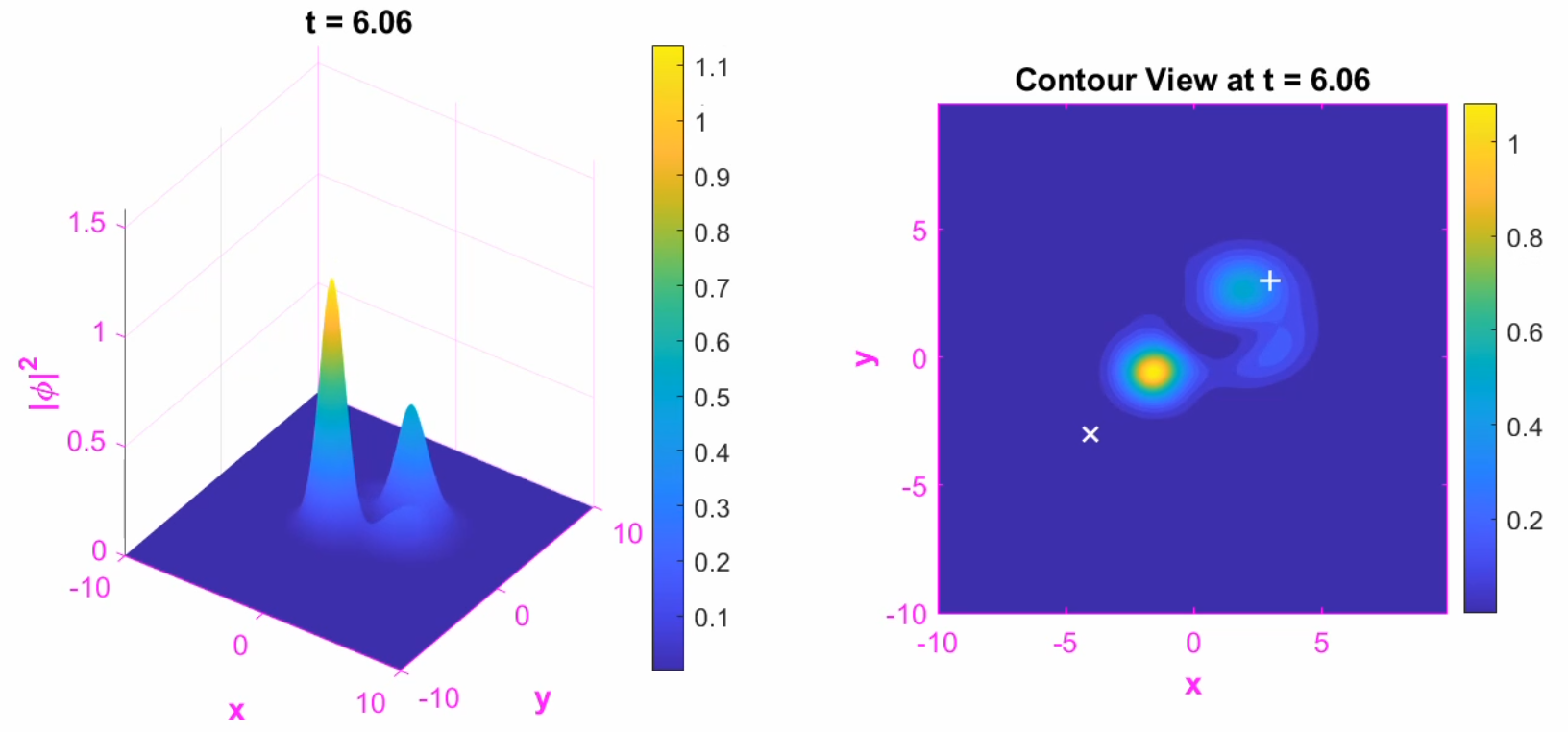}
	\includegraphics[width=12cm,height=3.3cm]{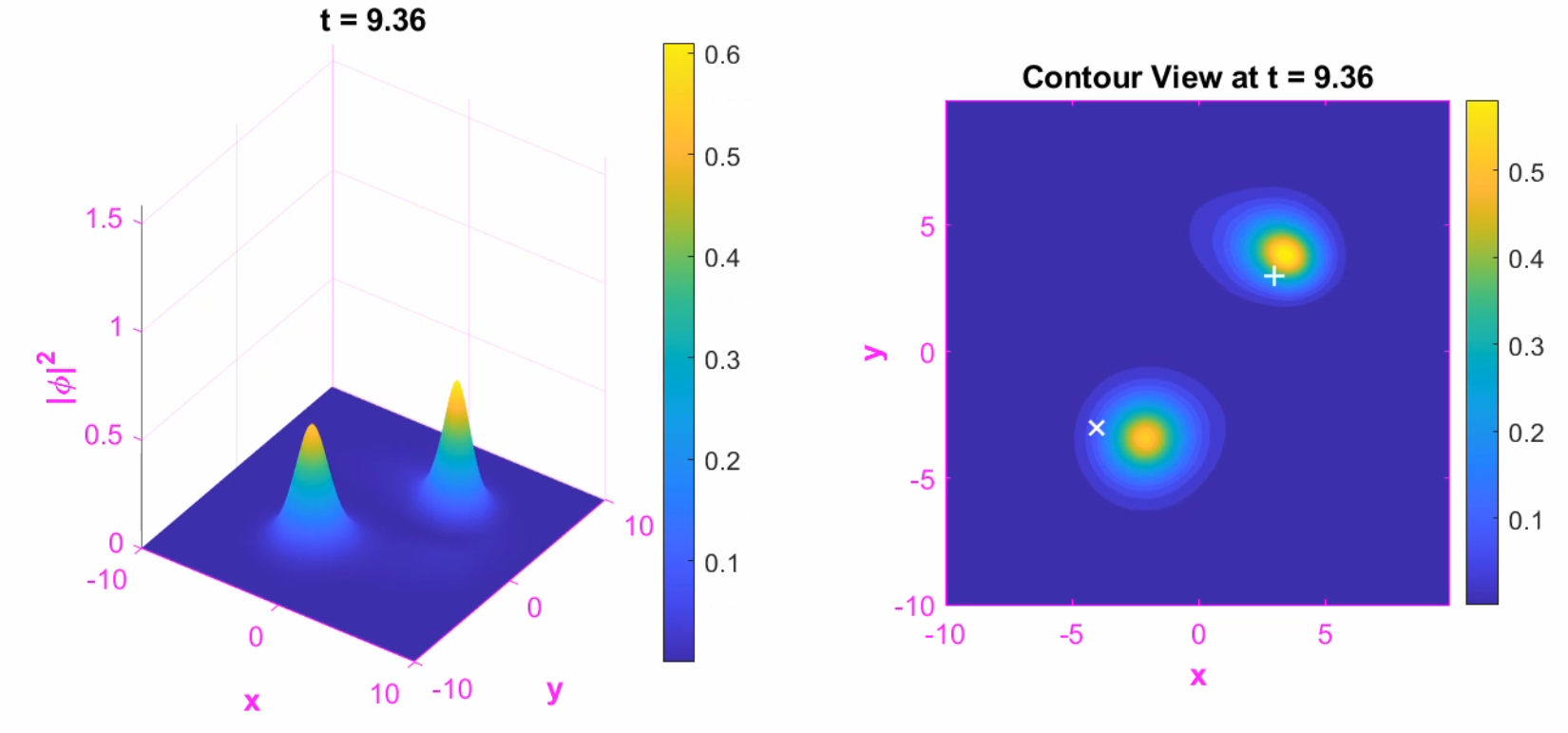}
	\caption{ Numerical dynamics of two-soliton solutions through numerical integration of Eq.(\ref{b1}). First column is the square module in 2D space, and second column is the corresponding contour view, all for time varying from time=0.05 (top) to time=9.36 (bottom). (The full dynamics of this figure is available as movie among the supplemental material)}
	\label{2sol2d}
\end{figure}

To  better study the interaction of two Line-solitons, we first start from two parallel Line-solitons as shown in Fig.(\ref{fig:Nk12}\textbf{a)} for the parameters   $\beta_{1}=2.5$, $\beta_{2}=2.2$, $\beta_{3}=2.2$, $\beta_4=2$, $g_{02}=-1$, $\chi_{0}=0.25g_{02}$, $\eta_0=-0.15$, $k_{01}=-3$, $k_{02}=2$, $M_0=1$ and $C(1)=C_+(1)$. By just changing the sign of parameter $\beta_3$ of the phase $\theta_1$ from $2.2$ to $-2.2$, we can observe a change in the propagation phase of one of the two Line-soliton. In this case, we therefore observe an interaction of the two Line-solitons as we can see in Fig.(\ref{fig:Nk12}\textbf{b)}. In order to better observe this interaction, we represent the interaction process as carried out on Figs.(\ref{fig:Nk12}\textbf{c)}, (\ref{fig:Nk12}\textbf{d)}, (\ref{fig:Nk12}\textbf{e)} and (\ref{fig:Nk12}\textbf{f)}. We see on the Figs.(\ref{fig:Nk12}\textbf{e)} and (\ref{fig:Nk12}\textbf{f)} that from $y=-1$ to $y=3$ the two solitons after interaction separate and continue their propagations while keeping their amplitudes constant. This would be most benefical for possible future experimental applications. The role of the coefficient of the HO interaction on the interaction of the two Line-solitons can be observed in Fig.(\ref{fig:Nk18}). We find that when decreasing the value of this coefficient from $\eta_0=-0.15$ to $\eta_0=-20.15$, the amplitude of the two Line-soliton increases and their latter undergo a displacement towards the increasing $x$.

A picture of direct numerical integration of Eq.(\ref{b1}) presented in Fig.(\ref{2sol2d}) shows how the 2D two-soliton solutions are obtained and evolve in time varying from $t=0.05$ to $t=9.36$. It appears that for some values of the set of parameters, the one-soltion moves according to anylitical predictions and hence the confirmation on the acuracy of our theoretical results. It is not worthy mentionning here the elastic interaction process that occur during propagation.

	\section{Conclusion}\label{concl} \label{sec6}
		
In conclusion, we investigated the two-dimensional modified Gross-Pitaevskii equation incorporating gain/loss effects and a time-independent isotropic confining potential. Using Hirota's bilinear method, we constructed families of one-soliton and multi-soliton solutions. By first fixing the spatial coordinate $y=1$, we generated second-order rogue matter waves with spatio-temporal localization from the one-soliton solution. In the presence of a time-independent confining potential ($K<0$, with $K=-0.05\sqrt{2}$), we demonstrated that this potential amplifies the amplitude of the second-order rogue matter wave over time.
Next, we fixed the temporal coordinate in the one-soliton solution, enabling to generate a line-soliton with double spatial localization. Our results revealed that gain/loss effects induce regions of collapse in the dynamics of the second-order rogue matter wave. However, for the line-soliton, gain/loss effects only lead to an increase of the wave amplitude without collapse. We established that controlling collapse due to gain/loss in trapped Bose-Einstein condensates requires selecting suitable values for the characteristic coefficient of the higher-order interaction, $\eta_0$.
By constructing multi-soliton solutions and varying parameters $k_{01}$, $k_{02}$, $\beta_{1}$,  $\beta_{2}$, $\beta_3$, $\beta_4$, $M_0$, $g_{02}$, $\chi_{0}$ and $\eta_{0}$, we explored different interaction scenarios involving second-order rogue matter waves and line-solitons. These interactions were found to be elastic, highlighting the intrinsic property of solitons to retain their characteristics after interacting with a system.
The findings indicate that gain and loss effects can trigger regions of collapse in second-order rogue wave dynamics while also facilitating the emergence of stable solitons. We further investigate the behavior of multi-solitons, delineating the conditions that facilitate the emergence of second-order rogue matter waves and line solitons by the regulation of interaction parameters. 
An in-depth examination of soliton propagation demonstrates that elastic-type collisions clearly conserve soliton properties after the collision. 
These results are confirmed by direct numerical integration of the problem where the one- and two-soliton solutions were obtained and the elastic interactions clearly evidenced.
Ultimately, our analyses demonstrate that, within the framework of Bose-Einstein condensates described by the two-dimensional modified Gross-Pitaevskii equation, higher-order interactions provide a means to control the amplitude of the generated matter waves. 
The exact analytical solutions derived in this study rigorously satisfy the original equation, which ensures their consistency with the numerical results and confirms their accuracy.
These findings hold promise for potential future applications.

\section*{	Credit authorship contribution statement}
\textbf{Cyrille Edgard Nkenfack} : Writing – original draft, Software, Investigation, Formal analysis, Data curation. \textbf{Olivier Tiokeng Lekeufack} : Writing – review and editing, Writing – original draft, Supervision, Project administration, Methodology, Conceptualization. 
\textbf{Subramaniyan Sabari} : Writing – review and editing, investigation, Visualization, Software,
Methodology. \textbf{Rene Yamapi} : Writing – review and editing, Visualization, Software,
Methodology. \textbf{Timoleon Crepin Kofane} : Writing – review and editing, Writing – original draft, Validation, Investigation, Formal analysis.
	
\section*{Declaration of competing interest}
The authors declare that they have no known competing financial interests or personal relationships that could have appeared to
influence the work reported in this paper.

\section*{Data availability}
Some data were used for the research described in the article and are available upon request.

\section*{Compliance with ethical standards}
\textbf{Conflict of interest} The authors declare that they have no conflict of interest.
	
\section*{Acknowledgments}
The authors are indebted to the referees for their positive and critical comments on the paper, which helped to improve the work immensely. S.S. thanks Funda\c c\~ao de Amparo \`a Pesquisa do Estado de S\~ao Paulo [Contracts 2020/02185-1 and 2024/04174-8].
	
	\bigskip


\begin{thebibliography}{99}

\bibitem{Anderson1995} M.H. Anderson, J.R. Ensher, M.R. Matthews, C.E. Wieman and E.A. Cornell, \textit{Observation of Bose-Einstein Condensation in a dilute atomic vapor}, Science \textbf{269} (1995) 198.
		
\bibitem{Morsch2006}  O. Morsch and M. Oberthaler, \textit{Dynamics of Bose-Einstein condensates in optical lattices}, Rev. Mod. Phys. \textbf{78} (2006) 179.
		
\bibitem{Lenz1993} G. Lenz, P. Meystre, and E. M. Wright, \textit{Nonlinear atom optics} Phys. Rev. Lett. \textbf{71} 3271 (1993) 3271.
		
\bibitem{McKeever2004}   J. McKeever, A. Boca, A. D. Boozer, R.
Miller, J. R. Buck, A. Kuzmich, and H. J. Kimble,  \textit{Deterministic generation of single photons from one atom trapped in a cavity}, Science \textbf{303}  (2004)   1992. 
		
\bibitem{Gross1961}
E.P. Gross, \textit{ Structure of a quantized vortex in boson systems}, Il Nuovo Cimento Series \textbf{10} (1961) 454.
		
\bibitem{Pitaevskii1961}
L.P. Pitaevskii, \textit{Vortex lines in an imperfect bose gas}, Soviet Physics JETP-USSR, \textbf{13} (1961) 2.
		
\bibitem{Daifovo1999}
F. Daifovo, S. Giorgini, L.P. Pitaevskii and S. Stringari, \textit{Theory of Bose-Einstein condensation in trapped gases},  Rev. Mod. Phys. \textbf{71} (1999) 463.	
       
\bibitem{Kohler2006}
T. K\^ohler, K. G\'ora and P.S. Julienne, \textit{Production of cold molecules via magnetically tunable Feshbach resonances}, Rev. Mod. Phys. \textbf{78} (2006) 1311.

\bibitem{Gammal2000}
A. Gammal, T. Frederico, L. Tomio and Ph. Chomaz, \textit{Atomic Bose-Einstein condensation with three-body interactions and collective excitations}, J. Phys. B: At. Mol. Opt. Phys. \textbf{33} (2000) 4053.

		
\bibitem{Lekeufack2015} 
O.T. Lekeufack, S. Sabari, S.B. Yamgoue, K. Porsezian and T.C. Kofane, \textit{Quantum corrections to the modulational instability of Bose-Einstein condensates with two- and three-body interactions}, Chaos, Solitons and Fractals \textbf{76}  (2015) 111.
		
\bibitem{Chen2014} J. Chen, J. Yang and L. Zhang, \textit{Dynamics and matter-wave solitons in Bose-Einstein condensates with two- and three-Body interactions}, Advances in Condensed Matter Physics \textbf{2014} (2014) 1. 

\bibitem{Sabari2015} S. Sabari, K. Porsezian, R. Murali \textit{Modulational and oscillatory instabilities of Bose-Einstein condensates with two- and three-body interactions trapped in an optical lattice potential}, Phys. Lett. A \textbf{379}  (2015) 299.

\bibitem{Zinner2009} N.T. Zinner and M. Thøgersen, \textit{Stability of a Bose-Einstein condensate with higher-order interactions near a Feshbach resonance}, Phys. Rev. A \textbf{80} (2009) 023607.
		
\bibitem{Ping2009}
P. Ping and L. Guan-Qiang, \textit{Effects of three-body interaction on collective excitation and stability of Bose-Einstein condensate}, Chin. Phys. B \textbf{41} (2009) 3221.
		
\bibitem{He2010}
J. He and Y. Li, \textit{Designate integrability of the variable coefficient nonlinear Schr\"odinger equations}, Stud. Appl. Math. \textbf{126} (2010) 1.
		
\bibitem{Strecter2002} K.E. Strecker, G.B. Patridge, A.G. Truscott, and R.G. Hulet, \textit{Formation and propagation of matter-wave soliton trains}, Nature  (London), \textbf{417} (2002) 150.
		
\bibitem{Atre2006} 
R. Atre, P.K. Panigrahi, G.S. Agarwal, \textit{Class of solitary wave solutions of the one-dimensional Gross-Pitaevskii equation}, Phys. Rev. E \textbf{73} (2006) 056611.
		
\bibitem{Nkenfack2025} C.E. Nkenfack, O.T. Lekeufack, F. Kenmogne, R. Yamapi and E. Kengne, \textit{Elastic interaction of second-order rogue matter waves for the modified Gross–Pitaevskii equation with time-dependent trapping potential and gain/loss}, Chaos, Solitons and Fractals \textbf{191} (2025) 115820.
		
\bibitem{Burger1999} S. Burger, K. Bongs, S. Dettmer, W. Ertmer, K. Sengstock, A. Sanpera, G. V. Shlyapnikov, and M. Lewenstein, \textit{Dark solitons in Bose-Einstein condensates}, Phys. Rev. Lett. \textbf{83} (1999) 5198.
		
\bibitem{Liang2005} Z.X. Liang, Z.D. Zhang, and W M. Liu,  \textit{Dynamics of a bright coliton in Bose-Einstein condensates with time-dependent atomic scattering length in an pxpulsive parabolic potential} Phys. Rev. Lett. \textbf{94} (2005) 050402. 


\bibitem{Radha2007} R. Radha and V.R. Kumar, \textit{Bright matter wave solitons and their collision in Bose–Einstein condensates} Phys. Lett. A, \textbf{370} (2007) 46.

\bibitem{Petrov2016} D.S. Petrov and G.E. Astrakharchik, \textit{Ultradilute Low-Dimensional Liquids}, Phys. Rev. Lett.  \textbf{117} (2016) 100401.

\bibitem{Edmonds2023} M. Edmonds, \textit {Dark quantum droplets and solitary waves in beyond-mean-field Bose-Einstein condensate mixtures}, Phys. Rev. Research \textbf{5} (2023) 023175.

\bibitem{Parit2021} M.K. Parit, G. Tyagi, D. Singh and P.K. PanigrahiJ, \textit {Supersolid behavior in one-dimensional self-trapped Bose–Einstein condensate}, Phys. B: At. Mol. Opt. Phys. \textbf{54} (2021) 105001.

\bibitem{Mithun2020} T. Mithun and K. Kasamatsu, \textit {Modulation instability associated nonlinear dynamics of spin–orbit coupled Bose–Einstein condensates}, J. Phys. B: At. Mol. Opt. Phys. \textbf{52} (2020) 045301.

\bibitem{Sasi2023} R. Sasireka, S. Sabari, A. Uthayakumar, L. Tomio, \textit {Domain formation of modulation instability in spin-orbit-Rabi coupled Gross-Pitaevskii equation with cubic-quintic interactions}, Phys. Lett. A \textbf{480} (2023) 128987.

\bibitem{Sabari2021} S. Sabari, R. Tamilthiruvalluvar, R. Radha, \textit {Modulational instability of spin-orbit coupled Bose-Einstein condensates in discrete media}, Phys. Lett. A \textbf{418} (2021) 127696.

\bibitem{Cabrera2018} C.R. Cabrera, L. Tanzi, J. Sanz, B. Naylor, P. Thomas, P. Cheiney and L. Tarruell, \textit{Quantum liquid droplets in a mixture of Bose-Einstein condensates}, Science \textbf{359},  (2018) 301.
		
\bibitem{Agrawal2013} G.P. Agrawal, \textit{ Nonlinear Fiber Optics}, $5^{th}$ ed., Academic Press: Oxford, UK (2013).
		
\bibitem{Yu2019} W. Yu, Q. Zhou, M.  Mirzazadeh, Liu W and A. Biswas, \textit{ Phase shift, amplification, oscillation and attenuation of solitons in nonlinear optics}, J. Adv. Res. \textbf{15} (2019) 69.
		
\bibitem{Yan2021} Y.Y. Yan and W.J. Liu, \textit{Soliton rectangular pulses and bound states in a dissipative system modeled by the variable-coefficients complex cubic-quintic Ginzburg-Landau equation}, Chin. Phys. Lett. \textbf{38} (2021) 094201.
		
\bibitem{Gu1995} C. Gu , editor, \textit{Soliton Theory and Its Applications}, Berlin, Heidelberg: Springer Berlin Heidelberg, (1995).
		
\bibitem{Kuznetsov1986} E. Kuznetsov, A. Rubenchik and V. Zakharov, \textit{Soliton stability in plasmas and hydrodynamics}, Phys. Rep. \textbf{142} (1986) 103.
		
\bibitem{Congy2021} T. Congy, G. El and G. Roberti, \textit{Soliton gas in bidirectional dispersive hydrodynamics},  Phys. Rev. E. \textbf{103} (2021) 042201.
		
\bibitem{Wang2020} L.L. Wang  and W.J. Liu, \textit{Stable soliton propagation in a coupled (2+1) dimensional Ginzburg-Landau system}, Chin. Phys. B, \textbf{29} (2020) 070502.
		
\bibitem{Chai2020} X. Chai , D. Lao , K. Fujimoto, R. Hamazaki, M. Ueda and  C. Raman, \textit{Magnetic solitons in a spin-1 Bose-Einstein condensate}, Phys. Rev. Lett. \textbf{125} (2020) 030402.
		
\bibitem{Potter2012} T. Potter, \textit{ Effective dynamics for N-Solitons of the Gross-Pitaevskii equation}, Jour. of Nonlinear Sci. \textbf{22} (2012) 351.
		
\bibitem{Lannig2020} S. Lannig, C.-M. Schmied , M. Pr\"ufer, P. Kunkel, R. Strohmaier, H. Strobel,  T. Gasenzer, P. Kevrekidis and M. Oberthaler,\textit{Collisions of three-component vector solitons in Bose-Einstein condensates}, Phys. Rev. Lett. \textbf{125} (2020) 170401.
		
\bibitem{Liu2008} W.-J. Liu, B. Tian, H.-Q. Zhang, L.-L. Li and Y.-S. Xue, \textit{Soliton interaction in the higher-order nonlinear Schr\"odinger equation investigated with Hirota's bilinear method}. Phys. Rev. E, \textbf{77} (2008) 066605.
		
\bibitem{Nkenfack2024} C.E. Nkenfack, O.T. Lekeufack, R. Yamapi, E. Kengne, \textit{Bright solitons and interaction in the higher-order Gross-Pitaevskii equation investigated with Hirota's bilinear method}, Phys. Lett. A, \textbf{511} (2024) 129563.

\bibitem{Wong2015} P. Wong, L.-H. Pang, L.-G. Huang, Y.-Q. Li, M. Lei and W.-J. Liu, \textit{Dromion-like structures and stability analysis in the variable coefficients complex Ginzburg-Landau equation}, Annals of Physics \textbf{360} (2015) 341.
	
\bibitem{Khaykovich2002} L. Khaykovich, F. Schreck, G. Ferrari, T. Bourdel, J. Cubizolles, L.D. Carr, Y. Castin and C. Salomon  \textit{ Formation of a matter-wave bright soliton}, Science \textbf{296} (2002) 1290.	
		
\bibitem{Cornish2006} S.L. Cornish, S.T. Thompson and C.E. Wieman, \textit{Formation of bright matter-wave solitons during the collapse of attractive Bose-Einstein condensates}, Phys. Rev. Lett. \textbf{96} (2006) 170401.

\bibitem{Radha2010} R. Radha, V. Ramesh Kumar, and Miki Wadati, \textit{Line-soliton dynamics and stability of Bose–Einstein condensates in $(2+1)$ Gross–Pitaevskii equation}, Journal of Mathematical Physics \textbf{51} (2010) 043507.
		
\bibitem{Wan2020} H.-T. Wang and X.-Y. Wen, \textit{Soliton elastic interactions and dynamical analysis of a reduced integrable nonlinear Schr\"odinger system on a triangular-lattice ribbon}, Nonlinear Dyn. \textbf{100} (2020) 1571.
		
\bibitem{Li2021} Y.M. Li, H.M. Baskonus and A.M. Khudhur, \textit{Investigations of the complex wave patterns to the generalized Calogero-Bogoyavlenskii-Schiff equation}, Soft Comput \textbf{10} (2021) 6999.
		
\bibitem{Jena2020} R.-M. Jena, S. Chakraverty and D. Baleanu, \textit{A novel analytical technique for the solution of time-fractional Ivancevic option pricing model}, Phys. A. \textbf{550} (2020) 124380.
		
\bibitem{Su2016} C.Q. Su, Y.T.  Gao, L. Xue and Q.M. Wang, \textit{Nonautonomous solitons, breathers and rogue waves for the Gross-Pitaevskii equation in the Bose-Einstein condensate}, Commun Nonlinear Sci. Numer. Simulat. \textbf{36} (2016) 457.
		
\bibitem{Guo2020}  H. Guo H, Y.J. Wang, L.X. Wang and X.F. Zhang, \textit{Dynamics of ring dark solitons in Bose-Einstein condensates}, Acta Phys. Sin. \textbf{69} (2020) 010302.

\bibitem{Kopycinski2023} J. Kopycinski, M. Lebek, W. Gorecki and K. Pawlowski, \textit{Ultrawide Dark Solitons and Droplet-Soliton Coexistence in a Dipolar Bose Gas with Strong Contact Interactions}, Phys. Rev. Lett.  \textbf{130} (2023) 043401.

\bibitem{Saito2003} H. Saito and M. Ueda, \textit{Dynamically Stabilized Bright Solitons in a Two-Dimensional Bose-Einstein Condensate}, Phys. Rev. Lett. \textbf{90} (2003) 040403.
		
\bibitem{Sabari2017} S. Sabari, K. Porsezian, and P. Muruganandam, \textit{Dynamical stabilization of two-dimensional trapless Bose–Einstein condensates by three-body interaction and quantum fluctuations}, Chaos, Solitons and Fractals, \textbf{103}  (2017) 232.
		
\bibitem{Wang2022} H. Wang, Q. Zhou and W. Liu, \textit{Exact analysis and elastic interaction of multi-soliton for a two-dimensional Gross-Pitaevskii equation in the Bose-Einstein condensation}, Jour. Adv. Res. \textbf{38} (2022) 179.

\bibitem{Mani2024}
K. Manikandan, K. Sakkaravarthi, S. Sabari, Dispersion management and optical soliton engineering in nonuniform inhomogeneous PT-symmetric nonlinear media, Physica D: Nonlinear Phenomena {\bf 470} (2024) 134388.

\bibitem{Tamil2019a}
R. Tamilthiruvalluvar, E. Wamba, S. Sabari, K. Porsezian,  Impact of higher-order nonlinearity on modulational instability in two-component Bose-Einstein condensates, Phys. Rev. E {\bf 99} (2019) 032202.

\bibitem{Sabari2013}
S. Sabari, E. Wamba, K. Porsezian, A. Mohamadou, T.C. Kofané, A variational approach to the modulational-oscillatory instability of Bose–Einstein condensates in an optical potential, Phys. Lett. A {\bf 377} (2013) 2408.

\bibitem{Tamil2019} R. Tamilthiruvalluvar and S. Sabari, \textit{Stabilization of trapless Bose-Einstein condensates without any management}, Phys. Lett. A, \textbf{383} (2019) 2033.

\bibitem{Sabari2022} S. Sabari, O.T. Lekeufack, S.B. Yamgoue, R. Tamilthiruvalluvar and R. Radha, \textit{Role of higher-order interactions on the modulational instability of Bose-Einstein condensate trapped in a periodic optical lattice}. Int. J. Theor. Phys. \textbf{61} (2022) 222.
		
\bibitem{Sabari2020} S. Sabari, O.T. Lekeufack, R. Radha and T.C. Kofane, \textit{Interplay of three-body and higher-order interactions on the modulational instability of Bose-Einstein condensate}, J. Opt. Soc. Am. B \textbf{37}  (2020)  A54.
	
\bibitem{Qi2012} X.-Y. Qi and J.-K. Xue, \textit{Modulational instability of a modified Gross-Pitaevskii equation with higher-order nonlinearity}, Phys. Rev. E, \textbf{86} (2012) 017601.
		
\bibitem{bib47} A. Collin, P. Massignan, and C.J. Pethick, \textit{Energy-dependent effective interactions for dilute many-body systems}, Phys. Rev. A \textbf{75} (2007) 013615.

\bibitem{Tamil2018} R. Tamilthiruvalluvar, S. Sabari, K. Porsezian, \textit{Stabilization of repulsive trapless Bose–Einstein condensates}, J. Phys. B: At. Mol. Opt. Phys. \textbf{51}  (2018) 165202.
        
\bibitem{Wu2020} X.Y. Wu, B. Tian, Q.X. Qu, Y.Q. Yuan and X.X. Du, \textit{Rogue waves for a (2+1)-dimensional Gross-Pitaevskii equation with time-varying trapping potential in the Bose-Einstein condensate}, Comput. Math. Appl. \textbf{79} (2020) 1023.
		
\bibitem{bib18} R. Hirota, \textit{Exact solution of the Korteweg-de-Vries equation for multiple collisions of solitons}, Phys. Rev. Lett.\textbf{27} (1971) 1192.

\bibitem{bib10} R. Hirota, \textit{The direct method in soliton theory}, Cambridge University Press (2004).
		
\bibitem{Kengne2020} E. Kengne, \textit{Rogue waves of the dissipative Gross–Pitaevskii equation with distributed coefficients}, Eur. Phys. J. Plus {\bf 135} (2020) 622.
		
\end{thebibliography}
\end{document}